\documentclass[12pt]{iopart}

\usepackage{graphicx}
\usepackage{verbatim}
\usepackage{subfigure}
\newcommand{\eqref}[1]{{(\ref{#1})}}

\begin{document}
\title{Classical versus quantum dynamics of the atomic Josephson junction}

\author{G.J. Krahn and D.H.J. O'Dell}

\address{Department of Physics and Astronomy, McMaster University,
Hamilton, Ontario L8S 4M1, Canada}

\begin{abstract}
We compare the classical (mean-field) dynamics with the quantum dynamics of atomic Bose-Einstein condensates in double-well potentials. The quantum dynamics are computed using a simple scheme based upon the Raman-Nath equations.  Two different methods for exciting a non-equilbrium state are considered: an asymmetry  between the wells which is suddenly removed, and a periodic time oscillating asymmetry. The first method generates wave packets that lead to collapses and revivals of the expectation values of the macroscopic variables, and we calculate the time scale for these revivals.  The second method permits the excitation of a single energy eigenstate of the many-particle system, including Schr\"{o}dinger cat states. We also discuss a band theory interpretation of the energy level structure of an asymmetric double-well, thereby identifying analogies to Bloch oscillations and Bragg resonances. Both the Bloch and Bragg dynamics are purely quantum and are not contained in the mean-field treatment.
 \end{abstract}

\section{Josephson Hamiltonian}
\label{sec:JosephsonHamiltonian}
The Josephson effect is a paradigm of macroscopic quantum mechanics that first arose in the context of superconductors. Josephson \cite{b_josephson} predicted that a coherent current $I \propto \sin \phi$ 
would tunnel between two superconductors separated by a thin layer of insulator if there was a difference $\phi$ in the macroscopic quantum phase between the order parameters in the two superconducting regions (see \cite{josephson_eff} for a review). 
Josephson-type effects have also been realized in superfluid $^3$He \cite{pereverzev}, superfluid $^4$He \cite{sukhatme}, and most recently in Bose-Einstein condensates (BECs) formed in atomic vapours \cite{cataliotti01,shin04,albiez05,levy07}. Trapped atomic BECs are well suited to fundamental studies of macroscopic quantum mechanics because almost everything about them can be controlled to a very high degree e.g.\ shape of the trapping potential, interatomic interaction strength, type of measurements performed etc., see  \cite{bose_ein} for a general review.  This means that a wide range of parameter regimes can be achieved in a single experimental setup. In this paper we are interested in comparing and contrasting the `classical' regime where mean-field theory provides an excellent description and a more quantum regime where quantum fluctuations play a role.

The ac Josephson effect, which is driven by a difference in chemical potential between two sides of a tunnelling barrier, can be realized in an atomic BEC by trapping the atoms in a double-well potential. Numerous theoretical studies,  e.g.\ \cite{javanainen86,dalfovo96,jack96,smerzi97,zapata98,ruostekoski98,giovanazzi00,leggettreview01}, have analyzed this setup. Unlike the situation found in strongly interacting systems, such as superconductors or quantum liquids, the Josephson equations for BECs can be derived using the mean-field Gross-Pitaevskii theory for which the underlying microscopic hamiltonian is well understood.
Labelling the two wells by $l$ (left) and $r$ (right), the Josephson equations governing the evolution of the macroscopic phase difference $\phi \equiv\phi_{r}-\phi_{l}$ and the atom number difference $k \equiv (N_{l}-N_{r})/2$ between two weakly coupled BECs 
in a symmetric double-well potential can be written \cite{smerzi97}
\begin{eqnarray}
   \dot{\phi} & = & \frac{E_C}{\hbar}k + \frac{4 k/N^2}{\sqrt{1-4 k^2/N^2}}\frac{E_{J}}{\hbar} \cos \phi \label{phase_ev} \\
     \dot{k} & = & -\frac{E_J}{\hbar}\sqrt{1-4k^2/N^2}\sin\phi   \label{k_ev}
\end{eqnarray}
where the dot represents a time derivative and $N$ is the total number of atoms $N=N_{l}+N_{r}$.
Following the notation used for superconductors, the parameters $E_{J}$ and $E_{C}$ are called the tunnelling and charging energies, respectively.  The tunnelling energy determines the maximum current $I_{J}=E_{J}/ \hbar$. By conservation of the total number of atoms, the current obeys $I = \dot{N}_{r}=-\dot{N}_{l}=I_{J} \sqrt{1-4k^2/N^2} \sin \phi$. The charging energy arises from  interatomic interactions and, providing $k \ll N$, it can be evaluated from the chemical potential at static equilibrium as
\begin{equation}\label{Ec}
    E_C=2\left.\left(\frac{d\mu_l}{dN_l}\right)\right|_{N_l=N/2}  \ .
\end{equation}
Expressions for $E_{J}$ and $E_{c}$ in terms of microscopic quantities can be obtained from the Gross-Pitaevskii theory in both the tight-binding and Thomas-Fermi regimes \cite{giovanazzi08}.

By identifying $\phi$ and $k$ as canonically conjugate variables, Hamilton's relations
\begin{equation}\label{H_eqns}
    \dot{\phi}=\frac{1}{\hbar}\frac{\partial H}{\partial k} \quad , \quad
     \dot{k}=-\frac{1}{\hbar}\frac{\partial H}{\partial\phi}
\end{equation}
imply that the effective hamiltonian governing the dynamics of the macroscopic variables takes the form \cite{smerzi97,bose_ein}
\begin{equation}
    H=\frac{E_C}{2}k^2-E_{J}\sqrt{1-4k^2/N^{2}}\cos\phi.
    \label{Ejamp_Ham}
\end{equation}
In this paper we will concentrate on the regime where the atom number difference between the wells is always much smaller than the total atom number $k \ll N$.  Furthermore, we assume the parameters obey $E_{c} \gg E_{J}/N^2$ which can always be satisfied for a large enough total atom number providing $E_{c} \neq 0$. Under these circumstances the hamiltonian (\ref{Ejamp_Ham}) reduces to \cite{smerzi97}
\begin{equation}
    H_J=\frac{E_C}{2}k^2-E_J\cos\phi  \ .
    \label{jH}
\end{equation}
and this is the form we shall work with from now on.
This hamiltonian is analogous to that of a pendulum or, equivalently, to that of a 
classical particle moving on a sinusoidal ``washboard'' potential.  In the (former) latter case $\phi$ is the (angular) displacement and $k$ the (angular) velocity.

The relative phase $\phi$ and number difference $k$ that appear in (\ref{jH}) are classical variables in the sense that their values  are simultaneously well defined. This is
in accordance with the derivation \cite{smerzi97} of the Josephson equations  (\ref{phase_ev}) and (\ref{k_ev}) from the Gross-Pitaevskii theory which is a mean-field theory that assumes all atoms share the same macroscopic wave function (the condensate order parameter). The Gross-Pitaevskii theory has proved enormously successful as a description for atomic BECs trapped in single-well potentials. However, the double-well system provides a very simple and analytically tractable extension in which we can explore beyond mean-field effects essentially because the single-particle kinetic tunnelling/hopping energy represented by $E_{J}$ can easily be much smaller than the interaction energy $E_{c}$. Under these circumstances it is necessary to (second) quantize the Josephson hamiltonian (\ref{jH}). We do this by promoting $\phi$ and $k$ to operators which satisfy \cite{bose_ein}
\begin{equation}
[ \hat{\phi}, \hat{k} ]= \mathrm{i} \ .
\label{eq:commutator}
\end{equation} 
In the $\phi$-representation where $\hat{k}=- \mathrm{i} \, \mathrm{d} / \mathrm{d} \phi$ the quantum version of the Josephson hamiltonian is
\begin{equation}
    \hat{H}_J=-\frac{E_C}{2}\frac{\mathrm{d}^2}{\mathrm{d} \phi^2}-E_J\cos\phi \ .
    \label{qjH}
\end{equation}
To complete the quantization we also need to stipulate that the wave function $\psi(\phi)$ that the hamiltonian (\ref{qjH}) acts on is single valued, i.e.\ is periodic $\psi(\phi+2\pi)=\psi(\phi)$. This ensures that the eigenvalues of $\hat{k}$ are integers and is in accordance with the notion that $\phi$ is a phase. The difference between the dynamics generated by the classical \eqref{jH} and quantum \eqref{qjH} hamiltonians is the main theme of this paper.

The requirement that the wave function be $2 \pi$-periodic is very natural from the point of view of the pendulum analogy. However, from the point of view of the particle in a washboard potential analogy it is a more restrictive condition. The time-independent Schr\"odinger equation associated with the hamiltonian (\ref{qjH}) is the Mathieu equation $(\hat{H}_{J}-E)\Psi=0$ \cite{mathfunctions}.  According to the Floquet-Bloch theorem the general solutions of the Mathieu equation can be written $\psi(\phi)=\exp (\mathrm{i}q\phi) U_{q}(\phi)$ where $q$ is the quasi-momentum. However, because our wave function is $2\pi$-periodic we must set the quasi-momentum $q$ to zero. Thus, it appears that analogues of a number of phenomena familiar from the physics of waves in periodic potentials, such as Bragg scattering and Bloch oscillations,  must be absent from the Josephson problem because they require finite values of the quasi-momentum. On the contrary, we shall see in Sections \ref{sec:bragg} and \ref{sec:band} that there is a sense in which we can achieve finite $q$ values and hence realize Bragg scattering and Bloch oscillation analogues in double-well systems. 

The validity of the quantization procedure given above to obtain the hamiltonian (\ref{qjH}) is actually far from obvious \cite{leggettleshouches}. The original $N$-atom double-well system corresponds to a quantum many-body system which is then approximated by a Gross-Pitaevskii mean-field theory to give the Josephson equations (\ref{phase_ev}) and (\ref{k_ev}). The system is then re-second quantized by quantizing the mean-field theory to give the hamiltonian (\ref{qjH}). However, it turns out that in the regime $k \ll N$ and $E_{c} \gg E_{J}/N^2$ the hamiltonian (\ref{qjH}) agrees with that obtained from a treatment based upon the fully quantum Bose-Hubbard model, see, e.g.\ \cite{gati2}. 

The organization of the rest of this paper is as follows. In section \ref{solveJH} we introduce the Raman-Nath equations which provide a simple framework for calculations 
associated with the quantum Josephson problem. In sections 
\ref{apot} and \ref{sec_raman_asym} we generalize our treatment to include an asymmetry between the two wells. In section \ref{sec:classicalvsquantum} we compare the quantum and classical dynamics of the Josephson junction following an excitation created by taking a system which is at equilibrium in an asymmetric double-well and suddenly removing the asymmetry, which is the current standard experimental probe. Section \ref{sec:timemodulated} treats double-wells that have an asymmetry that is modulated periodically in time and Section \ref{sec:bragg} examines tunnelling resonances that occur at certain values of the asymmetry that are analogous to Bragg scattering. In Section \ref{sec:band} we present an analysis of the asymmetric double-well problem based on band structure theory and relate adiabatic sweeps of the asymmetry to Bloch oscillations. We also suggest a way to generate Schr\"{o}dinger cat states.

\section{Raman-Nath equation for a BEC in a double-well potential}\label{solveJH}

The $2 \pi$-periodic wave function $\Psi(\phi,t)$ which determines the values of the macroscopic variables $\phi$ and $k$ can be expanded as a Fourier series
\begin{equation}
\Psi(\phi,t)=\frac{1}{\sqrt{2\pi}}\sum_{n=-\infty}^{\infty}A_n(t)\exp{(\mathrm{i}n\phi)}
\label{periodicwavefunction}
\end{equation}
where the `plane wave' basis states $\exp{(\mathrm{i}n\phi)}/\sqrt{2 \pi}$ are eigenfunctions of the number difference operator $\hat{k}$ with integer eigenvalues $n$. The probability amplitudes $A_{n}$ obey the normalization $\sum_n|A_n|^2=1$. 
In the original problem with $N$ atoms (where without loss of generality we take $N$ to be even) the integer $n$ must lie in the range $-N/2 \le n \le N/2$, but since we are working in the regime where the number difference is always small in comparison to $N$ the amplitudes $A_{n}$ become vanishingly small long before $\vert n \vert = N/2$ and so we have extended the sum in (\ref{periodicwavefunction}) to $\pm \infty$. Substituting (\ref{periodicwavefunction}) into the Schr\"{o}dinger equation associated with the Josephson Hamiltonian (\ref{qjH}) 
\begin{equation}
\mathrm{i}\hbar \frac{\partial}{\partial t} \Psi(\phi,t) = \left[ -\frac{E_C}{2}\frac{\mathrm{d}^2}{\mathrm{d} \phi^2} - E_J\cos(\phi) \right] \Psi(\phi,t) 
\label{eq:schrodinger}
\end{equation}
yields the coupled infinite set of Raman-Nath (RN) differential-difference equations
\begin{equation}
\mathrm{i} \frac{\mathrm{d}}{\mathrm{d} \tau} A_{n}(\tau)= n^{2} A_{n}(\tau)-\frac{\Lambda}{2}\left[A_{n+1}(\tau)+A_{n-1}(\tau)\right]  \ .
\label{eq:TDRN}
\end{equation}
We have written the RN equations in dimensionless form by defining $\tau \equiv E_{c} t/2 \hbar$ and the ratio of the energy parameters as $\Lambda$
\begin{equation}
\Lambda \equiv\frac{2E_J}{E_C}.
\label{eq:lambdadef}
\end{equation}
The RN equations first arose in the context of the dynamical diffraction of light by ultrasonic waves in fluids \cite{RN35,berry66} and have subsequently found extensive use in the description of the diffraction of atoms by standing-waves of light \cite{martin87}. By numerically integrating in time a suitably truncated set of RN equations they provide a simple scheme for computing the dynamics of the macroscopic variables of the double-well problem. A novel feature of the atomic BEC realization of the Josephson junction is that great control can be exerted over $\Lambda$. Either by adjusting the intensity of the laser which forms the central tunnelling barrier, or by using a Feshbach resonance to manipulate the interactions, the magnitude of $\Lambda$ can be varied between essentially zero and infinity.  These experimental `knobs'  can also be turned during the course of an experiment leading to a time-dependent $\Lambda$ which can also be easily handled within the RN framework \eqref{eq:TDRN}.

An alternative method for describing the dynamics of the macroscopic variables of the double-well problem is to expand $\Psi(\phi,t)$ in terms of the eigenfunctions of the Josephson hamiltonian. Let 
\begin{equation}
\Psi(\phi,t)=\sum_{j} \alpha_{j} \Psi^{j}(\phi,t)=  \sum_{j} \alpha_{j} \psi^{j}(\phi) \exp(-\mathrm{i}\epsilon^jt/\hbar)
\label{eq:eigenfunctionexpansion}
\end{equation}
where $\psi^{j}(\phi)$ is the $j$th eigenfunction and has an energy $\epsilon^j$. Defining the scaled energy $E^{j} \equiv 2 \epsilon^{j} / E_{c}$, the eigenstates $\psi^{j}$ obey a Helmholtz equation which has the same general form as the Mathieu equation
\begin{equation}\label{bloch1}
\left[-\frac{\mathrm{d}^2}{\mathrm{d} \phi^2}-
            \Lambda\cos(\phi)\right] \psi^j=E^j\psi^j.
\end{equation}
Similarly to above, we can expand the eigenfunctions in a number state basis
\begin{equation}\label{wavefunction}
\psi^j(\phi)=\frac{1}{\sqrt{2\pi}}\sum_{n=-\infty}^{n=\infty}A^j_{n}\exp{(\mathrm{i}n\phi)}
\end{equation}
which leads to the coupled set of time-independent RN equations
\begin{equation}
E^jA_n^j=n^2A_n^j-\frac{\Lambda}{2}(A_{n+1}^j+A_{n-1}^j).
\label{recurrence}
\end{equation}
The time-independent RN equations thus take the form of recurrence relations describing a tridiagonal matrix whose $j$th eigenvalue $E^{j}$ corresponds to a column eigenvector  $\{ \ldots A_{-3}^{j},A_{-2}^{j},A_{-1}^{j},A_{0}^{j},A_{1}^{j},A_{2}^{j},A_{3}^{j} \ldots \}$ made up of specific values of the amplitudes $A^{j}_{n}$. 
It is, of course, exactly the recurrence relation obeyed by the Fourier components of the even and odd Mathieu functions $ce_{r}(\phi)$ and $se_{r}(\phi)$, respectively \cite{mathfunctions}. We shall stick to the notation $\psi^{j}(\phi)$ to cover both $ce_{r}(\phi)$ and $se_{r}(\phi)$ so that $\psi^{0}(\phi)=ce_{0}(\phi)$,  $\psi^{1}(\phi)=se_{1}(\phi)$,  $\psi^{2}(\phi)=ce_{1}(\phi)$,  $\psi^{3}(\phi)=se_{2}(\phi)$ etc. As is to be expected from a problem involving a one dimensional wave equation, the eigenfunctions alternate in parity as one goes up in energy (i.e.\ increases the index $j$), with the ground state even.
The Fourier space eigenvectors $\{ \ldots A_{-3}^{j},A_{-2}^{j},A_{-1}^{j},A_{0}^{j},A_{1}^{j},A_{2}^{j},A_{3}^{j} \ldots \}$ are localized in $k$-space, i.e.\ $A_{n}^{j} \rightarrow 0$ for large enough $n$. In fact, the decay of $A_{n}^{j}$ with $n$ is exponentially fast for large $n$ \cite{odell01}. Comparison with the tridiagonal matrix representing the exact Bose-Hubbard hamiltonian \cite{gati2} shows that corrections to (\ref{recurrence}) are of order $1/N$ and so become negligible as $N \rightarrow \infty$. 

From Equation (\ref{bloch1}) we see that in the scaled variables the parameter $\Lambda$ determines the height of the sinusoidal potential and thus the separatrix for the classical motion. The separatrix divides classical phase space into two qualitatively different types of motion. When the energy $E^{j}$ is smaller than $\Lambda$ then in the classical pendulum analogy we have librational motion meaning that the phase $\phi(t)$ and angular velocity $k(t)$ oscillate and so periodically reverse their signs. In the classical particle in a sinusoidal potential analogy this corresponds to the particle having an energy less than the barrier tops and so rolling around within a single well of the sinusoidal potential. In the physical double-well problem librational motion corresponds to Josephson plasmon excitations \cite{levy07}.
When $E^{j}$ is greater than $\Lambda$ then in the classical pendulum analogy we have rotation meaning that the phase $\phi(t)$ continuously winds up in only one direction (the phase lives on a torus with $\phi=-\pi$ and $\phi= \pi$ identified so that it lies in the range $-\pi < \phi \le \pi$) and $k$ does not reverse its sign. In the classical particle in a sinusoidal potential analogy the particle has enough energy to roll into the neighbouring wells of the sinusoidal potential. In the physical double-well problem rotational motion is known as macroscopic quantum self-trapping \cite{smerzi97} and corresponds to motion in which the system is locked in a state with a larger number of particles in one well despite the symmetry of the potential. However, quantizing the system means that quantum tunnelling though the sinusoidal potential barriers ($E^{j} < \Lambda$), and quantum reflection above the barrier ($E^{j}> \Lambda$) blurs the distinction between libration and rotation. We thus expect motion near the separatrix to be one place where quantum effects are particularly visible.

\begin{figure}[t]
\centering
\subfigure[]{\includegraphics[width=0.49\columnwidth]{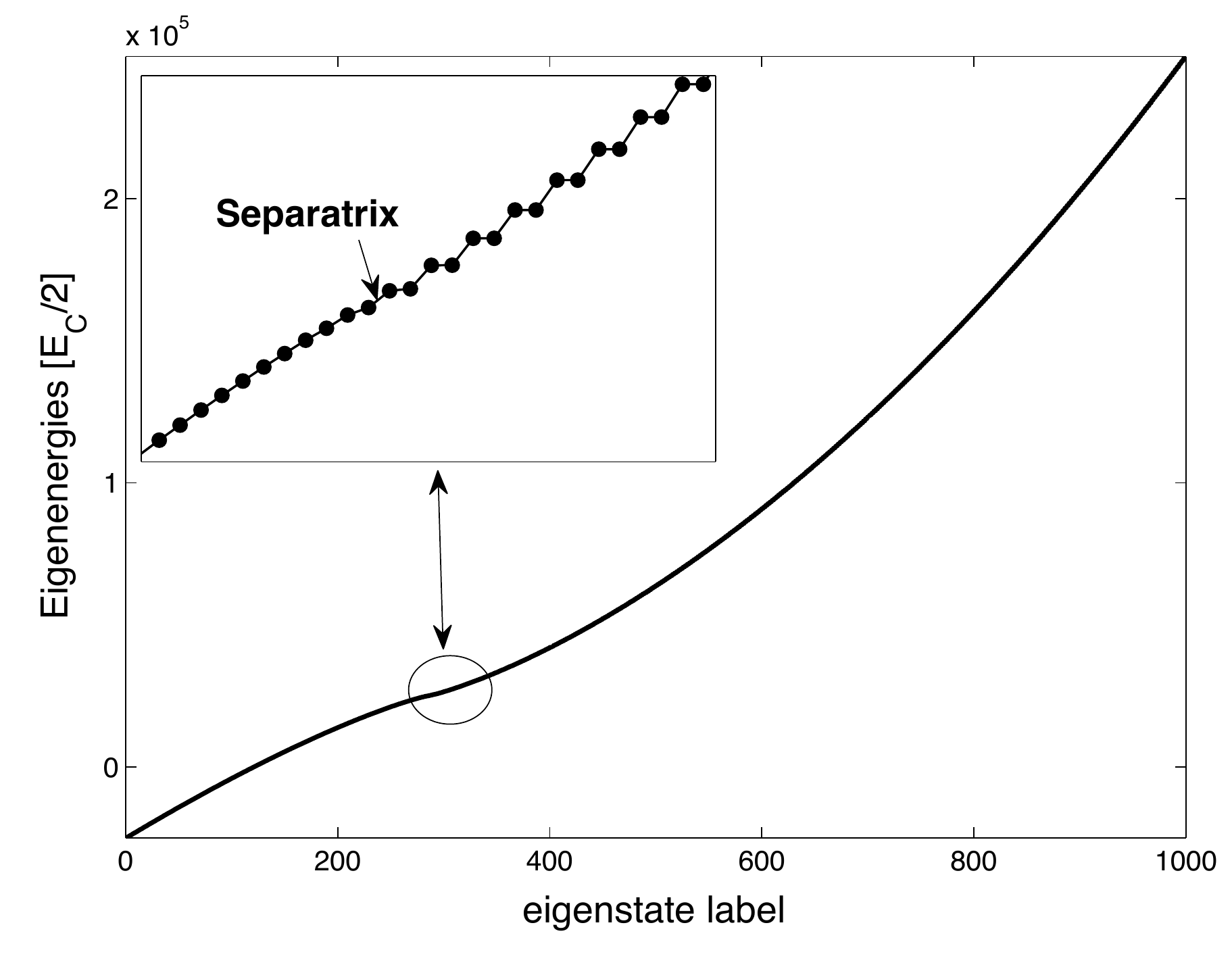}}
\subfigure[]{\includegraphics[width=0.49\columnwidth]{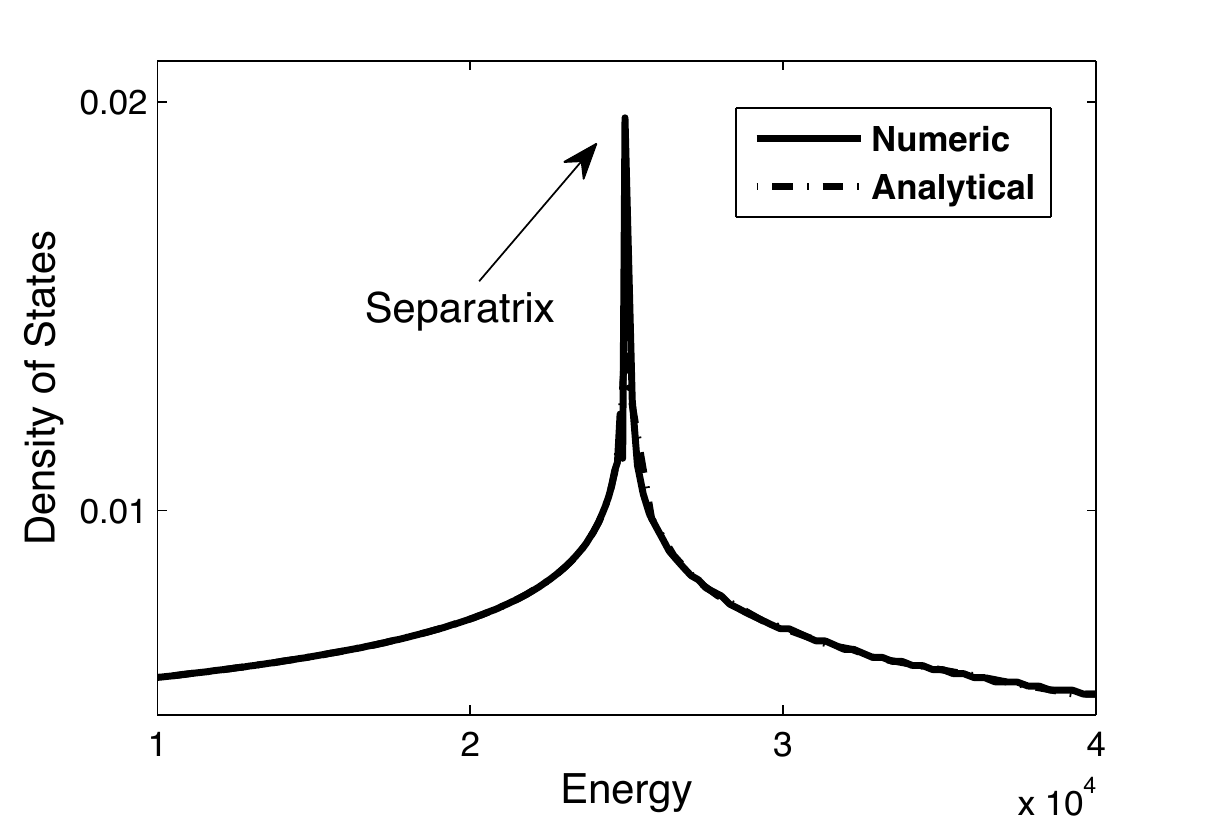}}
\caption{\scriptsize{$\Lambda=25000$.  (a) The energy eigenstates $E^j$ versus the eigenstate number $j$.  The energy spectrum is linear for energy states far below the classical separatrix (which lies inbetween eigenstates 285 and 286). (b) The density of states as a function of energy.  We show both the exact numerical result and the approximate analytic result given in \eqref{dstates_an1}. The agreement is excellent except very near the separatrix.  Both predict a sharp peak in the density of states at the separatrix.}}\label{eigenstates_sep}
\end{figure}

As $\Lambda$ is the only parameter in our hamiltonian, its magnitude plays an important role. In the literature three regimes are usually identified: (1) the Rabi regime $\Lambda \gg N^2$, (2) the Josephson regime  $N^2 \gg \Lambda \gg 1$, and (3) the Fock regime $\Lambda \ll 1$. In particular, the Rabi and Fock regimes correspond to the non-interacting and interaction-dominated limits, respectively, and the Josephson regime lies in between, see \cite{leggettreview01,bose_ein,gati2} for more discussion. In this paper our assumption $E_{c} \gg E_{J}/N^2$ excludes the Rabi regime. In fact, for the most part we shall be in the Josephson regime, with the exception of the discussion of the Bragg scattering and Bloch oscillation analogies in Sections \ref{sec:bragg} and \ref{sec:band} which concern the border between the Josephson and Fock regimes where $\Lambda \le 1$. Having reduced the problem to just two regimes,  from now on we take the view that $\Lambda$  \emph{plays a role analogous to the dimensionless ratio of the classical action to Planck's constant}. More precisely, $\Lambda=2 E_{J}/E_{c}=(S/\hbar)^2$, where $S$ is a constant having the units of action. The $\hbar^2$ comes from the kinetic energy $E_{c}$. With this identification we see that $\Lambda$ determines how quantum the system is:
\begin{itemize}
\item{\textbf{$\Lambda$ small.}} In this case the system may be viewed as being in a very quantum regime in the sense that there are only a few quantum states below the separatrix, i.e.\ `trapped' in the sinusoidal well. Only a small tridiagonal matrix (\ref{recurrence}) is required to capture the states having energies up to the separatrix and  the classical Josephson equations (\ref{phase_ev}) and (\ref{k_ev}) are expected to give a poor description of the dynamics below the separatrix. 
\item{\textbf{$\Lambda$ large.}} When $\Lambda$ is large we enter the semiclassical limit where there are many eigenstates inside the sinusoidal well and a large tridiagonal matrix is required. The classical Josephson equations are expected to give a good description of the dynamics (except close to the separatrix). 
\end{itemize}
In the semiclassical limit a large tridiagonal matrix (\ref{recurrence}) is required to capture the eigenstates up to the separatrix. An estimate of the required minimum dimensions $\mathcal{N} \times \mathcal{N} $ of the matrix size is given by \cite{odell01}
\begin{equation}
\mathcal{N}=\sqrt{2 \Lambda} \ .
\end{equation}
Eigenstates well below the separatrix (which are localised near the bottom of the sinusoidal wells) have a linear energy spectrum like the harmonic oscillator  (see Figure \ref{eigenstates_sep}). These states are known as Josephson plasmon excitations and occur at integer multiples of the energy \cite{bose_ein}
\begin{equation}
\hbar \omega_{\mathrm{pl}}=\sqrt{E_{c}E_{J}} \ . 
\label{eq:plasmafrequency}
\end{equation}
As shown in the inset in Figure \ref{eigenstates_sep}(a), above the separatrix the eigenstates rapidly tend to degenerate pairs.   Indeed, well above the separatrix (as $j \rightarrow \infty$) the sinusoidal potential becomes irrelevant and the hamiltonian tends to that of the quantum rotor. One of the eigenstates in each pair is $ce_{j}$ and has even parity and the other is $se_{j}$ and has odd parity: they are approximately the ($\pm$) superpositions of clockwise and anticlockwise rotor states.
Nevertheless, the sinusoidal potential does lead to a small energy splitting between the two states of each pair that scales as  $E^{j+1}-E^{j} =O(\Lambda^{j}/j^{j-1})$ as $j \rightarrow \infty$ \cite{mathfunctions}. Taking each pair as a single unit, the spectrum of the units is quadratic as $j \rightarrow \infty$ as expected for the quantum rotor.
 
 The energy spectrum flattens out near the separatrix (see Figure \ref{eigenstates_sep}).
   This can be understood in terms of the divergence of the period of the classical motion at the separatrix which in turn causes the density of states ($\propto 1/\mathrm{period}$) to have a peak there \cite{farrar}. The density of states can be calculated from the numerical eigenvalues as $D(E^j)=1/|E^{j+1}-E^{j}|$ and analytic expressions can be derived using Bohr-Sommerfeld quantization. The expressions valid below and above the separatrix are, respectively \cite{odell,hooley,farrar}
\begin{equation}\label{dstates_an1}
D(E)=\frac{1}{\pi}\sqrt{\frac{2}{\Lambda}}K\left(\frac{E+\Lambda}{2\Lambda}\right), \indent
D(E)=\frac{2}{\pi}\sqrt{\frac{1}{E+\Lambda}}K\left(\frac{2\Lambda}{E+\Lambda}\right),
\end{equation}
where K(x) is the complete elliptic integral of the first kind \cite{mathfunctions}.  As can be seen in Figure \ref{eigenstates_sep}, the numerical and analytic expressions are in excellent agreement except right at the separatrix where the Bohr-Sommerfeld method breaks down. Although we shall not make use of them here, analytic solutions to the RN equations in the semiclassical limit are available \cite{odell01}. These are based on uniform approximations that are valid right through the separatrix and so go beyond
WKB/Bohr-Sommerfeld quantization.

\section{Asymmetric double-well potential}
\label{apot}

In order to investigate Josephson oscillations in the double-well potential it is necessary to first excite the system into a non-equilibrium state so that its subsequent dynamics can be observed. One way to do this is to start from an equilibrium state in a slightly asymmetric (tilted) double-well potential and then to suddenly make the potential symmetric. The two key experiments \cite{albiez05} and \cite{levy07} have both used this method and we shall model this situation in this section. 

The Gross-Pitaevskii equation for a BEC in an asymmetric double-well leads to the following Josephson equations \cite{smerzi97}
\begin{eqnarray}\label{coupled_eqnsphi}
    \hbar\dot{\phi}&=&E_Ck+E_J\frac{4k}{N^2}\frac{1}{\sqrt{1-4k^2/N^2}}\cos{\phi}+\Delta\epsilon,  \\
    \hbar\dot{k}&=&-E_J\sqrt{1-4k^2/N^2}\sin{\phi}
    \label{coupled_eqnsk}
\end{eqnarray}
where $\Delta\epsilon$ is the difference between the zero-point energies of the two wells, i.e.\ magnitude of the tilt.
The equilibrium state is defined by $\dot{\phi}=0$ and $\dot{k}=0$,  and so from (\ref{coupled_eqnsk}) we see that the equilibrium phase difference in the asymmetric potential is still zero: $\phi_{\mathrm{eq}}=0$. However, there will be an unequal number of atoms on the two sides, i.e. $k_{\mathrm{eq}} \neq 0$, with more atoms sitting in the lower well. When the potential is suddenly changed to being symmetric our initial conditions are therefore $\phi \vert_{t=0}=0$ and $k \vert_{t=0} \neq 0$. In the pendulum analogy this corresponds to the pendulum starting at the instant where it is pointing vertically downwards but with a finite angular velocity. 

\begin{figure}[t]
\centering
\subfigure[]{\label{keq_graph1}\includegraphics[width=0.45\columnwidth]{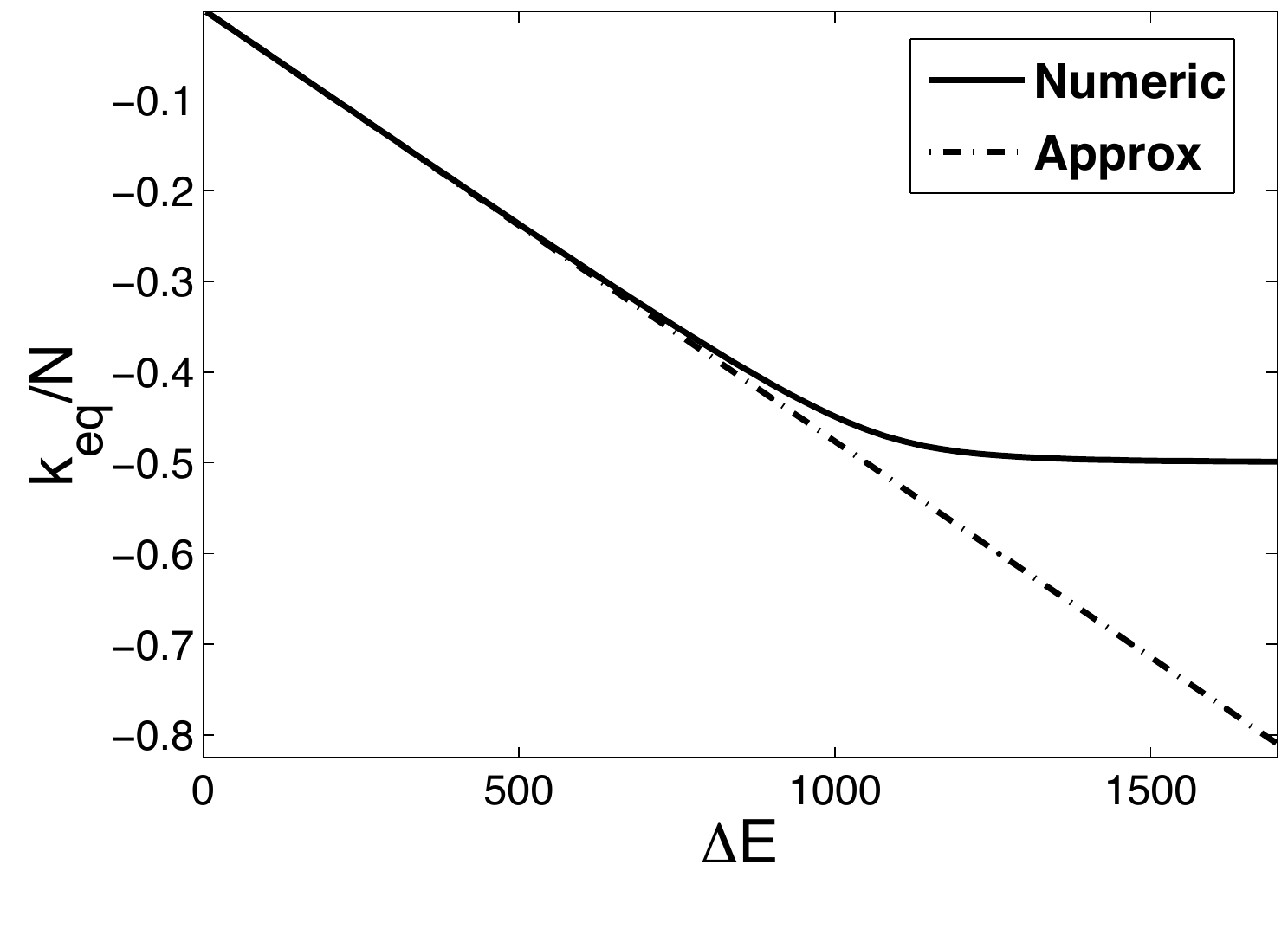}}
\subfigure[]{\label{keq_graph2}\includegraphics[width=0.45\columnwidth]{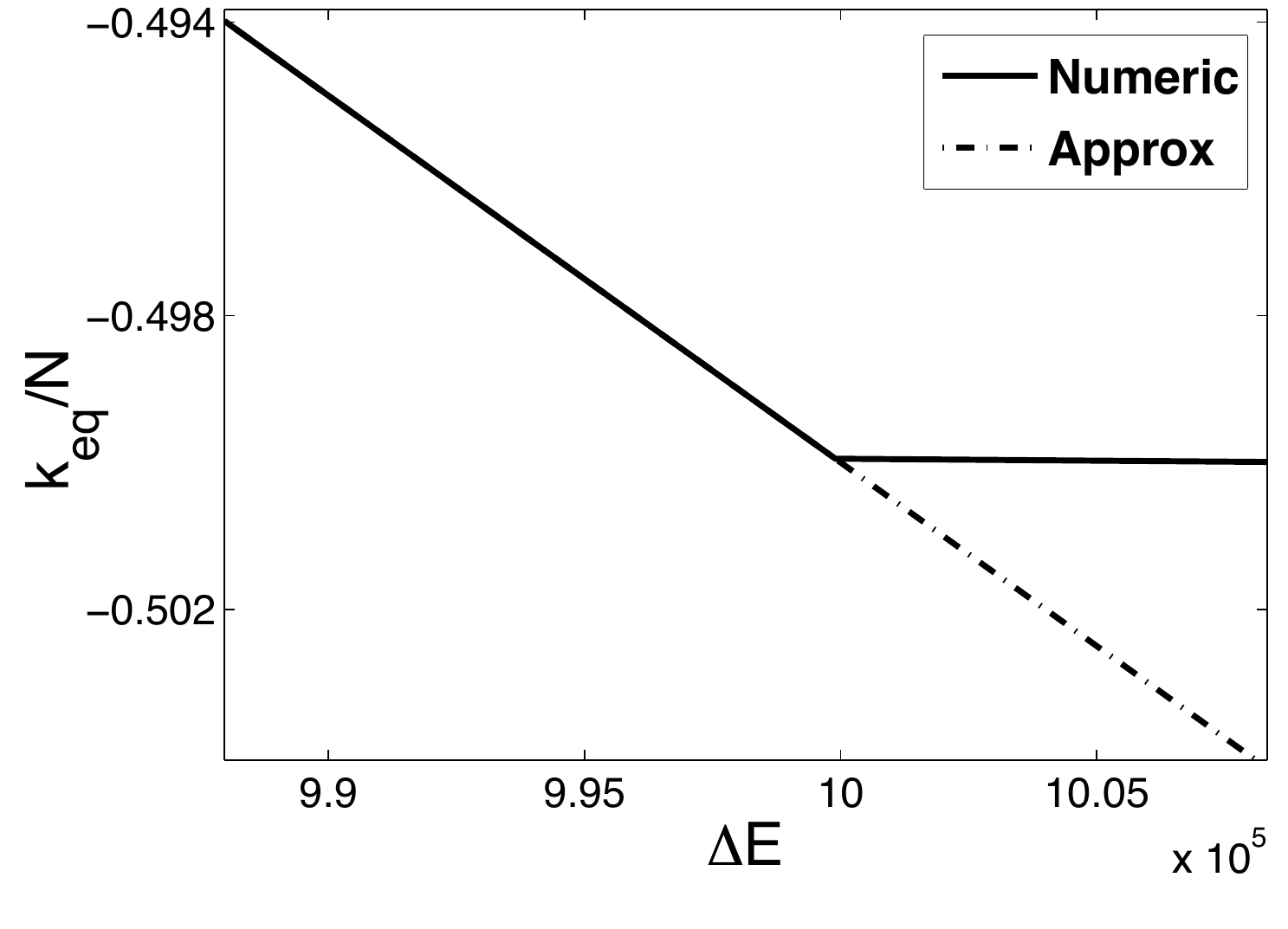}}
\caption{\scriptsize{$\Lambda=25000$. Here we plot the equilibrium value of the relative population imbalance as a function of the energy asymmetry.  The exact result from solving \eqref{keq1} numerically is compared with the approximate result given by Equation \eqref{keqf}.  (a) The number of particles is $N=10^3$.  The agreement is reasonable until $\Delta E\approx N$.  (b) $N=10^6$. The agreement is extremely good until $\Delta E=N$.}}
  \label{keq_graph}
\end{figure}

The equations of motion (\ref{coupled_eqnsphi}) and (\ref{coupled_eqnsk}) are generated by the classical hamiltonian
\begin{equation}\label{jHaa}
    H_{a}=\frac{E_C}{2}k^2-E_J\sqrt{1-\frac{4k^2}{N^2}}\cos\phi+\Delta\epsilon \, k \ .
\end{equation}
In the regime $k/N\ll1$ and $E_{c} \gg E_{J}/N^2$ this takes the simplified form
\begin{equation}\label{jHa}
    H_{Ja}=\frac{E_C}{2}k^2-E_J\cos\phi+\Delta\epsilon  \, k
\end{equation}
which will be referred to as the asymmetric Josephson hamiltonian.  If the asymmetry is too large we risk violating the condition $k/N\ll1$ even for the equilibrium state. It is therefore important to establish this extra condition of validity upon the hamiltonian (\ref{jHa}). At equilibrium Equation  (\ref{coupled_eqnsphi}) becomes
\begin{equation}\label{keq1}
    \left(1-\frac{4k_{\mathrm{eq}}^2}{N^2}\right)\left(\frac{\Delta E}{2}+k_{\mathrm{eq}}\right)^2=4\Lambda^2\left(\frac{k_{\mathrm{eq}}}{N^2}\right)^2
\end{equation}
where $\Delta E=2\Delta\epsilon/E_C$ is the dimensionless tilt asymmetry parameter. Assuming that $k_{\mathrm{eq}}/N$ is small we can expand as
\begin{eqnarray}\label{keq2}
    \left(\frac{\Delta E}{2}+k_{\mathrm{eq}}\right)&=&-2\Lambda\frac{k_{\mathrm{eq}}}{N^2}\left(1+\frac{1}{2}\cdot\frac{4k_{\mathrm{eq}}^2}{N^2}+\cdots\right) 
     \approx -2\Lambda\frac{k_{\mathrm{eq}}}{N^2} \ .
\end{eqnarray}
This gives
\begin{equation}\label{keqf}
    \frac{k_{\mathrm{eq}}}{N} \approx -\frac{\Delta E}{2}\frac{N}{2\Lambda+N^2}  \quad\quad (\mbox{when} \ k_{\mathrm{eq}}/N \ll 1) \ .
\end{equation}
In Figure \ref{keq_graph} we compare \eqref{keqf} with the exact result obtained by numerically solving \eqref{keq1}.  We see that the two are in excellent agreement almost all the way up to $\Delta E = N$ which is the saturation point where all $N$ atoms have moved into a single well. Our earlier assumption $N^2 \gg \Lambda$ (exclusion of the Rabi regime) means that we can further approximate (\ref{keqf}) as
\begin{equation}\label{keqs}
    k_{eq} \approx -\frac{\Delta E}{2} \ .
\end{equation}
We therefore see that the extra condition that the pendulum hamiltonian \eqref{jHa} is valid in the asymmetric case is that $\Delta E  \ll N$. Note that the result (\ref{keqs}) is actually the exact prediction given by the pendulum hamiltonian \eqref{jHa}.

\section{Raman-Nath equation for an asymmetric double-well}
\label{sec_raman_asym}

\begin{figure}[t]
\centering
\includegraphics[width=0.90\columnwidth]{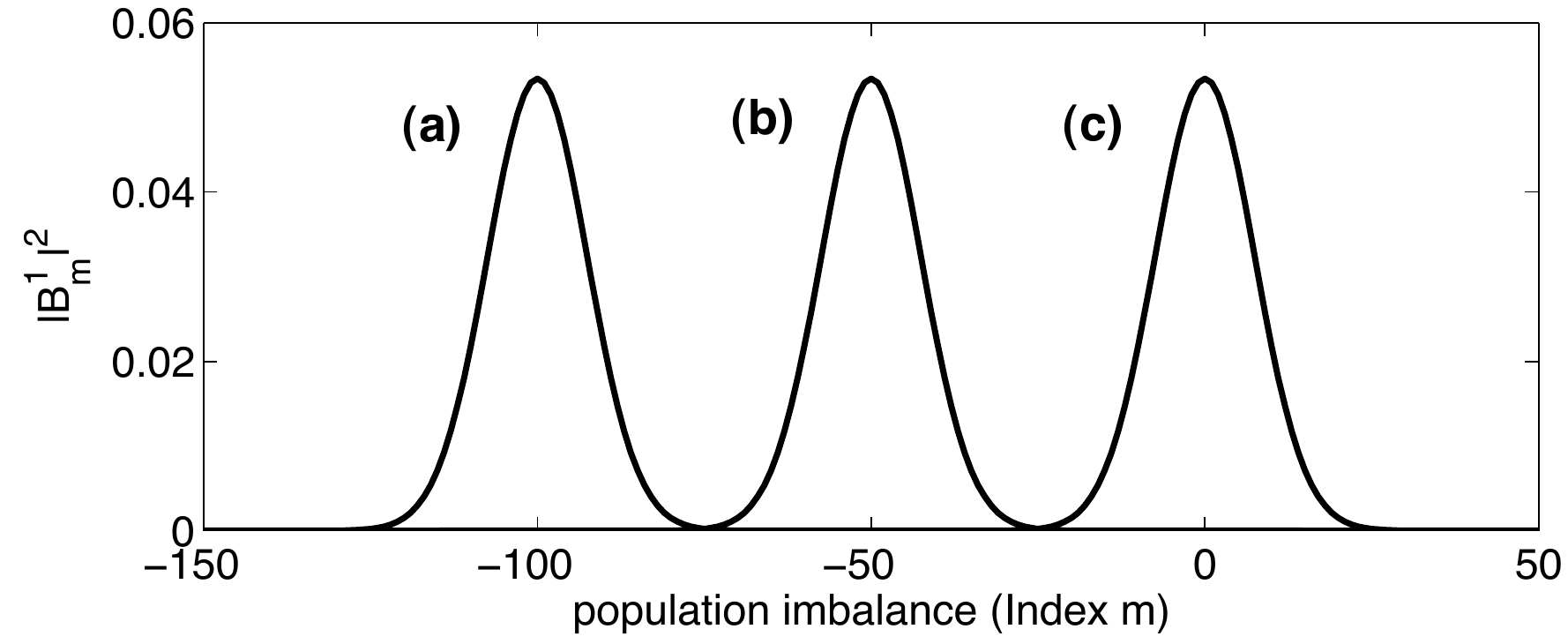}
\caption{\scriptsize{$\Lambda=25000$.  The ground state probability densities (as a function of population difference) for three different asymmetric double-well potentials:
$\Delta E=200,100,$ and $0$ for (a), (b), and (c) respectively.  For zero asymmetry the expectation value of $m$ is zero meaning the analogue quantum pendulum will remain motionless (apart from zero-point fluctuations).  When the asymmetry takes some finite value the system has the greatest probability of being found in a state where $m$ is equal to $-\Delta E/2$, which is the same as the classical prediction \eqref{keqs}. }}
\label{graph_dynam}
\end{figure}

\begin{figure}[t]
\centering
\subfigure[]{\includegraphics[width=0.32\columnwidth]{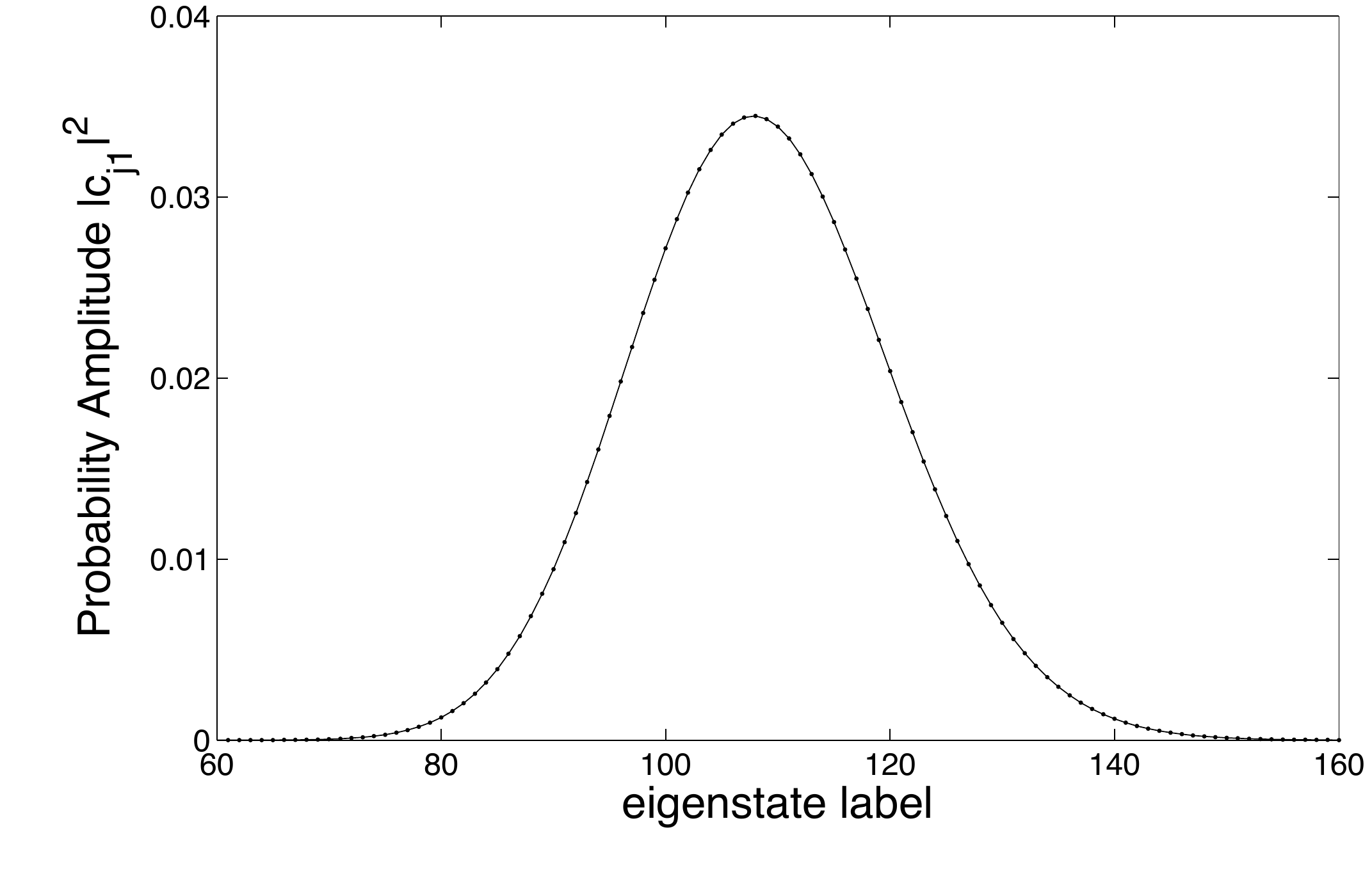}}
\subfigure[]{\includegraphics[width=0.32\columnwidth]{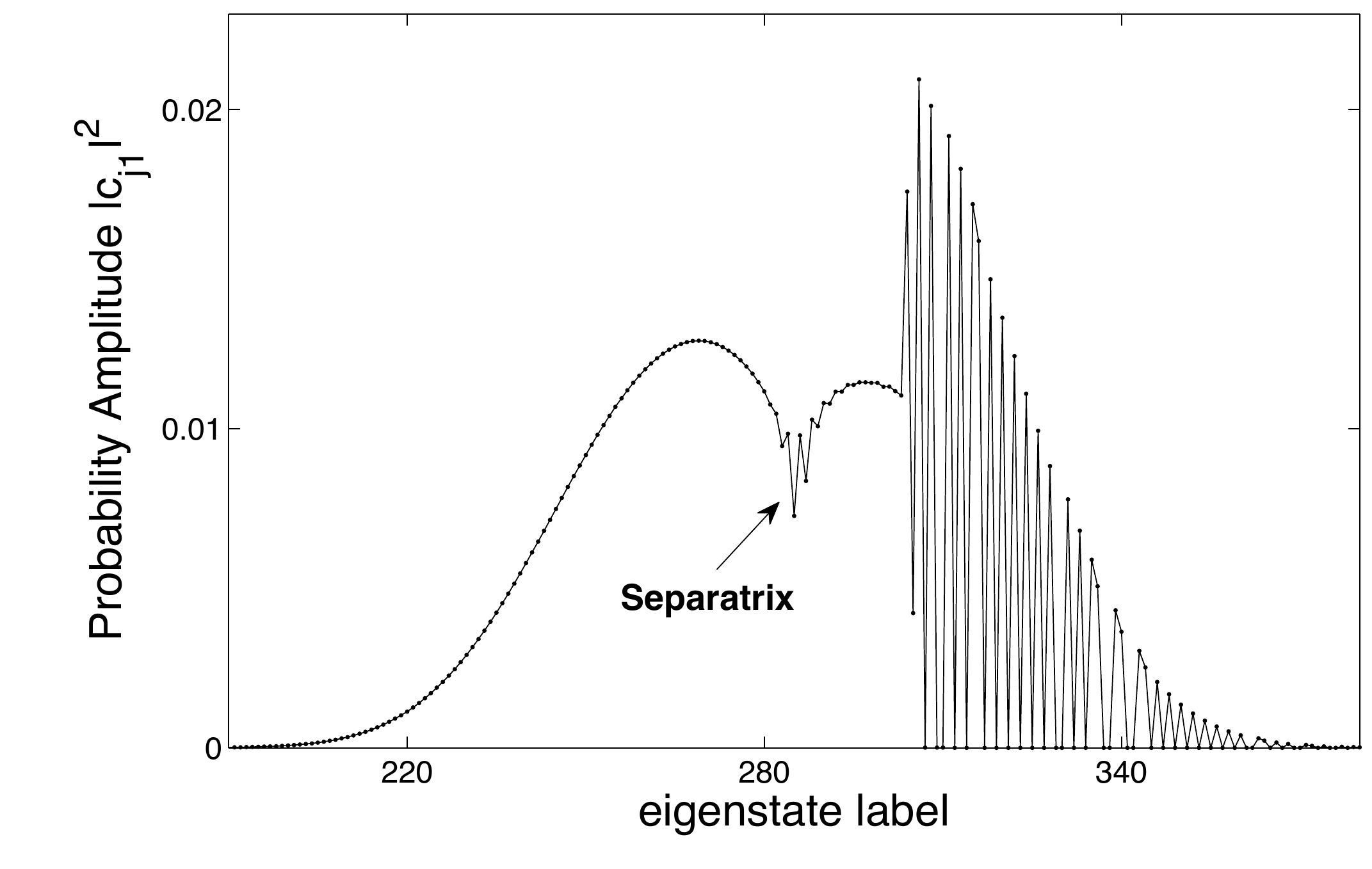}}
\subfigure[]{\includegraphics[width=0.32\columnwidth]{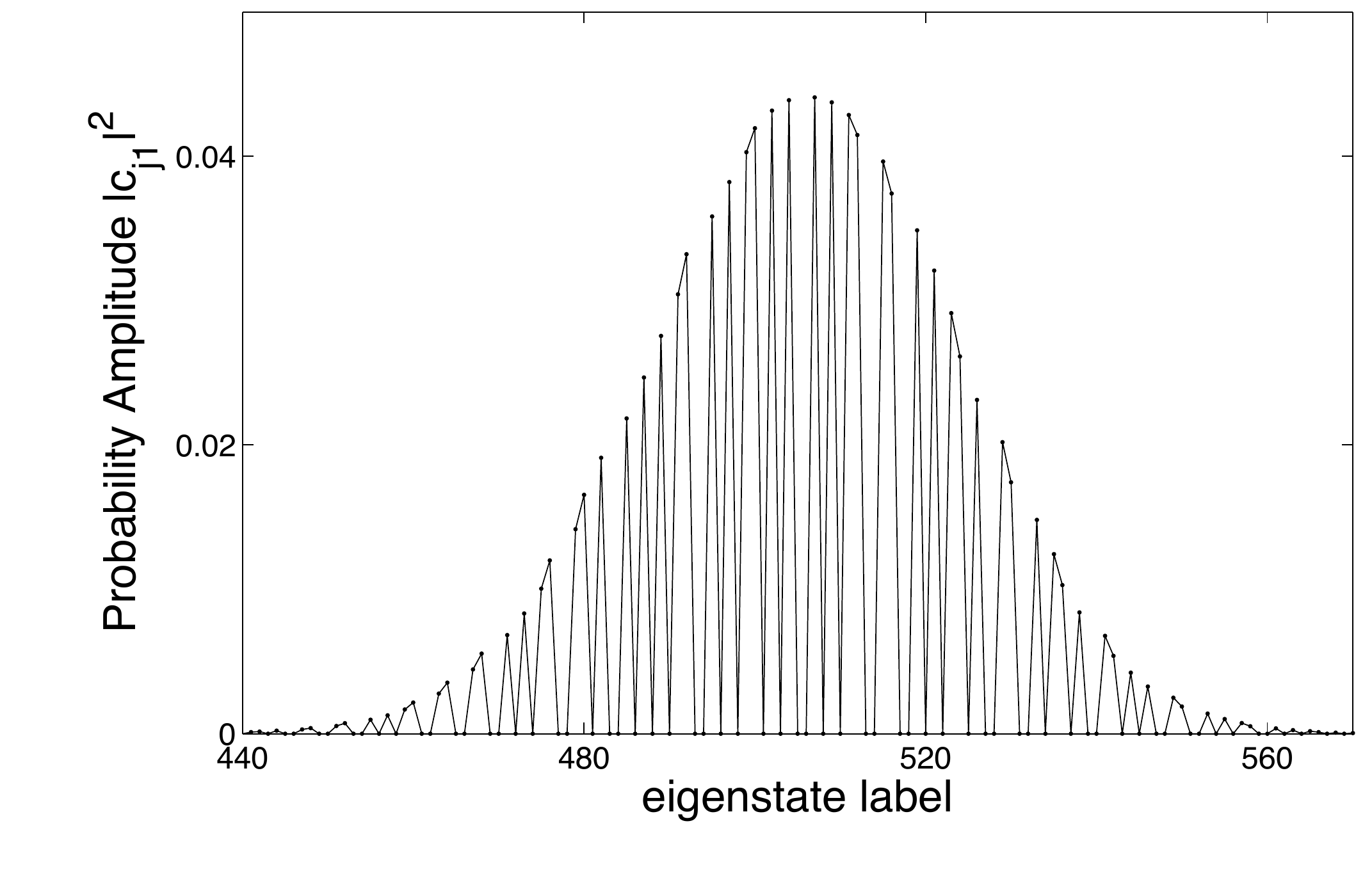}}
\caption{\scriptsize{$\Lambda=25000$.  
The initial state expressed in terms of the eigenstates of the symmetric double-well hamiltonian for 3 different energy shifts, i.e.\ immediately following `symmetrization' of the wells. (a) $\Delta E=400\rightarrow E_x=80.2$.  This corresponds to exciting a set of eigenstates entirely below the separatrix.  (b) $\Delta E=445\rightarrow E_x=99.2$. This corresponds to exciting a set of eigenstates in both regimes centered around the separatrix.  (c) $\Delta E=600\rightarrow E_x=180.4$. This corresponds to exciting a set  of eigenstates above the separatrix.}}\label{exp_coeff}
\end{figure}

The wave function for the asymmetric double-well is still $2\pi$-periodic and so for a given value of $\Delta E$ we expand
the $j$th eigenstate of the system as
\begin{equation}\label{asymm_state}
\psi^{j}_{a}=\frac{1}{\sqrt{2\pi}}\sum_mB_m^{j}\exp({im\phi})
\end{equation}
where the subscript $a$ denotes ``asymmetric''.
The time-independent Schr\"{o}dinger equation with the Hamiltonian \eqref{jHa} gives \begin{equation}\label{rec_rel2}
E^j_{a} B_m^{j}=(m^2+\Delta E \ m)B_m^{j}-\frac{\Lambda}{2}(B_{m+1}^{j}+B_{m-1}^{j}).
\end{equation}
Figure \ref{graph_dynam} plots the ground states for various values of $\Delta E$.  An asymmetry produces a non-zero expectation value for the population imbalance, as we saw classically.  Suddenly switching off the asymmetry \cite{gati2} propels the system into motion and from the perspective of the macroscopic quantum mechanical variables we assume that this process can be modelled by a projection of the equilibrium quantum state in the asymmetric potential onto the eigenstates of the symmetric potential. 
For this purpose it is useful to relate the two sets of eigenstates via the matrix of coefficients $c_{mn}$ 
\begin{equation}
\psi^{n}_{a}=\sum_m c_{mn}\psi^m  \ .
\label{asymm_basis}
\end{equation}
Expanding the eigenstates in the number difference basis like in \eqref{wavefunction} and \eqref{asymm_state} we find,
\begin{equation}
  c_{mn} = \langle \psi^{m} \vert \psi_{a}^{n} \rangle =\frac{1}{2\pi}\sum_{p,q}A_p^m B_q^{n}\int_{-\pi}^{\pi}\exp[\mathrm{i}(q-p)\phi]\mathrm{d}\phi = \sum_{p} A_{p}^{m}B_{p}^{n} 
 \label{eq:cmndefinition}
  \end{equation}
  where we have used the fact that the amplitudes $A_{p}^{m}$ are real (as are $B_{p}^{m}$).
For simplicity we take the initial state to be the ground state in the asymmetric potential. The resulting projection coefficients $c_{j1}$ are plotted in
Figure \ref{exp_coeff} for different initial asymmetries $\Delta E$. To help gauge the degree of excitation generated by each value of $\Delta E$ we introduce the notation $E_x$. This is the expectation value of the excitation energy in the symmetric double-well expressed as a percentage of the separatrix energy. Thus $E_x=0$ corresponds to the ground state energy of the symmetric double-well and $E_x=100$ to the separatrix energy. We see from Figure \ref{exp_coeff} that when exciting below the separatrix the distribution is smooth and gaussian-like, roughly corresponding to a coherent state.  Excitations above the separatrix are no longer smooth but oscillate strongly.  These two distinct behavioural regimes are joined at the separatrix which has properties of both.

\section{Classical versus quantum dynamics}
\label{sec:classicalvsquantum}

We now consider the dynamics of the macroscopic variables $\phi$ and $k$ following excitation by the method described in Section \ref{apot}. The 
classical (mean-field) dynamics are governed by Josephson's equations 
\begin{eqnarray}
    \frac{\mathrm{d}\phi}{\mathrm{d} \tau}& = &2k,  \label{coupled_eqns_simp2a} \\
    \frac{\mathrm{d} k}{\mathrm{d} \tau}&=&-\Lambda\sin{\phi} \label{coupled_eqns_simp2b}
\end{eqnarray}
with initial conditions $\phi(0)=0$ and $k(0)=-\Delta E/2$, see Equation \eqref{keqs}. Analytical solutions to \eqref{coupled_eqns_simp2a} and \eqref{coupled_eqns_simp2b} with the specified boundary conditions can be found in terms of special functions. For example, when the motion is below the separatrix we have
\begin{eqnarray}
\phi  & = & - 2 \arcsin \left[  \mathrm{sn}\left( \Delta E  \ \tau/2 \ \vert \ 8 \Lambda/ (\Delta E)^2 \right) \right] \label{eq:phisol} \\
k  & = & - (\Delta E / 2) \  \mathrm{dn}\left( \Delta E  \ \tau/2 \ \vert \ 8 \Lambda/ (\Delta E)^2 \right) \label{eq:ksol}
\end{eqnarray}
where $\mathrm{sn}(\theta \vert m)$ and $\mathrm{dn}(\theta \vert m)$ are Jacobian elliptic functions \cite{mathfunctions}. When the motion is above the separatrix Equation \eqref{eq:ksol} for $k(t)$ remains the same but Equation \eqref{eq:phisol} for $\phi(t)$ must be adjusted so that when $\phi$ reaches either $\pm \pi$ it is then mapped to $\mp \pi$ so that the evolution on the phase torus is continuous.
The solutions \eqref{eq:phisol} and \eqref{eq:ksol} are periodic with period
\begin{equation}
\tau_{0}=(8/\Delta E) \   \Re \left\{ \mathrm{K}\left(8 \Lambda/ (\Delta E)^2 \right) \right\}
\label{eq:period}
\end{equation}
where $\Re$ denotes the real part. Above the separatrix the motion is still periodic but the expression \eqref{eq:period} for the period must be divided by two. When $\Lambda/ (\Delta E)^2 \gg 1$ we can asymptotically expand the complete elliptic integral $\mathrm{K}(m)$ to find the period of the low-lying Josephson plasmon excitations 
\begin{equation}
\tau_{0} \sim \pi \sqrt{\frac{2}{\Lambda}} \left( 1+ \frac{1}{4}\frac{\Delta E^2}{8 \Lambda}+\frac{9}{64} \left(\frac{\Delta E^2}{8 \Lambda} \right)^2 + \ldots \right) \quad  \quad  \left[\mbox{when} \  \frac{ \Lambda}{\Delta E^2} \rightarrow \infty \right]
\label{eq:plasmonperiod}
\end{equation}
where the first term corresponds to the harmonic approximation. 
The opposite limit, namely $\Lambda/\Delta E^2 \ll 1$, is relevant for the high-lying rotor excitations: well above the separatrix we have 
\begin{equation}
\tau_{0}= \frac{2 \pi}{\Delta E} \left(1+\frac{1}{4}\frac{8 \Lambda}{\Delta E^2}+\frac{9}{64} \left(\frac{8 \Lambda}{\Delta E^2}  \right)^{2} + \ldots \right) \quad \quad \left[\mbox{when} \  \frac{ \Lambda}{\Delta E^2} \ll 1 \right].
\label{eq:rotorperiod}
\end{equation}

The quantum dynamics are treated using the RN equations which we use to
calculate the expectation values of the operators $\hat{\phi}$ and $\hat{k}$.
Whether the time-dependent version  \eqref{eq:TDRN} or the eigenfunction version \eqref{recurrence} of the RN equations is more suitable depends upon the length of time we want to track the dynamics for. For short times it is more efficient to use \eqref{eq:TDRN}, but for longer times \eqref{recurrence} is in principle faster because in this case time evolution is accounted for purely by the phase factors attached to each eigenfunction. Starting with the eigenfunction version, we expand $\Psi(\phi,t)$ in terms of eigenfunctions as in  \eqref{eq:eigenfunctionexpansion} and find
\begin{eqnarray}\label{eigenstate_a}
  \langle\phi(\tau)\rangle & \equiv & \langle\Psi(\phi,\tau)|\hat{\phi}|\Psi(\phi,\tau)\rangle \nonumber \\
   &=& \sum_{j,k}(\alpha_k)^{*} \alpha_j\exp \left[\mathrm{i}(E^k-E^j)\tau\right]\int_{-\pi}^{\pi}(\psi^k)^{*} \ \phi \ \psi^j \ \mathrm{d} \phi \ . 
\end{eqnarray}
Substituting in the Fourier series \eqref{wavefunction} for the eigenfunctions $\psi^{j}$  we obtain
\begin{equation}\label{eigenstate_p}
\langle\phi(\tau)\rangle=\sum_{j,k}\sum_{m, n\neq m}\frac{(-1)^{n-m}}{n-m}(\alpha_k)^{*} \alpha_j A_m^k A_n^j\sin \left[(E^k-E^j)\tau\right].
\end{equation}
A similar calculation for the expectation value $\langle k(\tau) \rangle$ yields
\begin{equation}\label{eigenstate_k}
\langle k(\tau)\rangle=\sum_{j,k,m}m (\alpha_k)^{*} \alpha_j A_m^k A_m^j \cos\left[(E^k-E^j)\tau\right].
\end{equation}
To compute \eqref{eigenstate_p} and \eqref{eigenstate_k} we need to know the coefficients $\alpha_{j}$ of the eigenfunction expansion \eqref{eq:eigenfunctionexpansion}. These are precisely the coefficients \eqref{eq:cmndefinition}, i.e.\ $\alpha_{j}=c_{j1}$ for the case where the initial state is the ground state in the asymmetric double-well.

Turning to the time-dependent version of the RN equations as given by \eqref{eq:TDRN}, 
the time-dependent amplitudes $A_{n}(\tau)$ are evolved from their values at $\tau=0$ which are directly given by those of the ground state in the asymmetric double-well: $A_{n}(0)=B_{n}^{1}$, see Figure \ref{graph_dynam}. Using \eqref{periodicwavefunction} the expectation value of the relative phase is
\begin{eqnarray}
  \langle\phi(\tau)\rangle &=& \langle\Psi(\phi,\tau)|\hat{\phi}|\Psi(\phi,\tau)\rangle \nonumber \\
   &=& \frac{1}{2\pi}\sum_{n,m}A_n^*(\tau)A_m(\tau)\int_{-\pi}^{\pi}\phi \ \exp{[\mathrm{i}(m-n)\phi]} \ d\phi \\
   & = & -\mathrm{i}\sum_{n \neq m}\frac{A_n^*(\tau)A_m(\tau)(-1)^{m-n}}{m-n} \label{exp_p_time}
\end{eqnarray}
The time-dependent amplitudes $A_{n}(\tau)$ are in general complex numbers, so that 
the above expression is always found to be real.
The expectation value $\langle k \rangle$ is calculated similarly,
\begin{equation}\label{exp_k_time}
  \langle k(\tau)\rangle = -\langle\Psi(\phi,\tau)| \mathrm{i} \frac{\partial}{\partial\phi}|\Psi(\phi,\tau)\rangle 
   = \sum_{n}n|A_n(\tau)|^2.
\end{equation}
We have verified that the two versions of the RN equations give identical predictions for the expectation values. However, note that the summations appearing in \eqref{exp_p_time} and \eqref{exp_k_time} are over fewer indices than the equivalent summations in \eqref{eigenstate_p} and \eqref{eigenstate_k}. This tends to make computations based upon the time-dependent version of the RN equations faster, especially when working in the semiclassical limit where the number of amplitudes that need to be included becomes large.

\begin{figure}[t]
\centering

\includegraphics[width=0.30\columnwidth]{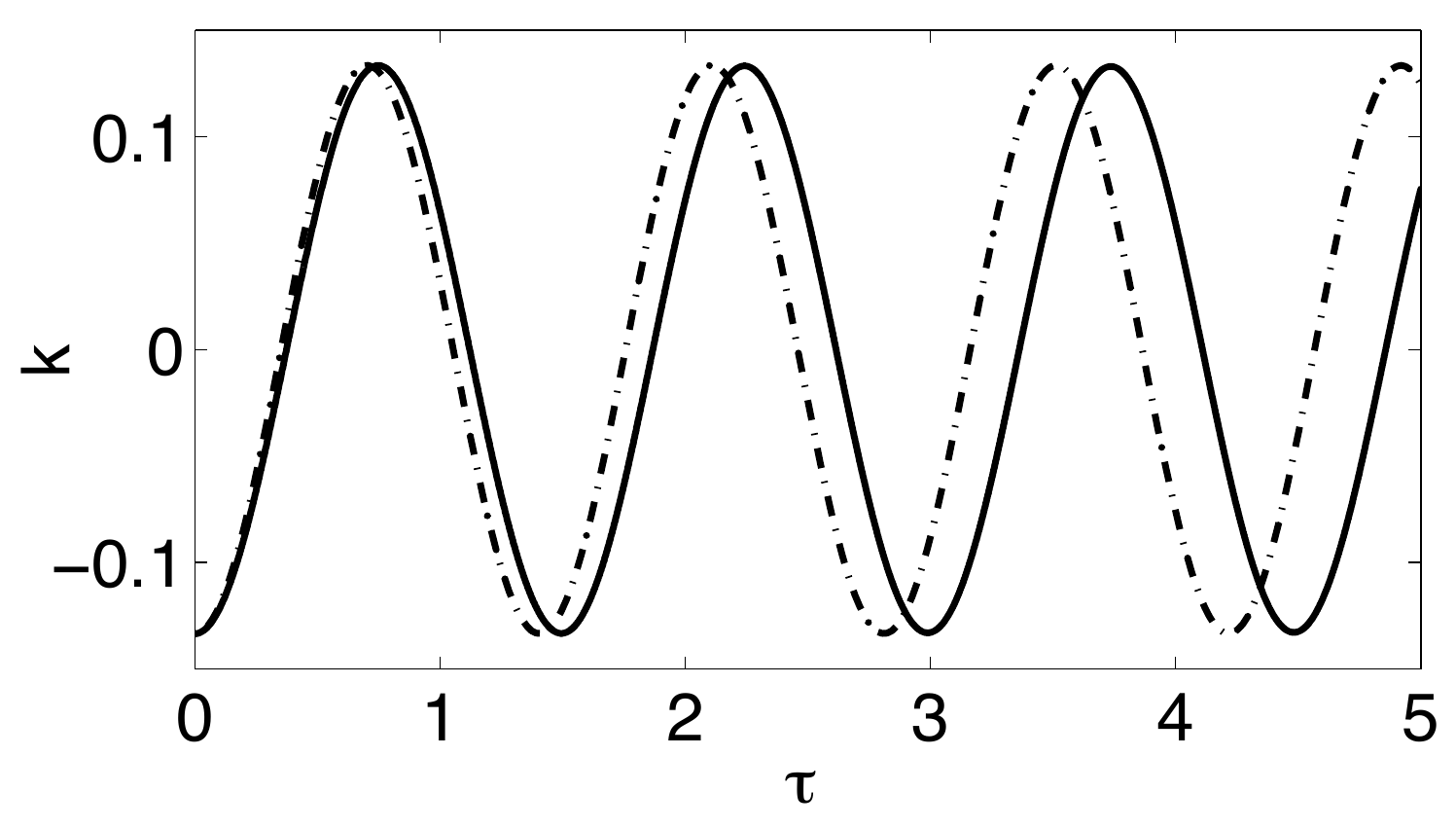}
\includegraphics[width=0.30\columnwidth]{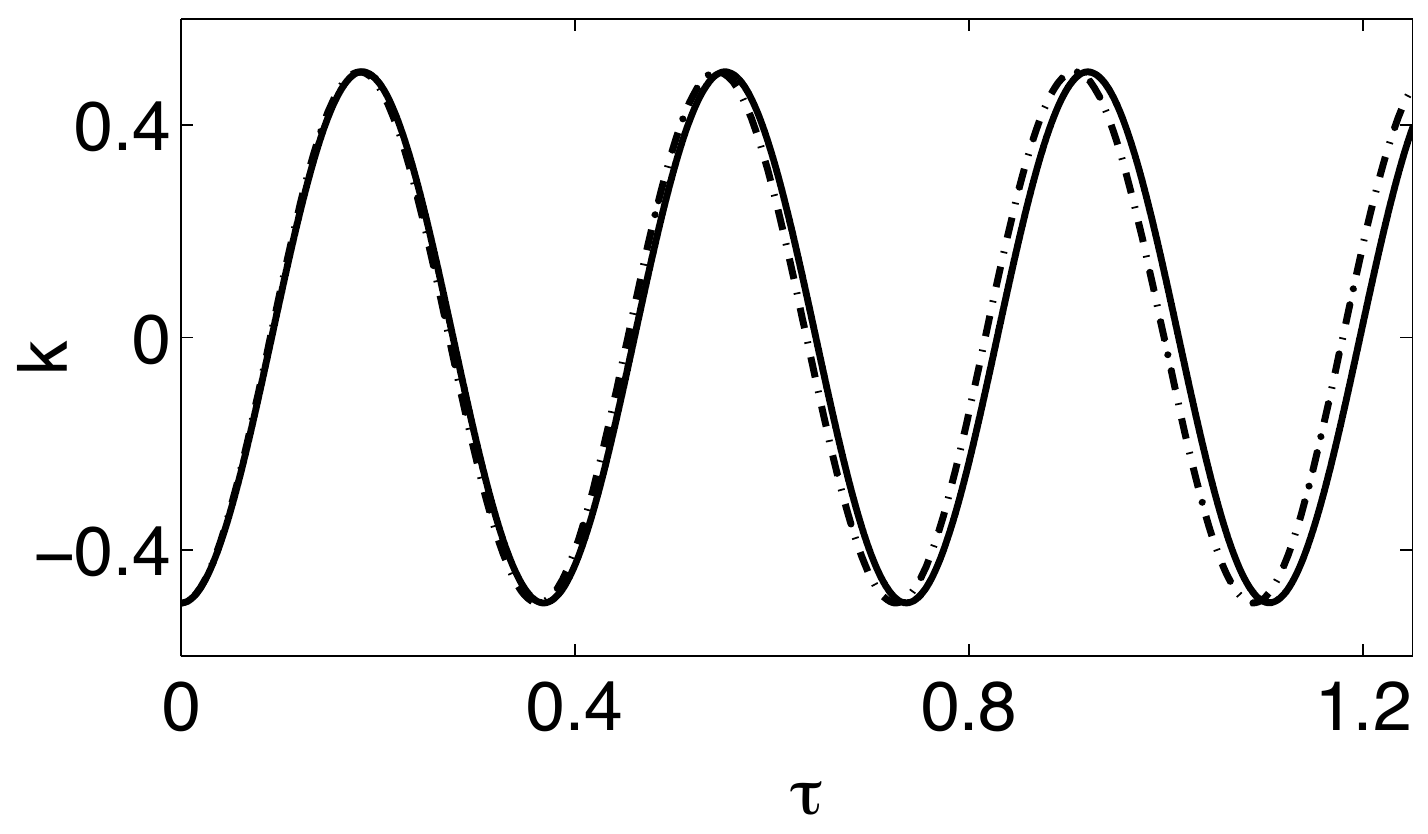}
\includegraphics[width=0.30\columnwidth]{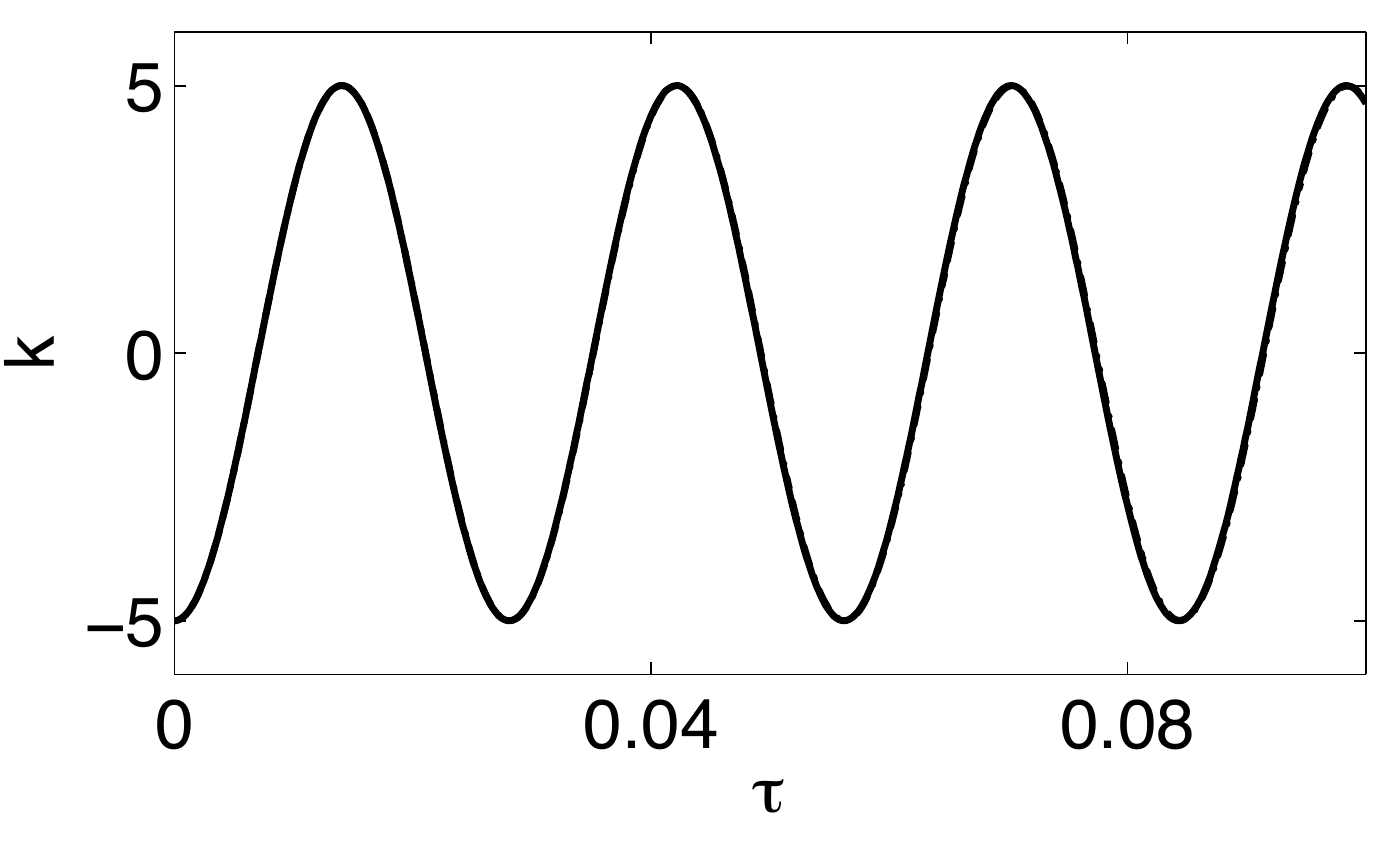}
\subfigure[$\Lambda=10$]{\includegraphics[width=0.30\columnwidth]{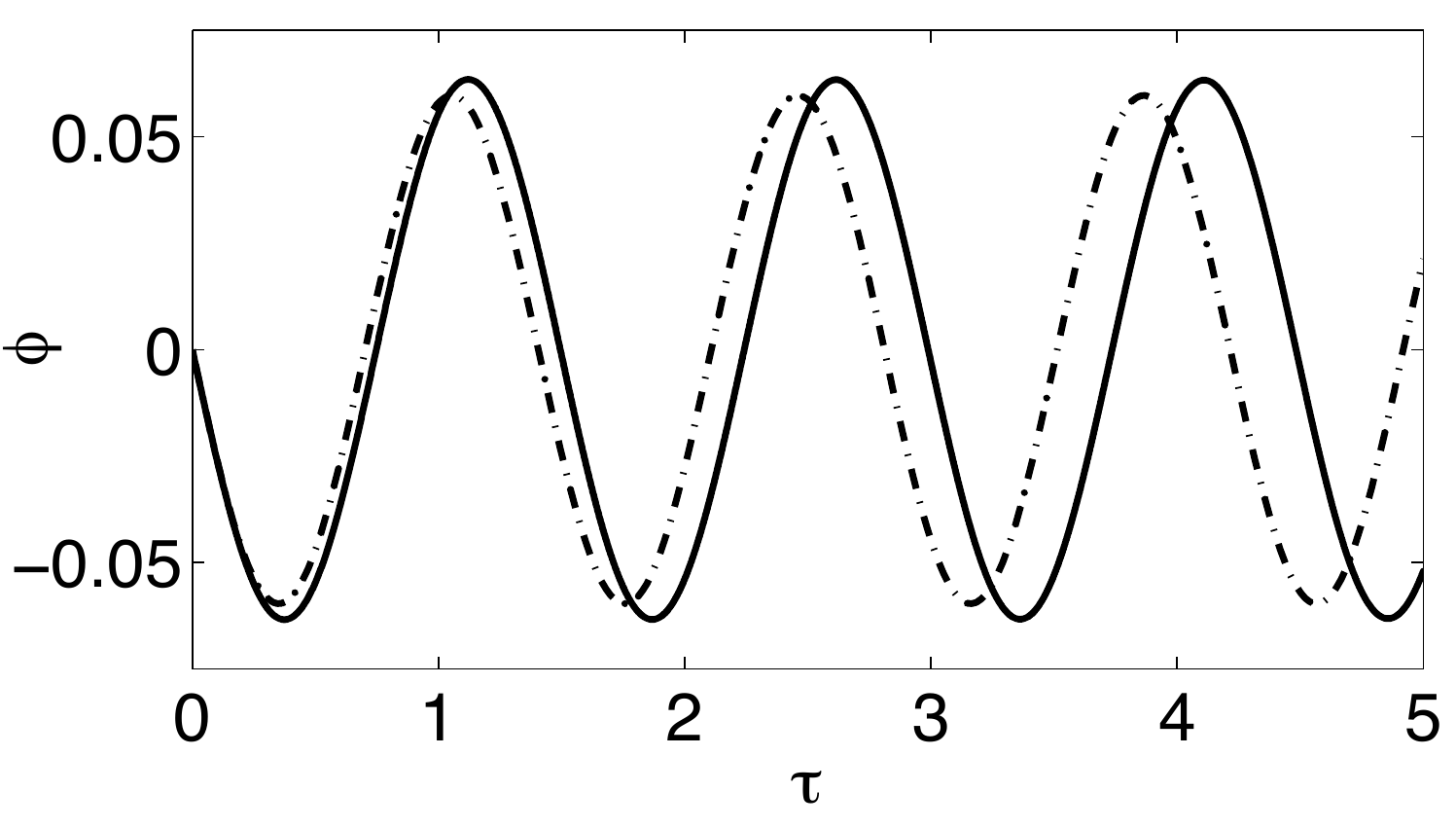}}
\subfigure[$\Lambda=150$]{\includegraphics[width=0.30\columnwidth]{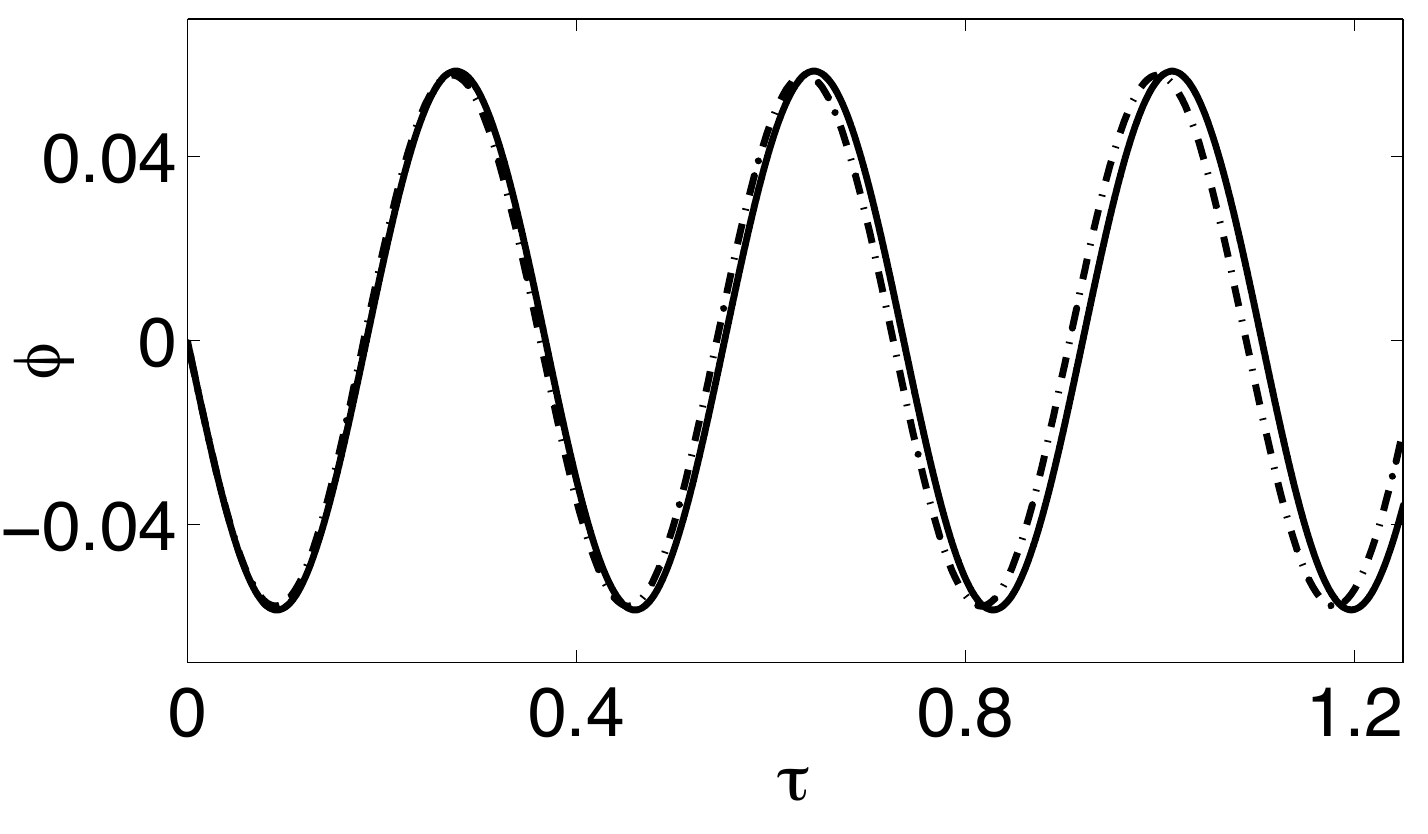}}
\subfigure[$\Lambda=25000$]{\includegraphics[width=0.30\columnwidth]{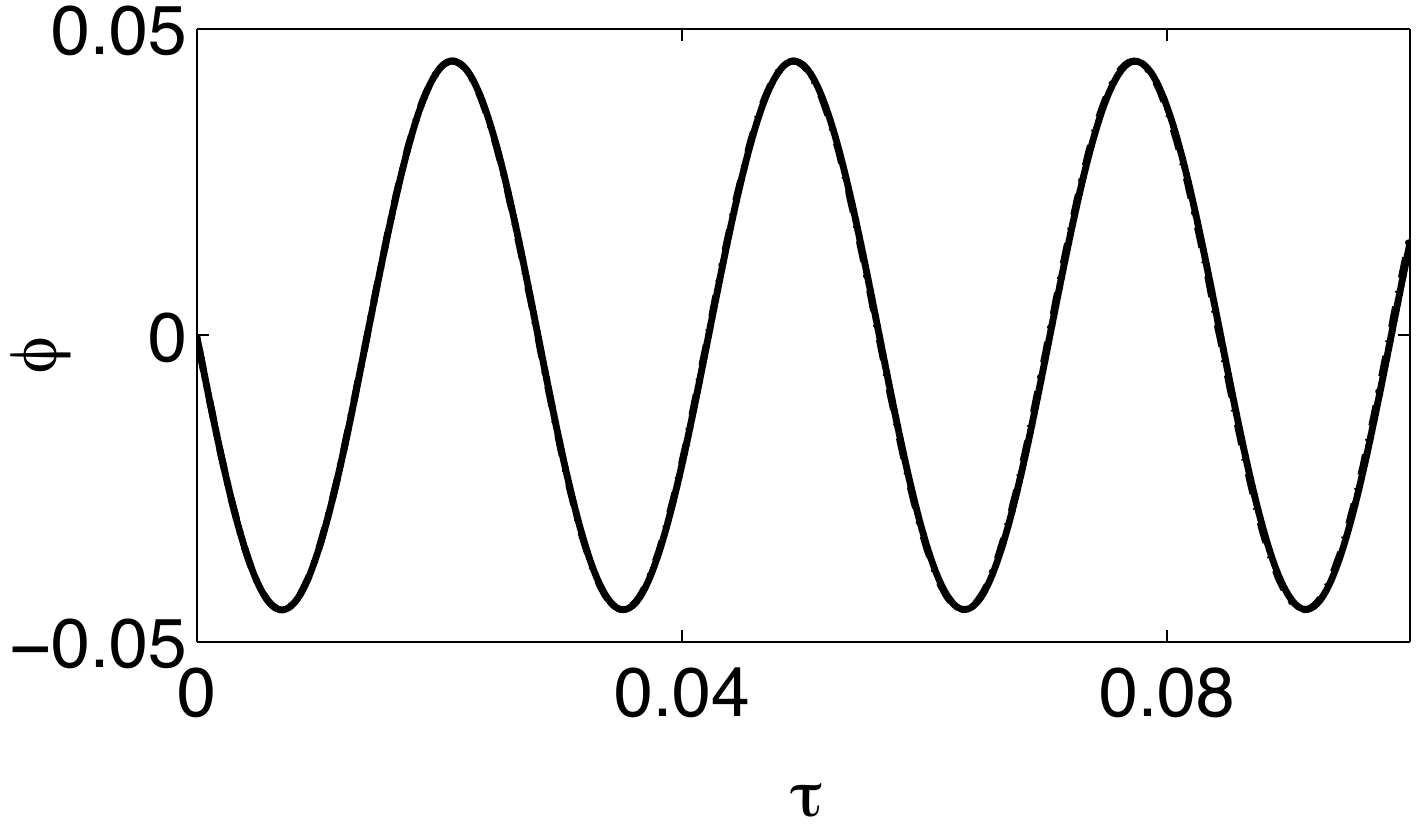}}
\caption{\scriptsize{Comparison of the classical (dashed curves) and quantum (solid curves) predictions for the
temporal evolution of the population imbalance (top row) and relative phase (bottom row) for three different values of  $\Lambda$ (which is related to Planck's constant as $\Lambda \propto \hbar^{-2}$).
 These dynamics are for excitations far below the separatrix.   (a) $\Delta E=0.267\rightarrow E_x=0.1$.  (b) $\Delta E=1\rightarrow E_x=0.1$.  (c) $\Delta E=10\rightarrow E_x=0.1$.  }}
\label{fig_wellbelow}
\end{figure}

\begin{figure}[t]
\centering
\includegraphics[width=0.30\columnwidth]{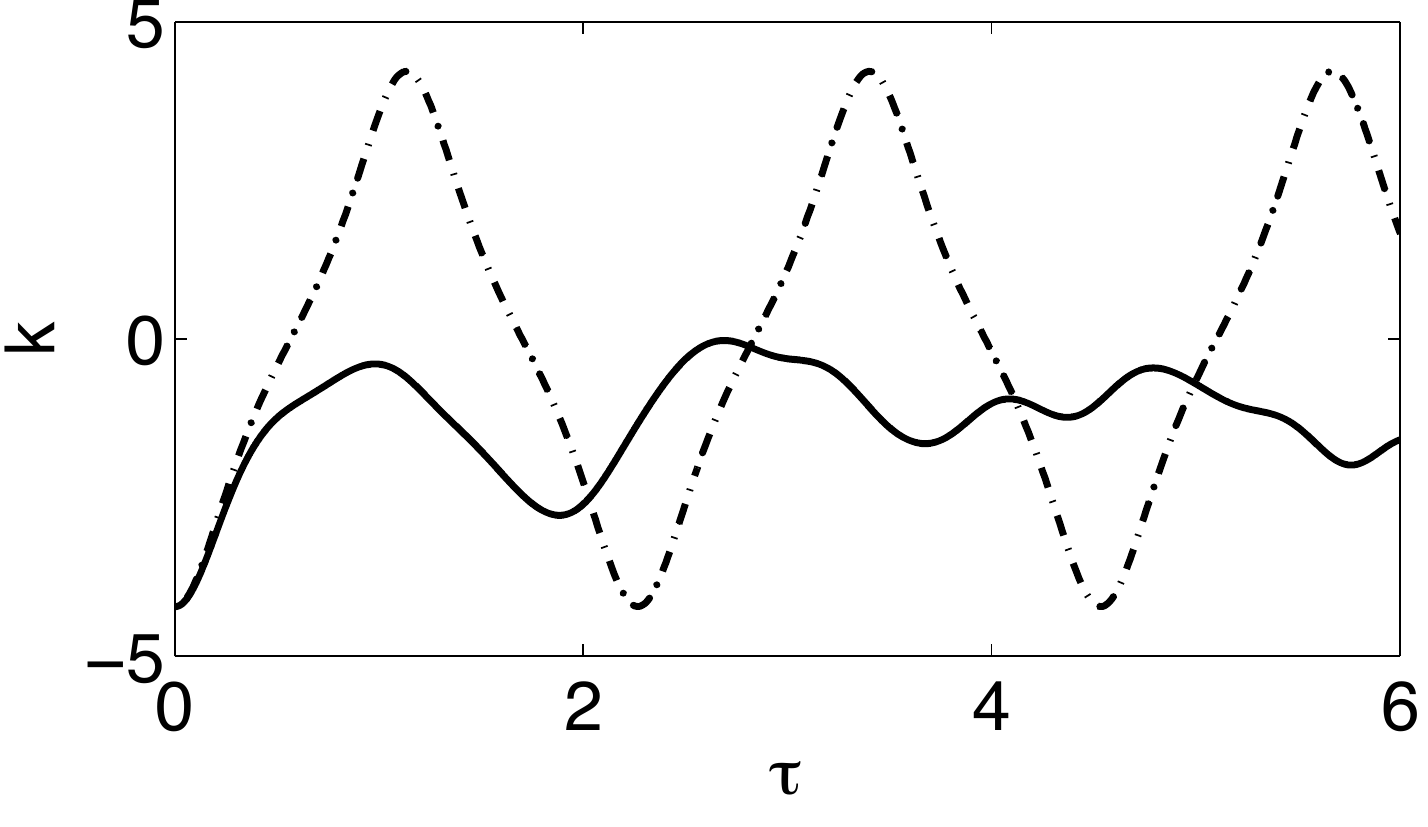}
\includegraphics[width=0.30\columnwidth]{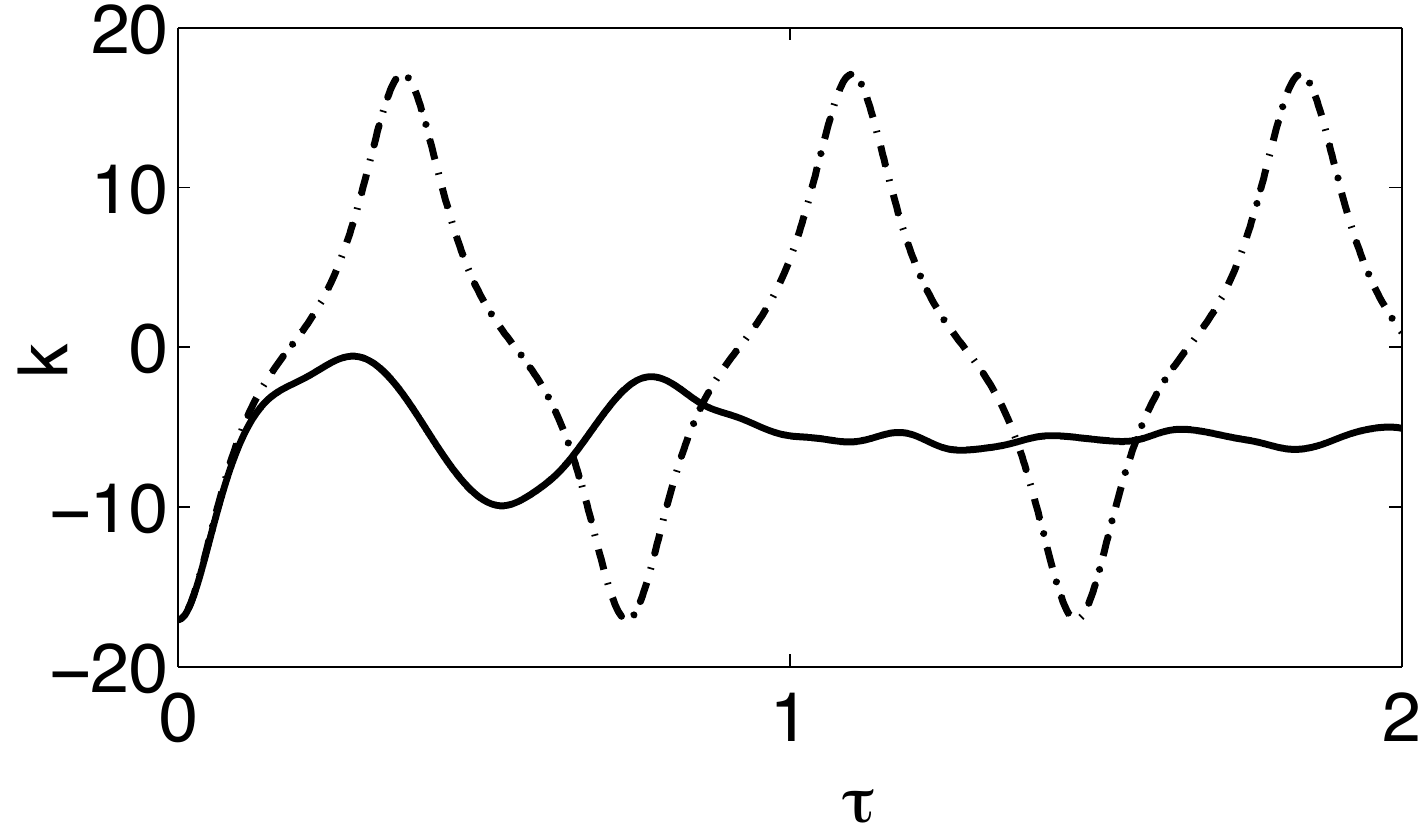}
\includegraphics[width=0.30\columnwidth]{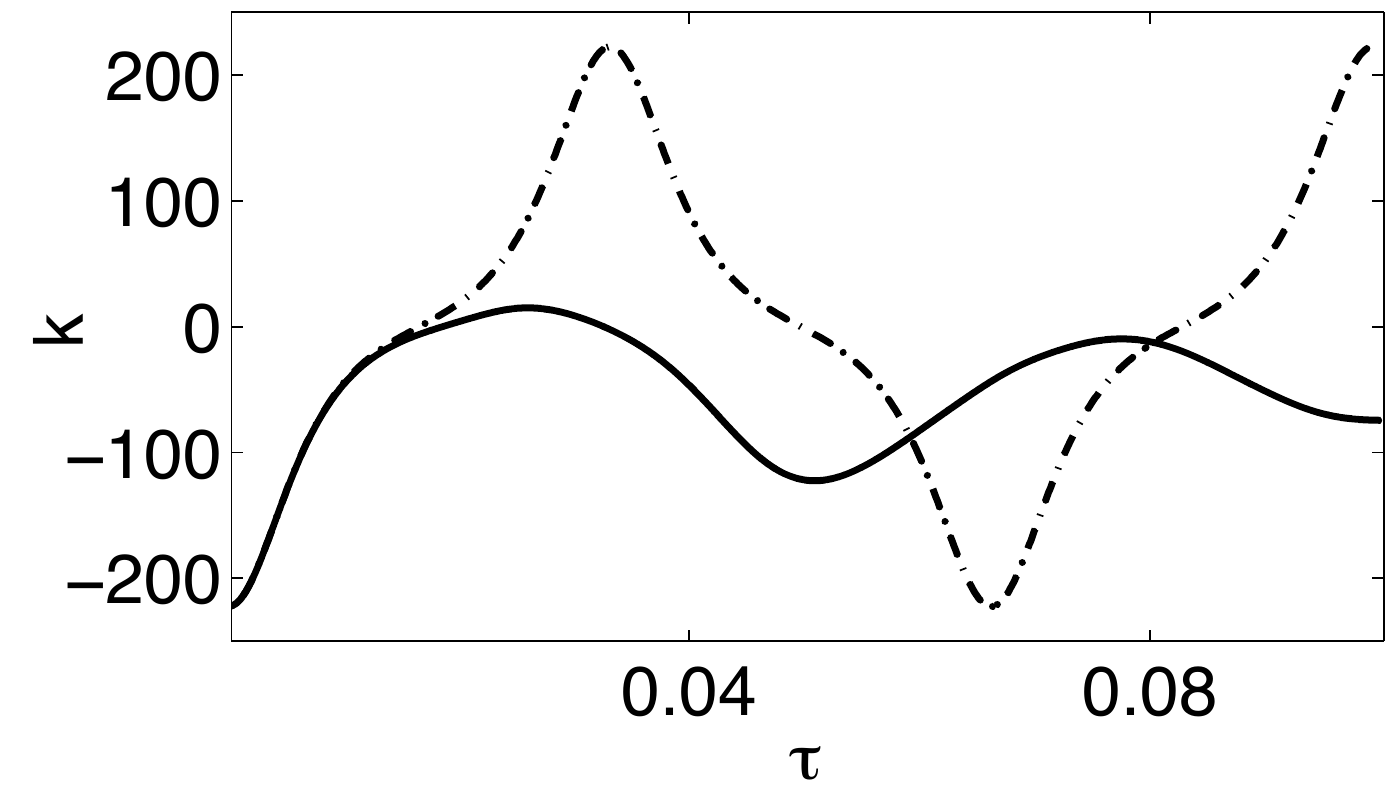}
\subfigure[$\Lambda=10$]{\includegraphics[width=0.30\columnwidth]{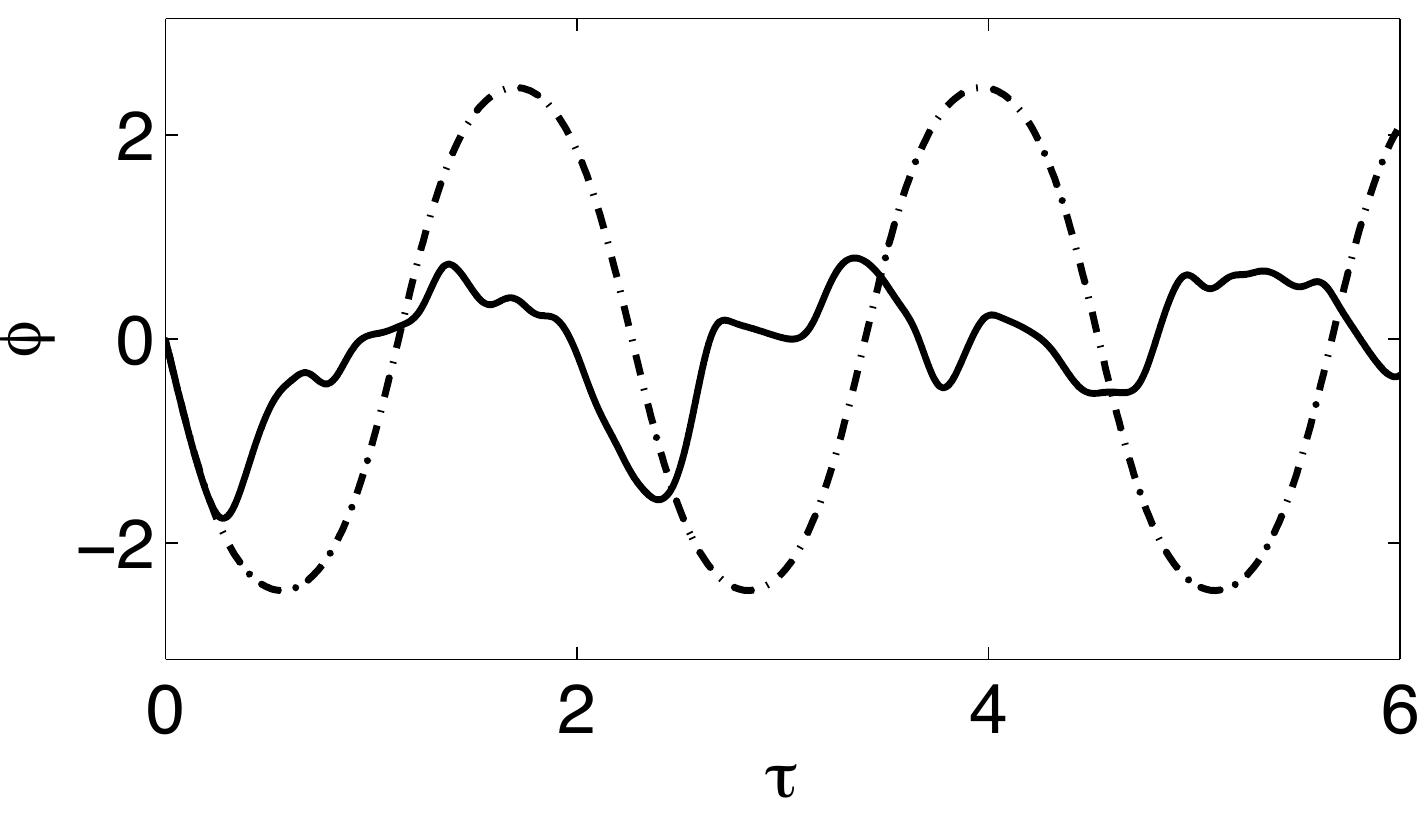}}
\subfigure[$\Lambda=150$]{\includegraphics[width=0.30\columnwidth]{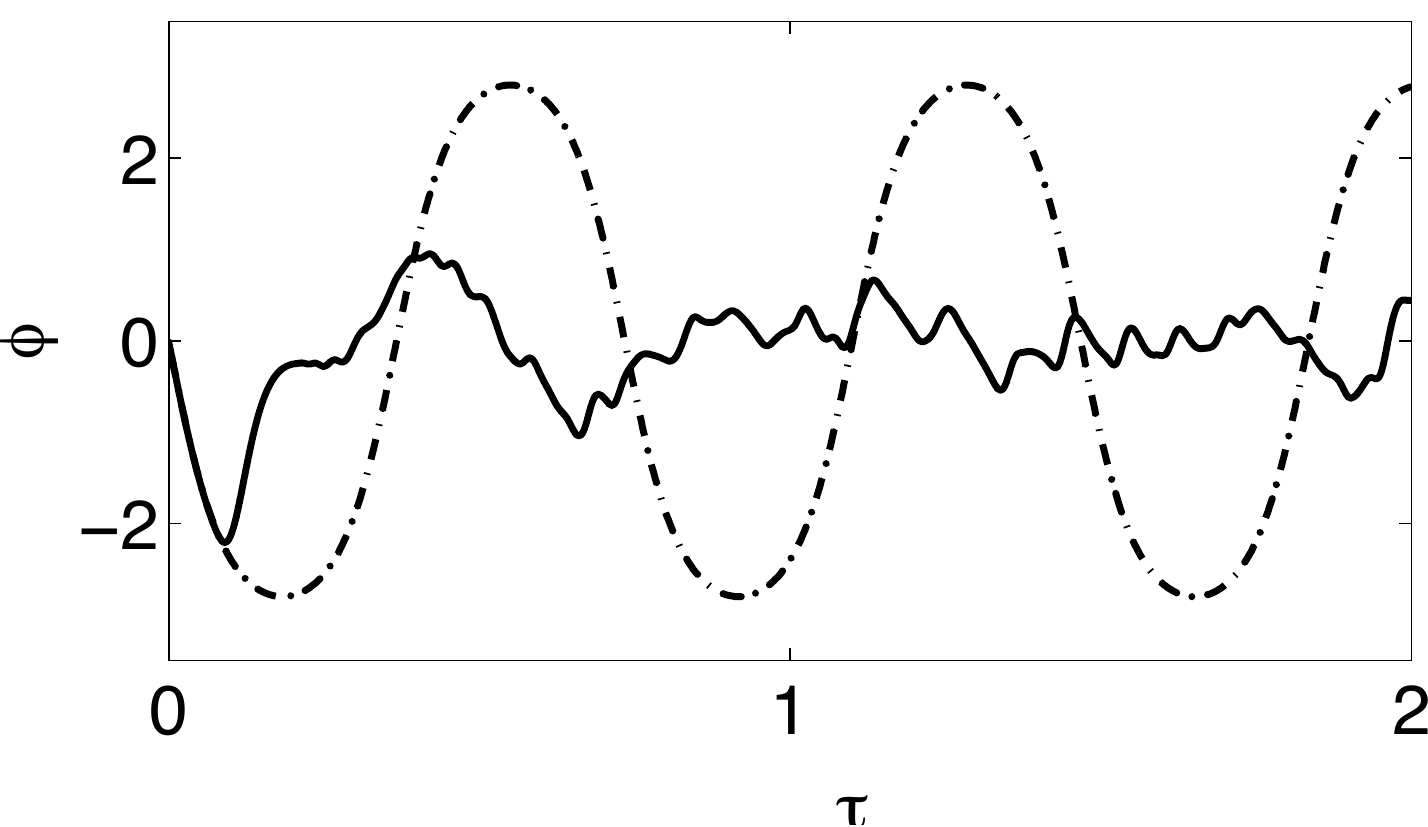}}
\subfigure[$\Lambda=25000$]{\includegraphics[width=0.30\columnwidth]{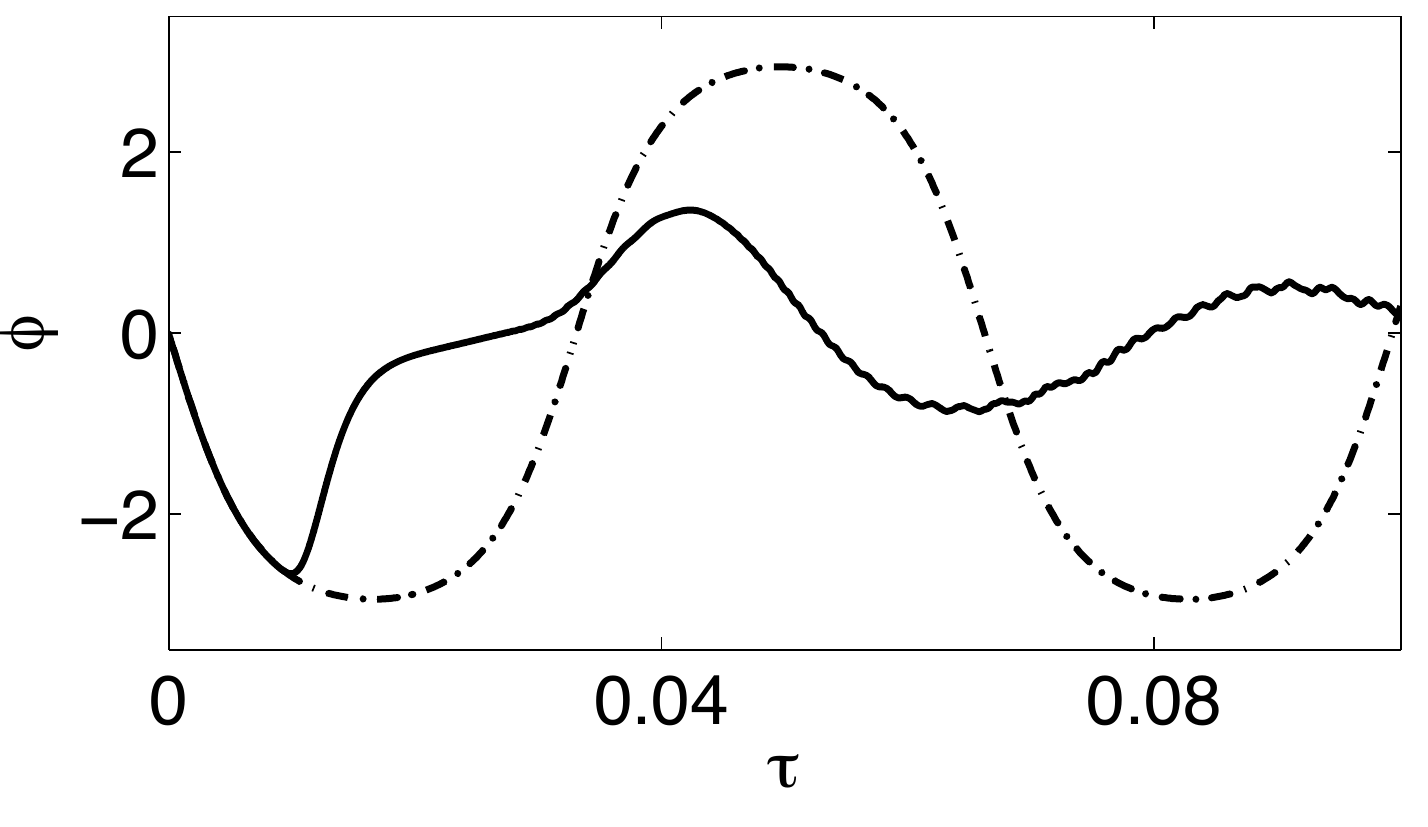}}
\caption{\scriptsize{The same as Figure \ref{fig_wellbelow} but for excitations near the separatrix.    (a) $\Delta E=8.44\rightarrow E_x=99.9$. (b) $\Delta E=34.141\rightarrow E_x=99.9$. (c) $\Delta E=445\rightarrow E_x=99.2$. Even in the semiclassical limit $\Lambda \rightarrow \infty$ the classical and quantum results only agree for the first $1/4$ period.}}
\label{fig_rightat}
\end{figure}

\begin{figure}[t]
\centering
\includegraphics[width=0.30\columnwidth]{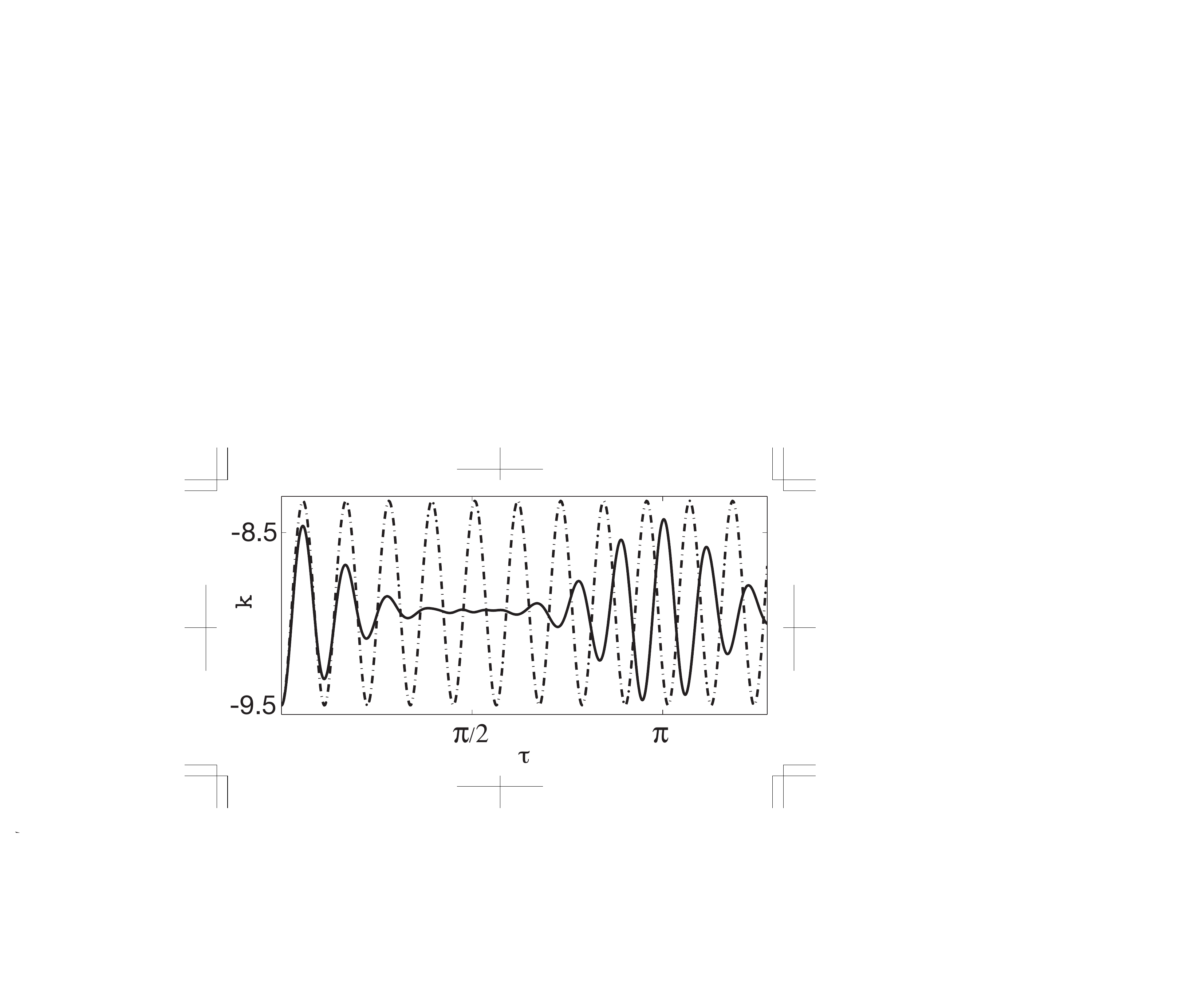}
\includegraphics[width=0.30\columnwidth]{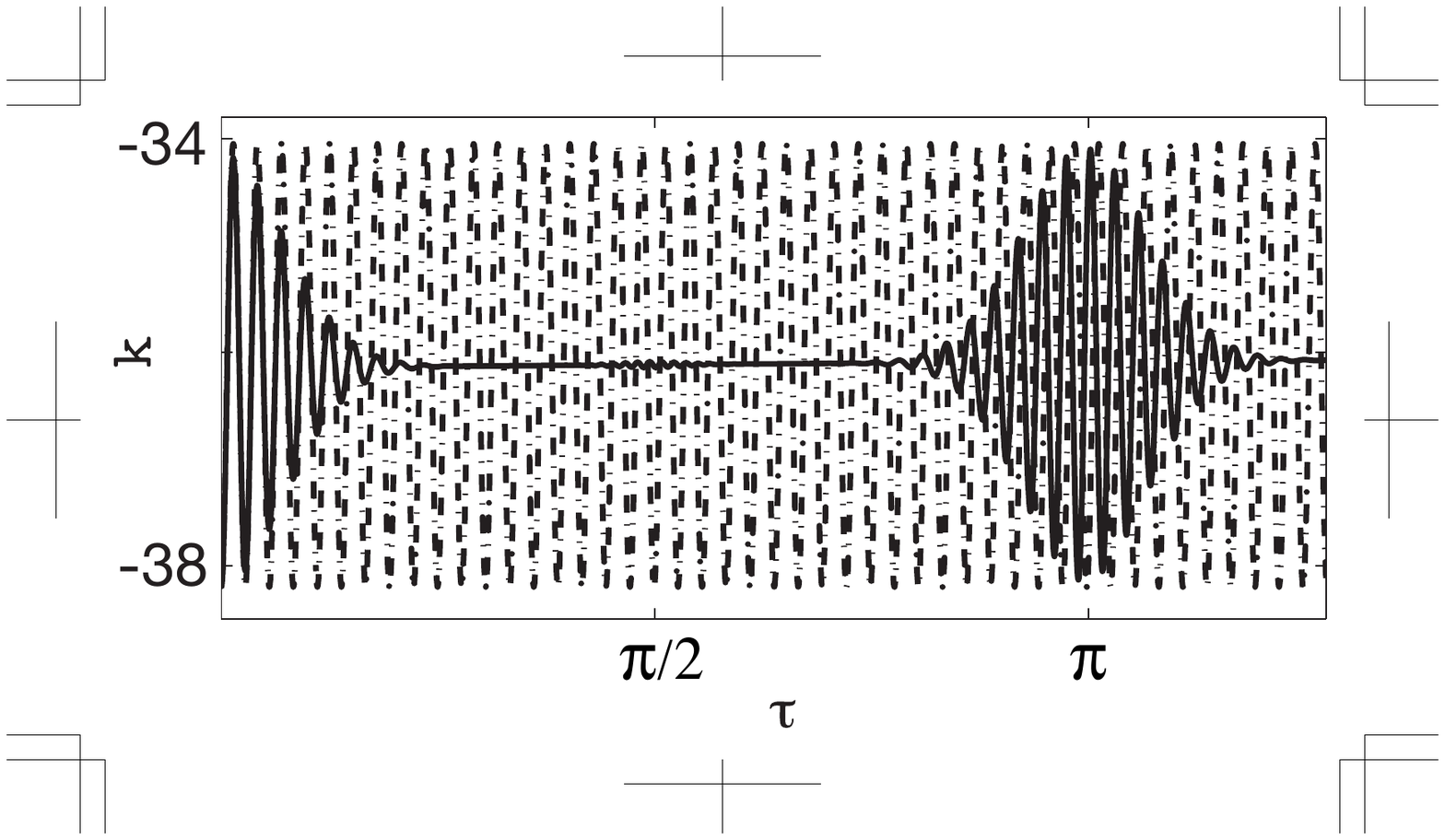}
\includegraphics[width=0.27\columnwidth]{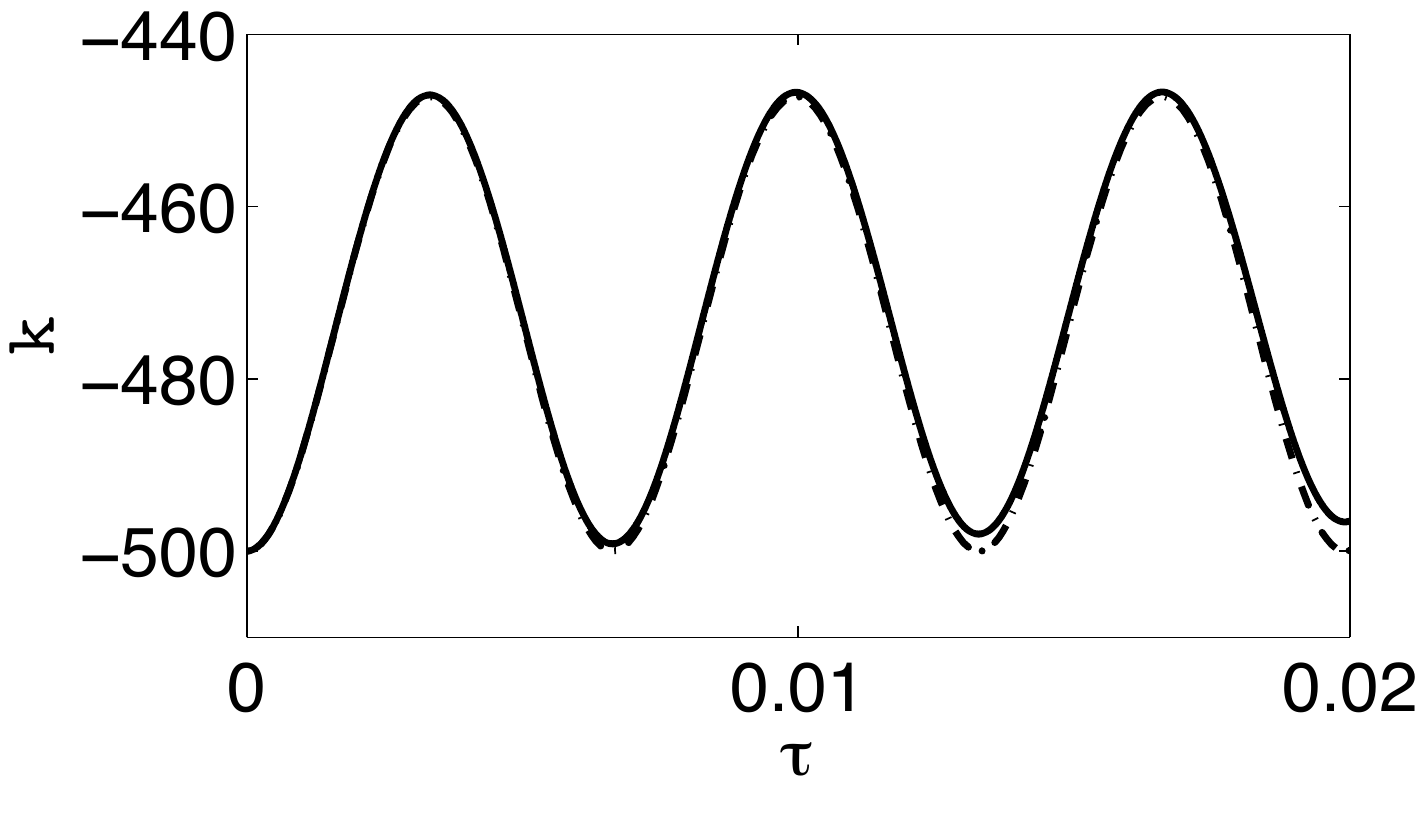}
\subfigure[$\Lambda=10$]{\includegraphics[width=0.30\columnwidth]{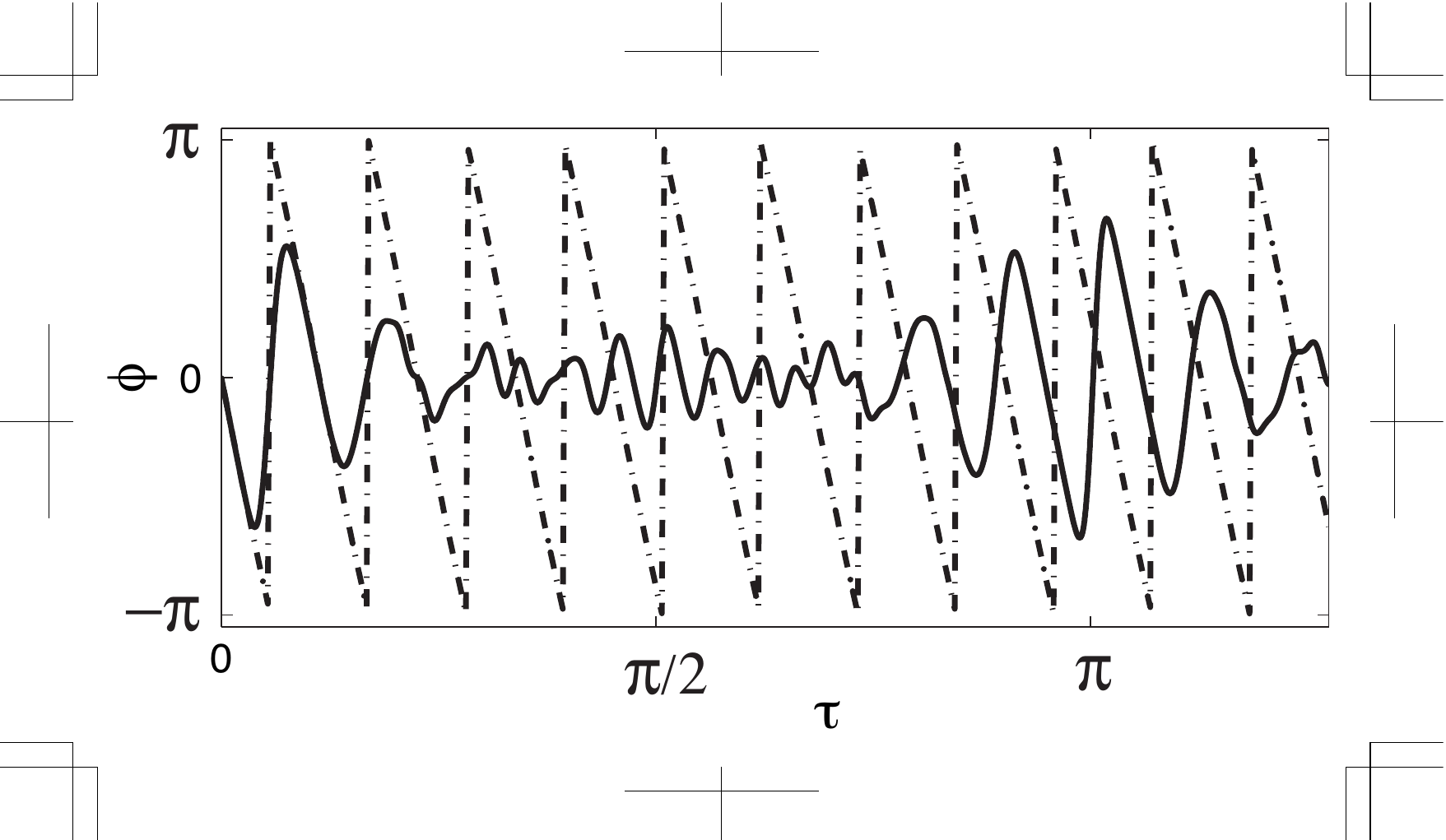}}
\subfigure[$\Lambda=150$]{\includegraphics[width=0.30\columnwidth]{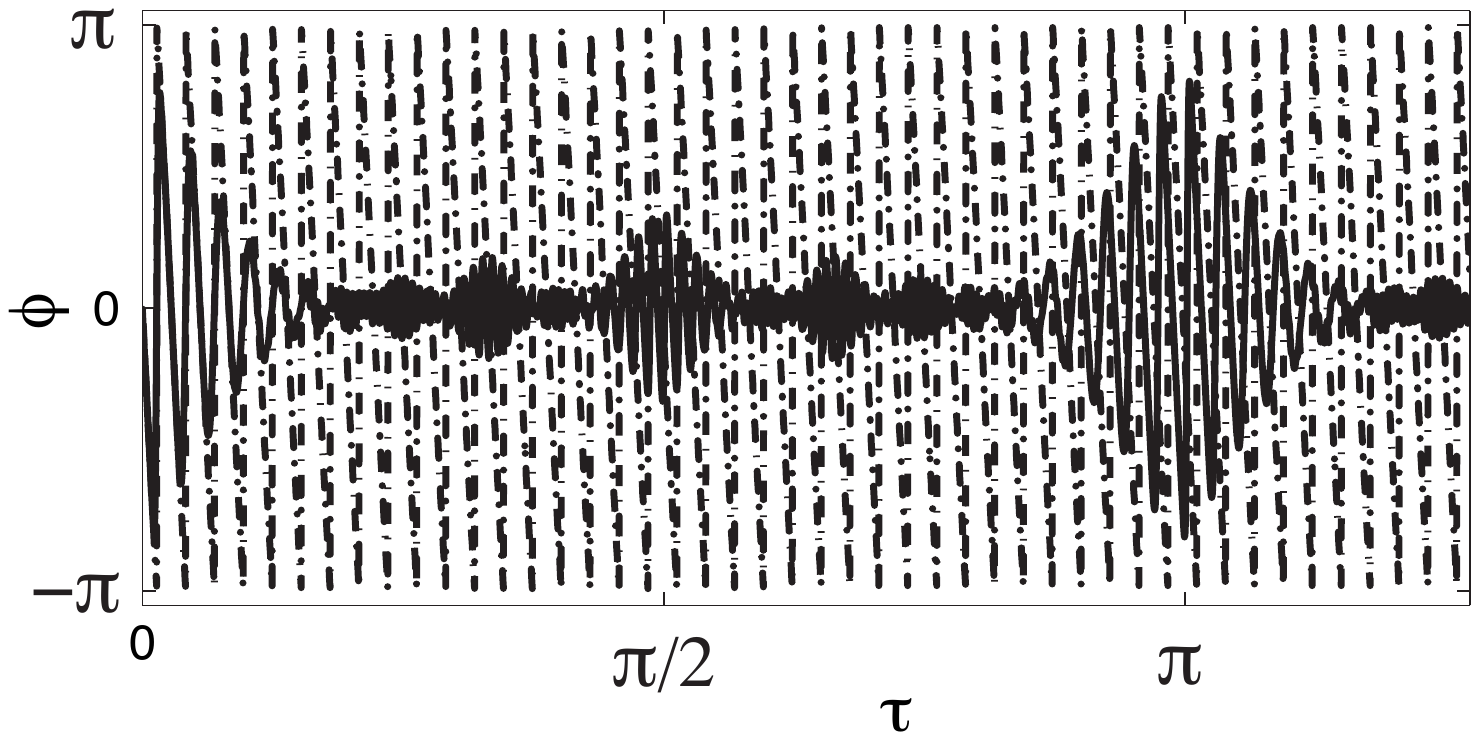}}
\subfigure[$\Lambda=25000$]{\includegraphics[width=0.27\columnwidth]{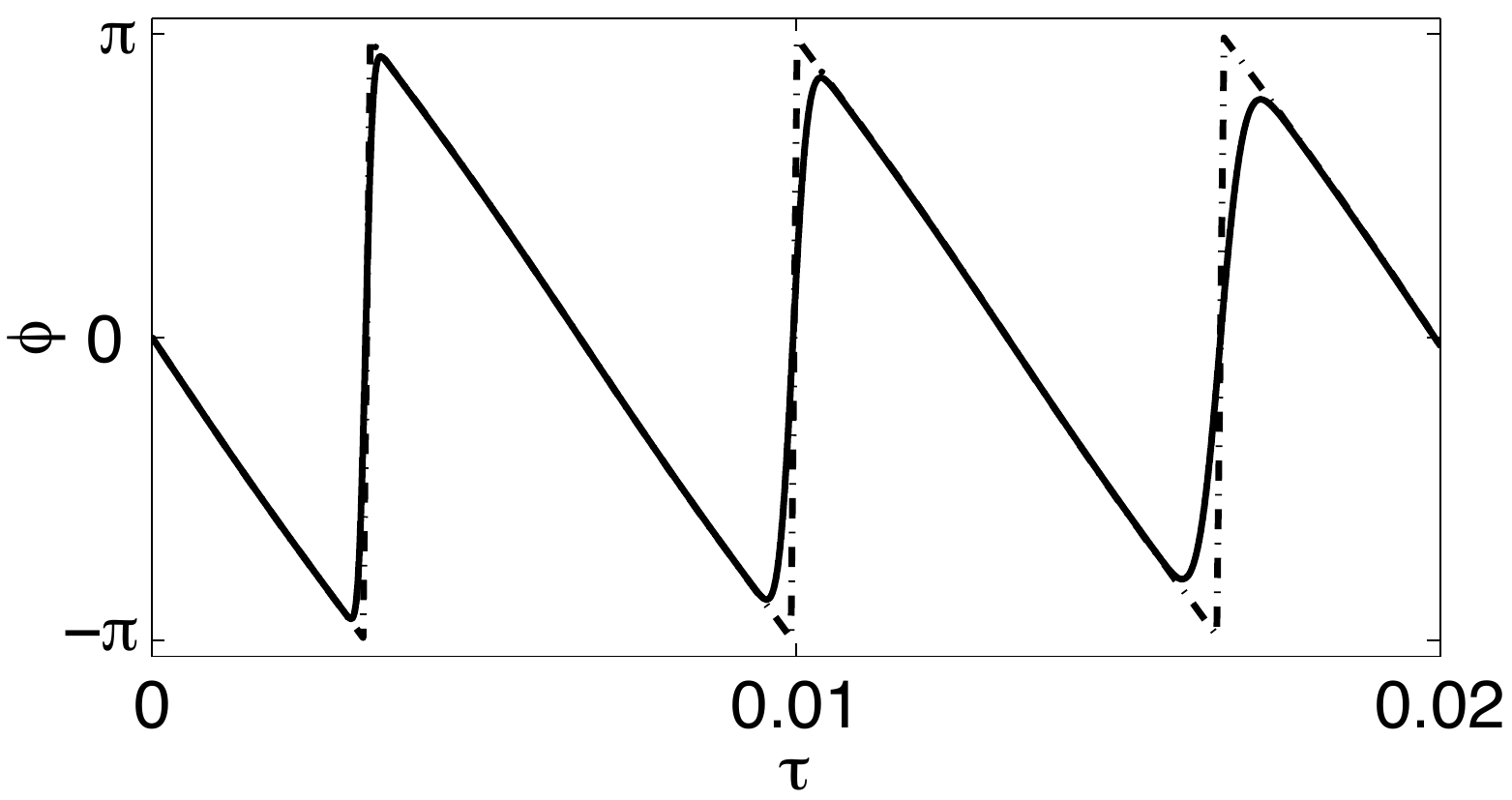}}
\caption{\scriptsize{The same as Figure \ref{fig_wellbelow} but for excitations far above the separatrix.   (a) $\Delta E=18.9\rightarrow E_x=501.1$.  (b) $\Delta E=76.4\rightarrow E_x=501.1$.  (c) $\Delta E=1000\rightarrow E_x=501.1$. Note the change of scale on the abscissa of the (c) panels: the oscillations become very rapid as a function of $\tau$ in the classical limit and so we have only shown the first few.}}
\label{fig_wellabove}
\end{figure}

In Figures \ref{fig_wellbelow}, \ref{fig_rightat}, and \ref{fig_wellabove}
we compare the classical and quantum predictions for $\phi(t)$ and $k(t)$ for three different regimes of excitation: far below, near, and far above the separatrix, respectively. All the upper graphs plot the temporal evolution of the population imbalance and the lower graphs plot the temporal evolution of the relative phase.
Like in Section \ref{sec_raman_asym}, the degree of initial excitation is specified by $E_x$.  
For excitations that are far below the separatrix we see that the quantum and classical predictions agree well for large $\Lambda$ even for times corresponding to many classical periods
 but when $\Lambda$ becomes small, corresponding to a more quantum system, there are observable differences in the expected frequency of oscillation. This renormalization of the classical frequency by quantum fluctuations has been discussed previously by Smerzi and Raghavan \cite{smerzi00}. In either case the classical period far below the separatrix is accurately given by the first few terms in the expansion \eqref{eq:plasmonperiod}. 
 
 For excitations very near the separatrix the classical prediction is valid only for short times for any value of $\Lambda$.  This is because the classical motion is qualitatively different above and below the separatrix whereas the quantum motion contains elements of both due to the combined effects of quantum tunnelling and the finite width in energy of the initial wave packet (see Figure \ref{exp_coeff}). We see in Figure \ref{fig_rightat} that near the separatrix the quantum and classical predictions diverge from each other on a time scale that is always shorter than one quarter of a classical period $\tau_{0}$ (the quantum result clings longest to the classical one as  $\Lambda \rightarrow \infty$).  This can be understood by noting that the initial state has $\langle \hat{\phi} \rangle=0$ and is localised around the bottom of the sinusoidal well where quantum and classical agree best. However, for motion below the separatrix (as in Figure \ref{fig_rightat}) the subsequent evolution always reaches the classical turning point $\phi_{\mathrm{tp}}=-\arccos [1-\Delta E^2 /(4 \Lambda)]$ at times equal to one quarter of the classical period: $\tau_{\mathrm{tp}} = \tau_{0}/4$. At the classical turning point the classical motion reverses direction but part of the quantum wave packet tunnels through the barrier (which is narrow near the separatrix) with the result that the quantum and classical predictions diverge at or slightly before this point. For the parameters chosen in Figure \ref{fig_rightat} neither of the expansions  \eqref{eq:plasmonperiod} and \eqref{eq:rotorperiod} provide particularly accurate approximations to the exact result \eqref{eq:period} for the period $\tau_{0}$ but \eqref{eq:rotorperiod} gives the right order of magnitude indicating that $\tau_{\mathrm{tp}} = O (\pi/ \Delta E)$.

Excitations  far above the separatrix are rotating states in the classical pendulum analogy. In Figure   \ref{fig_wellabove} we can clearly see the macroscopic self-trapping effect in the behaviour of the number difference $k$ in both the quantum and classical predictions: the amplitude of $k$ performs small (relative) oscillations about a particular fixed value and does not reverse sign. The striking difference between the quantum and classical predictions is that the quantum prediction undergoes periodic collapses and revivals due to the discrete energy spectrum. In general, the time scale for revivals of the quantum wave function is given by \cite{doncheski03}
\begin{equation}
T_{\mathrm{rev}}=\frac{4 \pi \hbar}{\vert  E''(j_{0})  \vert}
\label{eq:revivaltime}
\end{equation}
where $E''(j_{0})$ is the 2nd derivative of the energy spectrum with respect to the quantum number $j$ labelling the energies and is evaluated at the centre of the wave packet $j_{0}$. Far above the separatrix the quantum Josephson hamiltonian reduces to that of the quantum rotor 
\begin{equation}
\hat{H}_{\mathrm{rotor}}=-\frac{E_{c}}{2}\frac{\mathrm{d}^{2}}{\mathrm{d} \phi^{2}}
\label{eq:rotorenergy}
\end{equation}
which has the spectrum $E^{j}=E_{c} j^{2}/2$ where $j$ is an integer $j=0, \pm1, \pm2, \dots $. Thus, in our scaled time units, we obtain 
$\tau_{\mathrm{rev}}=2 \pi$. This is actually twice the revival time of $\pi$ that can be clearly observed in Figure \ref{fig_wellabove}(a). The discrepancy can be explained by noting that in  
Figure \ref{fig_wellabove} we have plotted the expectation value which contains the square of the wave function and thus we expect the expectation value to revive on a time scale which is half that of the wave function revival time \cite{styer01}. From Figure \ref{fig_wellabove} we also see that the revival time increases as the system becomes more classical in the sense that more oscillations occur during the time $\tau_{\mathrm{rev}}$.
We note in passing that
collapses and revivals can in principle also take place for wave packets excited below the separatrix but the time scale is typically much longer than above the separatrix. This is because below the separatrix the spectrum is to the first approximation linear and for a purely linear spectrum then Equation \eqref{eq:revivaltime} predicts $T_{\mathrm{rev}} \rightarrow \infty$ \cite{doncheski03}. Thus, only the small non-linear correction terms contribute to a finite collapse and revival time for excitations far below the separatrix.

\begin{figure}[t]
\centering
\includegraphics[width=0.90\columnwidth]{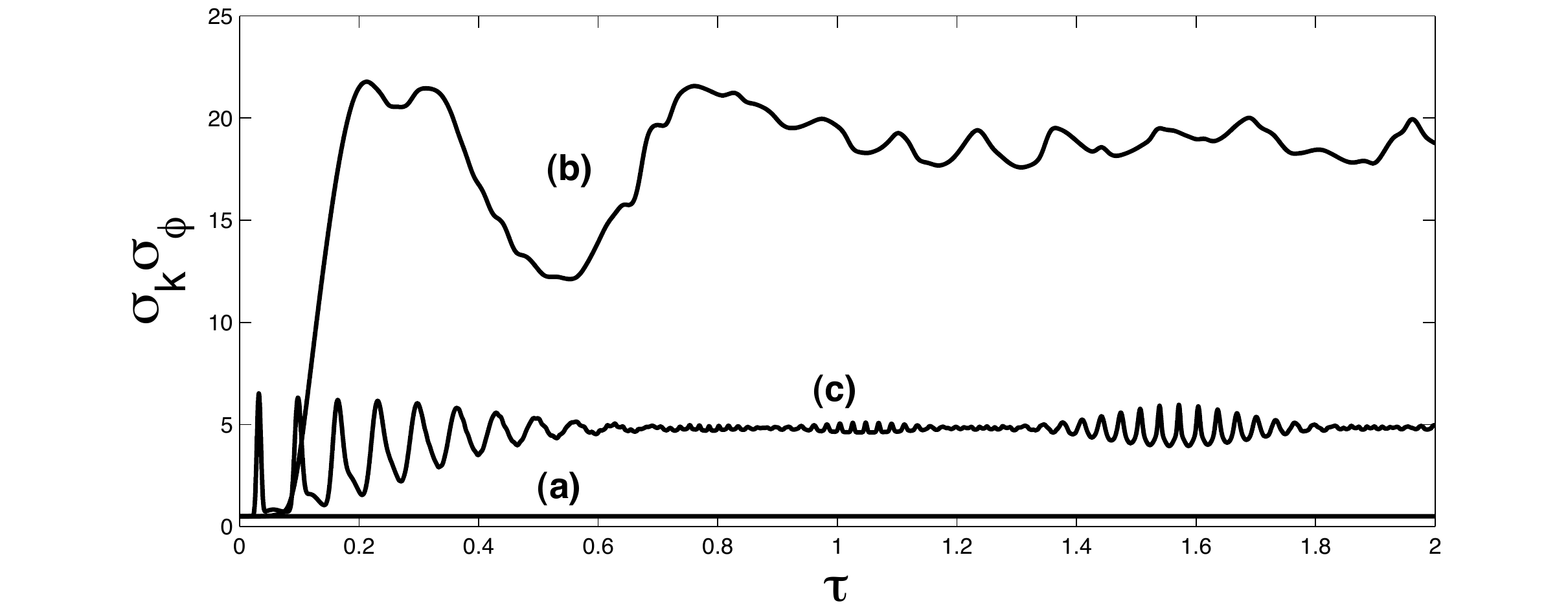}
\caption{\scriptsize{These 3 curves demonstrate the uncertainty relation for the dynamical variables $\phi$ and $k$.  (a) $\Delta E=1\rightarrow E_x=0.1$.  For excitations  far below the separatrix there are small fluctuations, but the uncertainty remains very near $1/2$.  (b) $\Delta E=34.141\rightarrow E_x=99.9$.  For excitations near the separatrix, the uncertainty begins at a minimum and then increases sharply after a short time.  (c) $\Delta E=100\rightarrow E_x=857.9$.  For excitations far above the separatrix the uncertainty begins to oscillate but centers itself around a finite value.}}
\label{uncert_fig}
\end{figure}

As a final point in this section we check that the uncertainty relation $\sigma_{\phi}\sigma_k\geq 1/2$ satisfied by the variances $\sigma_{\phi}$ and $\sigma_{k}$ of the phase and number difference is obeyed for the evolution we have computed.
 In Figure \ref{uncert_fig} we plot the uncertainty product for $\Lambda=150$ and for each excitation regime.  Excitations far below the separatrix remain very close to the minimum uncertainty, experiencing only small fluctuations, as expected for a gaussian (coherent) wave packet in a nearly harmonic potential.  Near the separatrix the uncertainty product starts at the minimum value of $1/2$, then after a short time increases very rapidly.  Far above the separatrix the uncertainty product starts as a minimum but then oscillates and centers itself around a finite value above $1/2$.

\section{Time-modulated double-well potentials}
\label{sec:timemodulated}

The method of excitation described in Section \ref{apot} generates `classical-like'  gaussian wave packets of eigenstates. However, to fully explore the quantum dynamics of the double-well system it is desirable to be able to excite individual eigenstates, for example to generate the Schr\"{o}dinger cat states discussed in Section \ref{sec:band} below. One way to excite an individual eigenstate is to start from the ground state and apply a time-modulated asymmetry which is resonant with a particular transition. As an example we will use this method to excite a single eigenstate near the separatrix. 
States near the separatrix are interesting because they are at the boundary between two qualitatively different classical regimes.  The special behaviour of these states can be hidden due to the population of the surrounding states if the method of Section \ref{apot} is used.  

\begin{figure}[t]
\centering
\includegraphics[width=\columnwidth]{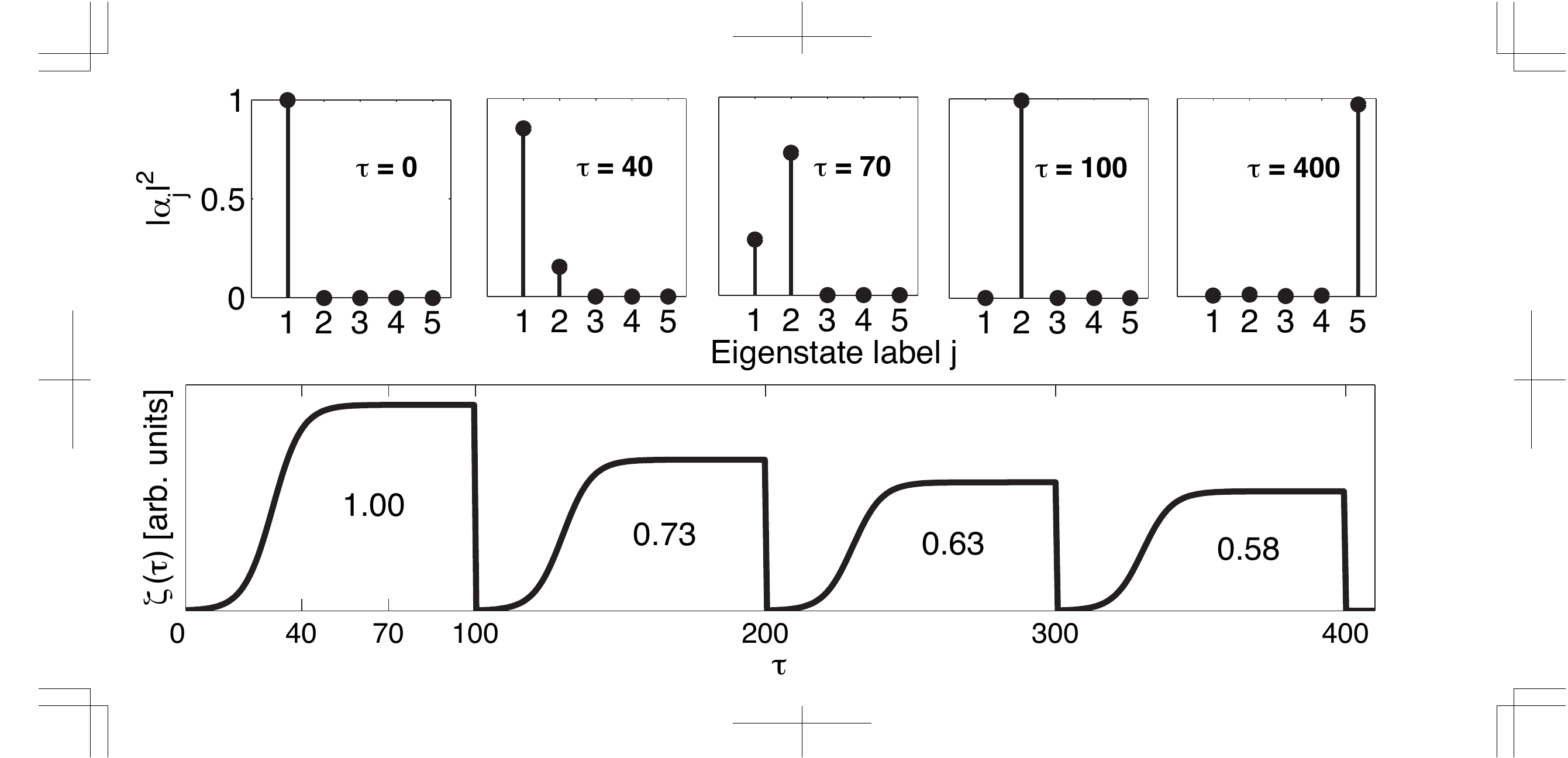}
\caption{\scriptsize{Excitation of a single eigenstate ($j=5$) from the ground state ($j=1$) by means of a time oscillating potential.  Upper panel: snap shots after various times of the probabilities $\vert\alpha_{j} \vert^{2}$ to occupy the jth eigensate of the balanced well system. 
 (b) In order to reach the state $j=5$, which lies immediately below the separatrix, we pump energy into the system in a series of four pulses (whose profile is of the form  $a\times (1+\tanh[\tau/b])$) and whose area is determined by  \eqref{rab_freq1}. Each pulse is resonant with its respective transition.
We set the total time of the process to be  $\tau=400$.  Notice that the final probability to be in state $5$ is not exactly 1 because other states have been marginally excited.  This result can be improved if we make the pulses more adiabatic, requiring smaller pulses over longer times but conserving the total area.}}
\label{fig_excite_up}
\end{figure}
 
To transfer the system from an eigenstate $a$ of the symmetric double-well with energy $E^{a}$ to another eigenstate $b$ of the symmetric double-well with energy $E^{b}$ we apply a time-varying asymmetry of the form $\Delta E(t)=\zeta(\tau)\sin(\Omega\tau)$ which is resonant with the transition so that $\Omega=(E^b-E^a)/\hbar$. The pulse envelope $\zeta(\tau)$, which is assumed to be slowly varying in comparison to  $\hbar/ \Omega$, determines the temporal shape of the pulse.
The evolution of the amplitudes is governed by the time-dependent RN equation
\begin{equation}
\label{coupled_amp_finite1}
\mathrm{i}\frac{\mathrm{d}A_n(\tau)}{\mathrm{d}\tau}=(n^2+\Delta E(\tau)n)A_n(\tau)-\frac{\Lambda}{2}(A_{n+1}(\tau)+A_{n-1}(\tau)).
\end{equation}
In what follows we shall choose the initial state to be the ground state of the symmetric double-well. 

A good estimate of the time required to excite the target eigenstate $b$ can be obtained by approximating our multi-level system as a two-level system consisting of just the states $a$ and $b$ and considering the effect of applying a time-modulated asymmetry with a constant amplitude $\tilde{\zeta}$.  The characteristic frequency at which the two-level system is driven between its two levels is then given by the Rabi frequency  $\omega_{R}^{ba}=  - \mathrm{i}  \tilde{\zeta} E_{c}/(2 \hbar)  \vert\langle \psi^b| \hat{k}|\psi^a\rangle \vert$ \cite{quantum_mech}. Introducing the dimensionless Rabi frequency $\Omega_{R}^{ba}=2\hbar \omega_{R}^{ba}/E_{c}$  we find
\begin{equation}
    \Omega_R^{ba} = \tilde{\zeta} \sum_m m A_m^b A_m^a \ .
   \label{rab_freq1}
\end{equation} 
The time to fully transfer the system from $a$ to $b$, i.e.\ to deliver a $\pi$-pulse, is given by $\tau_{\mathrm{R}} \equiv \pi/\Omega_R^{ba}$. 

In practice, the amplitude $\zeta(\tau)$ must be switched on smoothly from zero so that the frequency spread is small and only a single upper eigenstate is excited. Empirically we find that a single target state can be excited only if the maximum value of $\zeta(\tau)$ obeys the conditions $\zeta_{max}\ll\Lambda$ and $\zeta_{max}\ll\Omega_R$.
Nevertheless, by direct integration of the RN equations \eqref{coupled_amp_finite1} using the pulse shapes for $\zeta(\tau)$ shown in Figure \ref{fig_excite_up} we find that \eqref{rab_freq1} accurately predicts the total (integrated) pulse area required
to make the transfer $a \rightarrow b$, i.e.\ $\int \zeta(\tau) \mathrm{d} \tau \approx \tilde{\zeta} \tau_{R}$.

An important point to note is that the time-modulated asymmetry technique can only be used to directly connect states with opposite parity, as is evident from expression \eqref{rab_freq1}. Like for any one-dimensional Helmholtz equation, the eigenstates of the Mathieu equation alternate in parity as one goes up in energy. Starting from the ground state, which is even, we can therefore only directly excite odd parity states. Furthermore, $\Omega_{R}^{ba}$ rapidly becomes small when the eigenstates $a$ and $b$ are far apart in energy, i.e.\ their labels $j$ differ significantly. Crudely speaking, this is because in $k$-space the eigenstates below the separatrix resemble those of the simple harmonic oscillator (Hermite polynomials) in coordinate space and are strongly peaked at the classical turning points \cite{odell01}. The overlap integral between the initial and final state that occurs in \eqref{rab_freq1} therefore rapidly decreases in magnitude when the classical turning points differ i.e.\ for eigenstates $a$ and $b$ which are far apart in energy. For this reason we find the surprising result that it can be far quicker to excite up to the final level by a series of steps via intermediate levels rather than to directly excite the upper level (this also allows us to excite a final state that has the same parity as the initial state). In Figure \ref{fig_excite_up} we show the results of the stepping process $1\rightarrow 2 \rightarrow 3 \rightarrow 4 \rightarrow 5$, so that the initial and final states have the same parity,
 for the case $\Lambda=10$ (the separatrix occurs between states $5$ and $6$).  Figure \ref{fig_excite_up} also shows the pulse envelopes for each step: each envelope has the form  $a\times(1+\tanh[\tau/b])$, where $a$ and $b$ are constants. The total pulse ``area'' was $\tilde{\zeta} \, \tau_{R}=9.02$ which is the sum of the $\tau_{\mathrm{R}}$ values predicted by Equation \eqref{rab_freq1}, and to make the process adiabatic we set the total excitation time to be $\tau=400$. In order to compare the single step with the multi step method, consider the case $1\rightarrow 2 \rightarrow 3 \rightarrow 4 \rightarrow 5 \rightarrow 6$, i.e.\ to the state immediately above the separatrix, which has odd parity. Equation \eqref{rab_freq1} gives a total pulse ``area'' of 11 for the multi step method whereas it gives a pulse ``area'' of $777$ for the single step $1 \rightarrow 6$. The efficiency of the multi step method in comparison to a single step method becomes even more striking as the number of steps increases, albeit at the cost of greater experimental complexity.

\section{Bragg resonances  in the Josephson Junction}
\label{sec:bragg}

In this and the final Section we consider two types of dynamics,  which we shall refer to as Bragg scattering and Bloch oscillations, which are purely quantum effects (i.e.\ beyond mean-field) and so are not present at all  in the Josephson equations (\ref{phase_ev}) and (\ref{k_ev}). Rather, these types of motion only appear in the quantum treatment embodied by equations (\ref{eq:commutator}) and (\ref{qjH}). As their names suggest, these two phenomena can be viewed as analogues of well-known wave scattering effects in lattices.  Indeed, Haroutyunyan and Nienhuis \cite{nienhuis03} have previously given an analysis of the mathematical connections between atomic double-well systems and atomic diffraction, including the Bragg scattering analogy. Our purpose here is rather to present an intuitive physical discussion
and refer the interested reader to  \cite{nienhuis03} for more formal details.
  
  \begin{figure}[t]
\centering
{\includegraphics[width=8.0 cm]{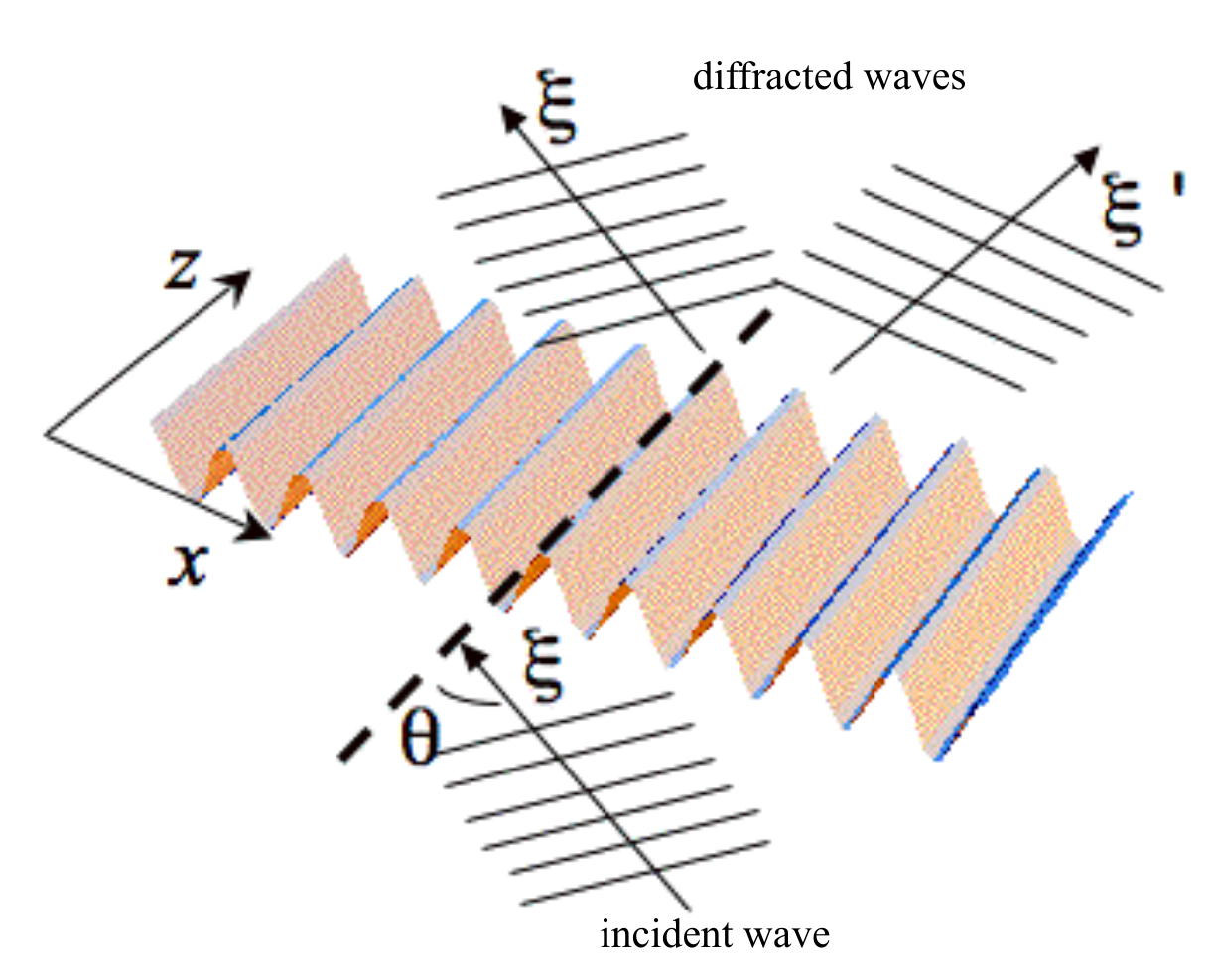}}
\caption{\scriptsize{Bragg scattering of a wave by a periodic potential}}
\label{fig:atomdiffraction}
\end{figure}

Consider a beam of non-interacting quantum particles that form a plane wave incident upon a lattice whose interaction with the particles is given by the periodic potential $V(x)=-V_{0} \cos K x$,  as depicted in Figure \ref{fig:atomdiffraction}. Such an interaction is realized in experiments where atoms are diffracted from a standing wave of laser light \cite{martin87,martin88,oberthaler96}, but the same basic form of refractive index also occurs in, for example, the description of the diffraction of light by ultrasonic waves, the original setting of the RN equations \cite{RN35,berry66}. If the total momentum of each particle is $\hbar \xi$, and the $x$-component is $\hbar q$, the atom beam travels at angle  $\sin \theta= q/\xi$ to the $z$ axis and is described by
the wave function $\Psi(x,z)=\exp [\mathrm{i}(\sqrt{\xi^2-q^2}z+qx)]$.  The periodicity of the lattice means that it can only transfer momentum to the particle wave function in discrete units of $\hbar K$ and so the diffraction of the particles is captured by the wave function  \cite{odell,odell98}
\begin{equation}
\Psi(x,z)=\exp [\mathrm{i}\sqrt{\xi^2-q^2}z] \sum_{n=-\infty}^{\infty} A_{n}(z) \exp [\mathrm{i}(n K +q)x]
\label{eq:atomicdiffractionwf}
\end{equation}
where $A_{n}(z)$ is the amplitude of the $n$th diffracted beam which travels at an angle $\tan \theta_{n}=(nK+q)/\sqrt{\xi^2-q^2}$ to the $z$ axis. The beam amplitudes $A_{n}(z)$ are functions of the depth $z$ through the diffracting medium: they describe dynamical diffraction in a thick grating. The wave function (\ref{eq:atomicdiffractionwf}) is only an approximate description of the true experimental situation because it assumes that the 
$z$-component of momentum, namely $\hbar \sqrt{\xi^2-q^2}$, is a constant of the motion unaffected by the entry and exit from the periodic potential. This holds when the particles' incident energy is much greater than the lattice potential $\hbar^{2} \xi^{2} / 2m \gg V_{0}$. Substitution of \eqref{eq:atomicdiffractionwf} into the Schr\"odinger equation  yields the equations for dynamical diffraction. In the paraxial approximation, which is valid when the diffraction angles are small enough that the term $\mathrm{d}^{2} A_{n}/ \mathrm{d} z^2$ can be neglected, these equations reduce to the RN equations. The RN equations for oblique incidence have exactly the same form as 
Equation \eqref{coupled_amp_finite1} that we used above to describe the asymmetric double-well Josephson junction, and this connection forms the basis of the analogy between the two cases. However, in the diffraction problem the parameters that appear in 
\eqref{coupled_amp_finite1} have a different physical origin: 
\begin{eqnarray}
\Lambda & = & \frac{2mV_{0}}{\hbar^2 K^2} \label{eq:lambadiffraction} \\
\Delta E & = & 2 \frac{q}{K} \label{eq:deltaEdiffraction} \\
 \tau & = & \frac{z K^2}{2 \sqrt{\xi^2-q^2}}. \label{eq:taudiffraction}
 \end{eqnarray}
  Thus, for the diffraction problem the parameter $\Lambda$ that determines whether the system is in the quantum ($\Lambda$ small) or classical  ($\Lambda$ large) regime depends upon whether the lattice is shallow or deep, respectively, in comparison to the recoil energy of the particles $\hbar^{2} K^2/ 2m$. The magnitude of the inbalance $\Delta E$ between wells in the double-well case is determined in the diffraction case by  the initial transverse wavenumber $q$ (i.e.\ the component of the initial momentum along the $x$-direction), or in other words, the angle of incidence. Finally, the dimensionless time parameter $\tau$ corresponds to the distance $z$ travelled  through the lattice potential. In atomic diffraction experiments the parameters (\ref{eq:lambadiffraction})--(\ref{eq:taudiffraction}) can all be tuned over large ranges, something which is much harder to do in conventional solid state X-ray or electron diffraction experiments. For instance, the dimensionless Planck's constant parameter $\Lambda$ can be controlled by the intensity of the laser beams forming the standing wave.
 However, one should not confuse in this discussion the interacting atoms in the double-well problem with the non-interacting atoms in the atomic diffraction problem since they play quite different roles in the models described here.

 
 Given the above the general treatment of the diffraction problem, we now specialize to Bragg scattering.  Bragg resonances occur when the conditions are met for constructive interference of waves reflected from the different planes of a lattice and are therefore a pure wave phenomenon that has no counterpart for particles.
 Bragg diffraction has been observed in a number of atomic diffraction experiments e.g.\ \cite{martin88} and \cite{oberthaler96}. The well known Bragg condition
states that the Bragg angles $\theta_{\mathrm{B}}$ satisfy
\begin{equation}
2d \sin \theta_{\mathrm{B}} = n \lambda 
\label{eq:bragg}
\end{equation}
where $d= 2 \pi /K$ is the period of the lattice, $\lambda= 2 \pi / \xi$ is the wavelength  of the incident wave and $n$ is a positive or negative integer. The Bragg condition (\ref{eq:bragg}) can be written in terms of the initial transverse wavenumber  as $q_{\mathrm{B}}= n K/2$. The Bragg scattering resonance couples the incident beam (the $A_{0}$ term in Equation \eqref{eq:atomicdiffractionwf}) to the $-n$th diffracted beam (the $A_{-n}$ term in Equation \eqref{eq:atomicdiffractionwf}) and this corresponds to specular reflection from the planes of the potential.  

When the periodic potential is weak , i.e.\ when $\Lambda$ is small, and we are close to a Bragg resonance, we can make the well known two-beam approximation and restrict our attention to only the $A_{0}$ and $A_{-n}$ beams, the amplitudes of the other diffracted beams being much smaller \cite{martin87,martin88,oberthaler96,odell98}. 
Taking, for example, the case where $n=1$, the two-beam solution to the RN equation  \eqref{coupled_amp_finite1} gives the following expressions for the intensities of the two beams as a function of time (equivalent to propagation depth through lattice)
\begin{eqnarray}
\vert A_{0}(\tau) \vert^{2}  & = & \cos^{2} \left[\sqrt{(\Delta E -1)^2+\Lambda^2} \ \tau/2 \right] \\ & & +\frac{(\Delta E -1)^2}{(\Delta E -1)^2 +\Lambda^2} \sin^{2}\left[\sqrt{(\Delta E -1)^2+\Lambda^2} \ \tau/2\right] \label{eq:twobeam1} \nonumber  \\
\vert A_{-1}(\tau) \vert^{2} &  = &  \frac{\Lambda^2}{(\Delta E -1)^2+\Lambda^2} \sin^{2}\left[\sqrt{(\Delta E -1)^2+\Lambda^2} \ \tau/2\right]. \label{eq:twobeam2}
\end{eqnarray}
We see that the width of the resonance as $\Delta E$ is varied is controlled by $\Lambda$. The acceptance angle (width of resonance) is wide for short times but narrows at longer times. This can be understood on the grounds of an energy-time uncertainty argument $\Delta \mathcal{E} \Delta t \ge \hbar$.
Notice also that the relative population of the two beams oscillates as a function of the time/depth through the grating $\tau$ with a period $2 \pi / \sqrt{(\Delta E -1)^2+\Lambda^2}$. In the field of electron diffraction these oscillations are known  as ``pendell\"osung'' (pendulum solutions). The cases where $\vert n \vert =2,3,4 \ldots$ are more complicated but analytic expressions can be obtained by adiabatically eliminating the intermediate amplitudes so that one still has a two-beam solution \cite{shorebook}. 

\begin{figure}[t]
\centering{
\includegraphics[width=0.45\columnwidth]{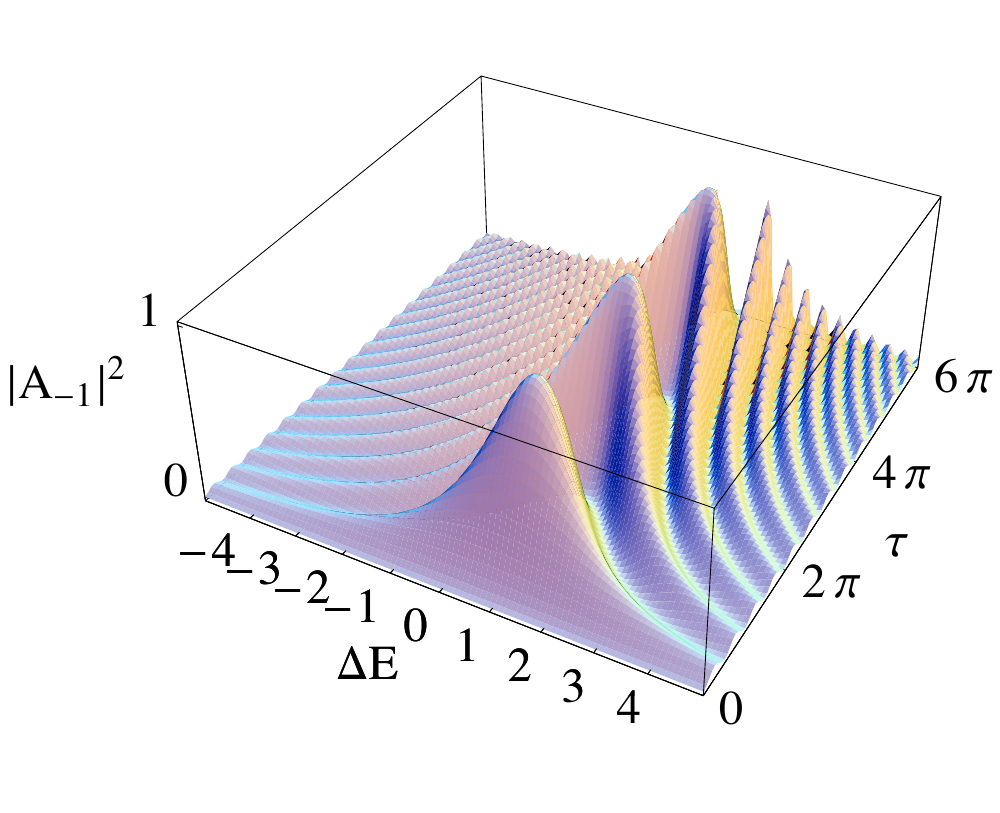}}
\caption{\scriptsize{$\Lambda=1$. Intensity of Bragg diffracted beam $\vert A_{-1} \vert^2$, as given by the analytic two-beam approximation (\ref{eq:twobeam2}), plotted as a function of $\Delta E=2 q /K$ (which is related to the angle of incidence via $\sin \theta = q / \xi$) and time $\tau$. For short times the resonance is very wide, much wider than $\Lambda$. For longer times   $\vert A_{-1} \vert^2$ varies rapidly as a function of $\Delta E$ and the central resonance narrows. The periodic oscillations in $\tau$  are known as pendell\"osung oscillations in X-ray diffraction theory.}}
\label{fig:twobeam}
\end{figure}

\begin{figure}[t]
\centering
\subfigure[]{\includegraphics[width=0.45\columnwidth]{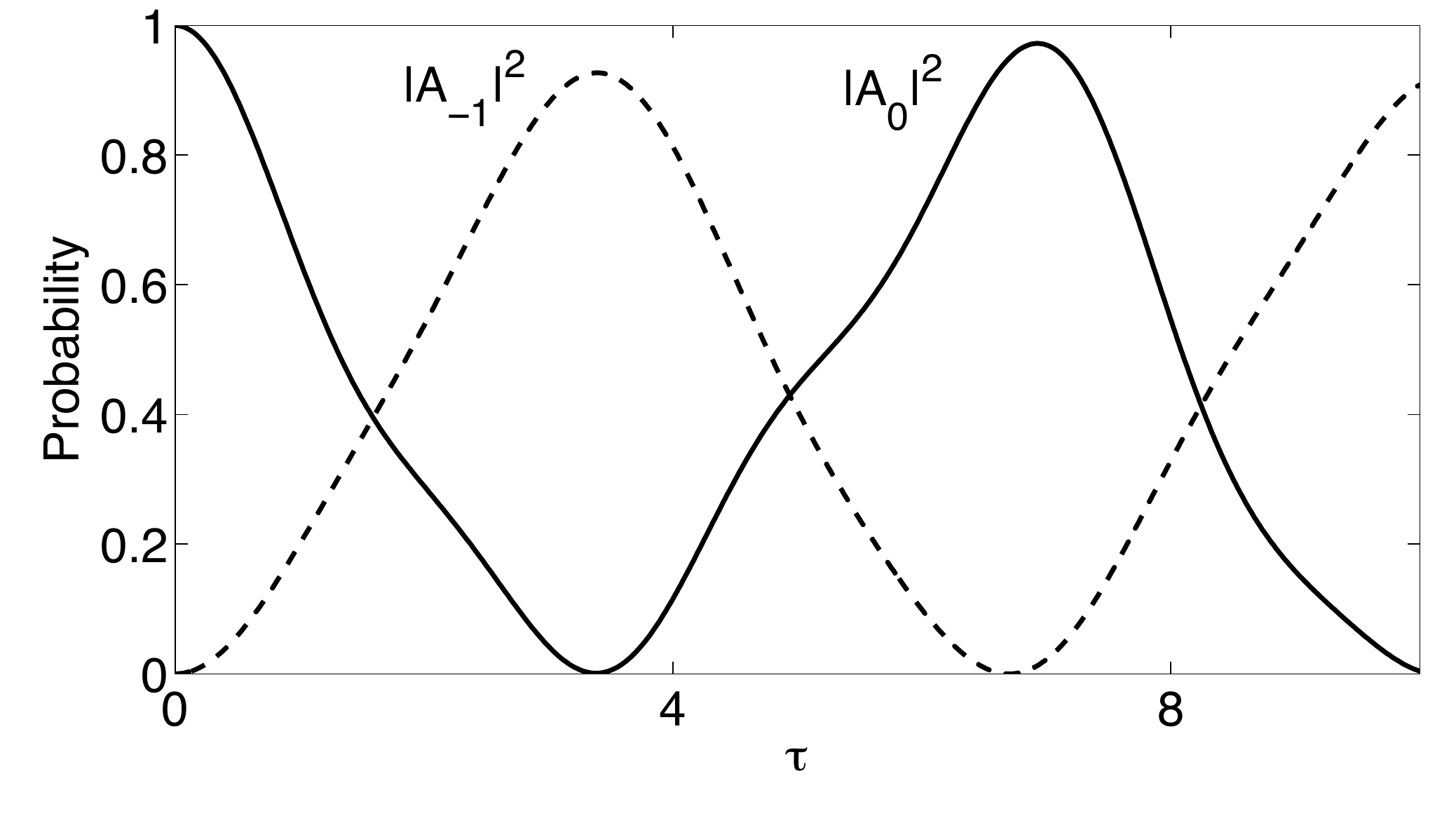}}
\subfigure[]{\includegraphics[width=0.45\columnwidth]{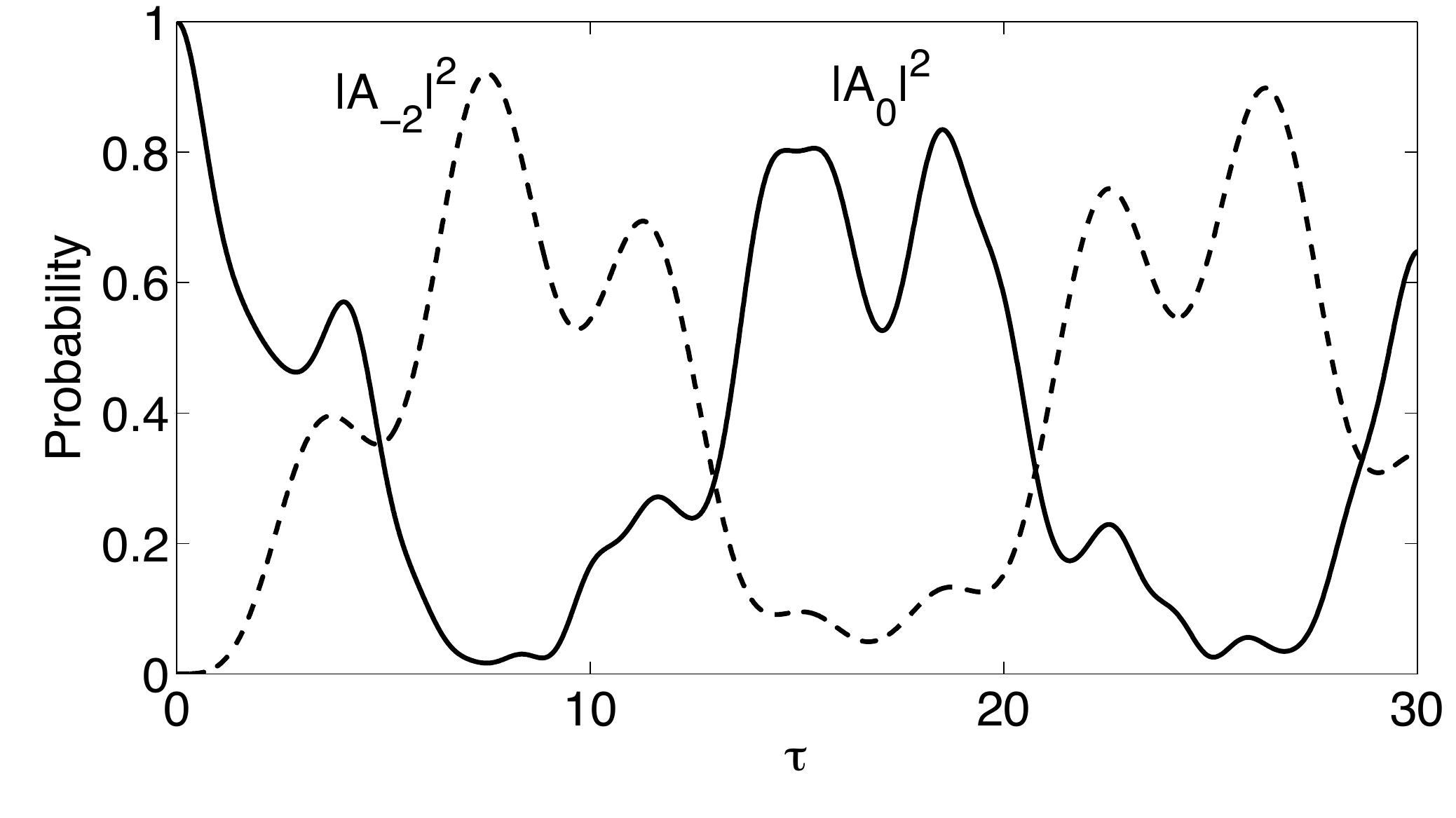}}
\subfigure[]{\includegraphics[width=0.45\columnwidth]{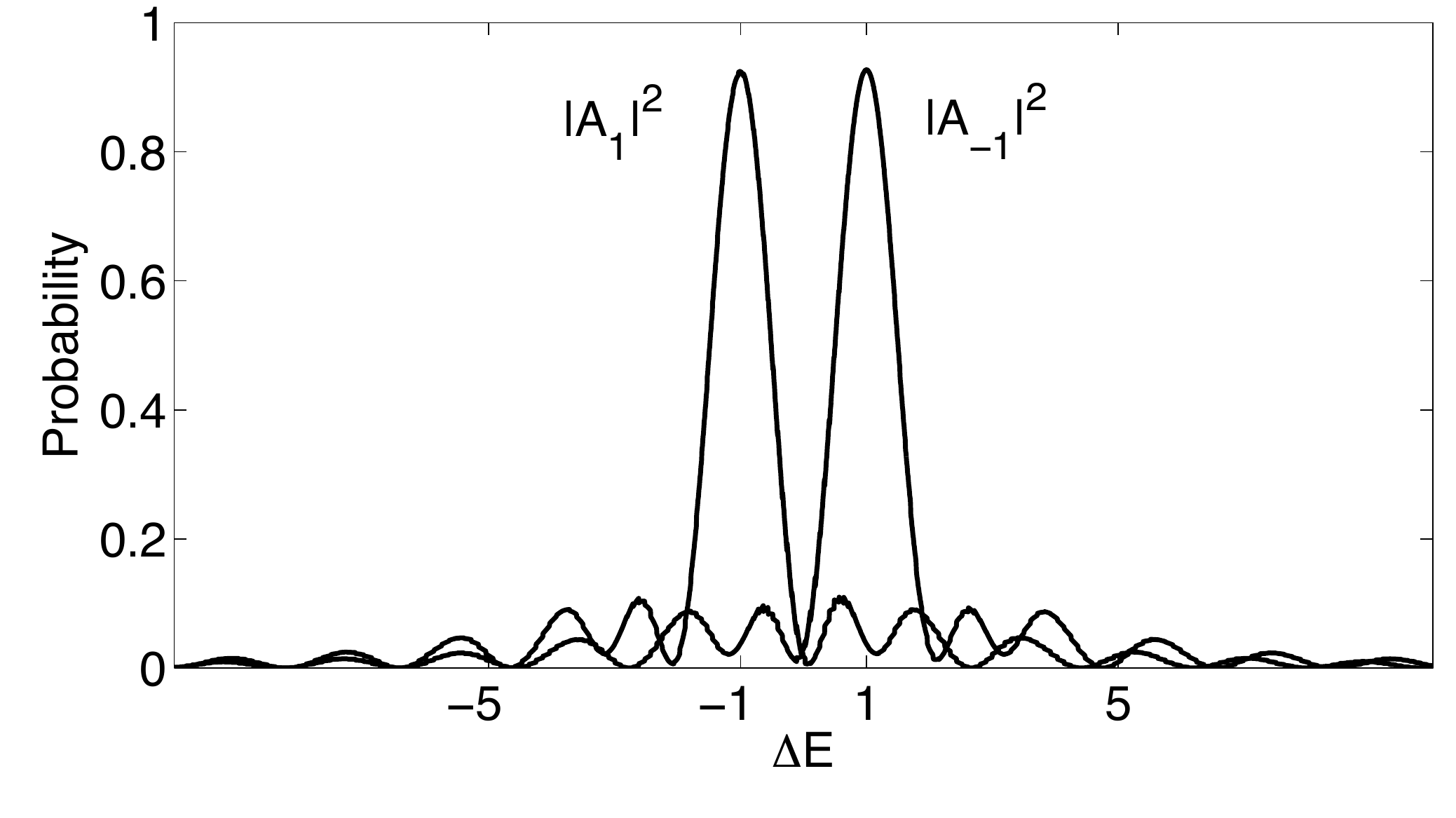}}
\subfigure[]{\includegraphics[width=0.45\columnwidth]{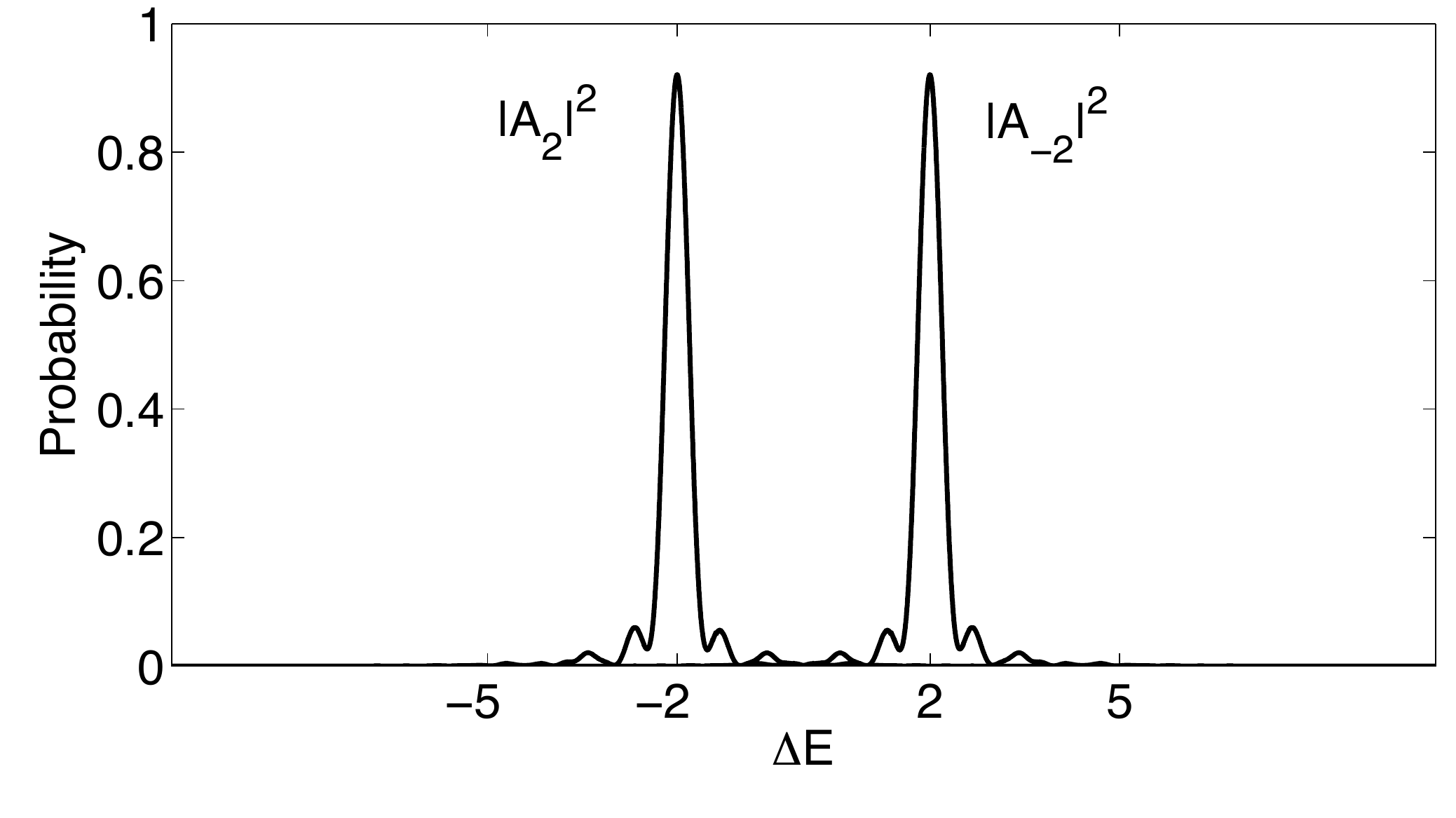}}
\caption{\scriptsize{$\Lambda=1$. Upper panels: Pendell\"osung oscillations found by numerical solution of the full RN equations \eqref{coupled_amp_finite1}. In the diffraction problem the pendell\"osung are between two diffracted beams as a function of depth through the lattice. In the Josephson problem the pendell\"osung become oscillations between two number difference states as a function of time. Which two diffracted beams (number states)  are involved depends on the value of the incident angle (energy asymmetry) $\Delta E=2 q/K$. In (a) $\Delta E =1$ and in (b) $\Delta E=2$. All the other beams (states) which are not shown have much smaller populations. Lower panels: the coupling between the two number difference eigenstates (diffracted beams) has a clear resonance structure as a function of $\Delta E$. Resonances occur at  $\Delta E=\ldots,-2,-1,1,2,\ldots.$ The lower panels are shown at times approximately equal to half  
their respective pendell\"osung periods: c) $\tau=3.4045$  (d) $\tau=7.1525$.}}
\label{fig_brag_res}
\end{figure}

The analytic result in the two-beam approximation (\ref{eq:twobeam2}) for the intensity of the first Bragg diffracted beam $\vert A_{-1}\vert^{2}$ is plotted in Figure \ref{fig:twobeam}. Meanwhile, Figure \ref{fig_brag_res} displays the results of an exact numerical solution of the full set of time-dependent RN equations \eqref{coupled_amp_finite1}. 
As expected, we find that the approximate analytical results agree with the numerical ones providing $\Lambda$ is small and we are close to a Bragg angle.
In particular, the top row of graphs in Figure \ref{fig_brag_res} show the temporal evolution of the two strongly-coupled beams at the Bragg resonances (a) $\Delta E=1$ and (b) $\Delta E =2$. The pendell\"osung are clearly visible, particularly in (a). In the bottom row we see the classic Bragg resonance structure as a function of the incident angle $\Delta E$.
  The value of $\tau$  chosen for the bottom row of Figure \ref{fig_brag_res} is such that the population of the relevant Bragg scattered beam is at a peak, i.e. half a pendell\"osung period.

We can now use these simple results from the theory of the diffraction of non-interacting particles to predict phenomena for the case of interacting particles in a double-well potential. Recalling that, according to Equation (\ref{periodicwavefunction}), 
the amplitudes $A_{0}$ and $A_{-n}$ refer to number difference states we see that
the analogue of a Bragg resonance in the Josephson problem is a tunnelling resonance between two number difference, i.e.\ the swapping of a precise number of atoms between the two wells \cite{nienhuis03} as the amplitudes $A_{0}$ and $A_{-n}$ oscillate. In order to achieve the analogous initial conditions as in the ``real'' Bragg scattering scenario described above, where the incident atom beam was well collimated so that at $\tau=0$ we have $A_{0}=1$ and $A_{n \neq 0}=0$, it is necessary to 
start in a single population difference state. For simplicity, in what follows we shall assume that the initial state is  the precisely balanced state with $k=0$ (this is the most likely result in a balanced double-well) and so the initial conditions are $A_0=1$ and $A_{n\neq0}=0$. We shall return at the end of the Section \ref{sec:band} to consider the  challenging demands this places on an experimental realization. At time $\tau=0$ we suddenly switch on an energy asymmetry $\Delta E$ which is held at a constant value. The temporal evolution of the amplitudes is then exactly that shown in Figures \ref{fig:twobeam} and \ref{fig_brag_res}. As shown in the top row of Figure \ref{fig_brag_res}, by tuning the energy asymmetry $\Delta E$ to a resonance we find  pendell\"osung between two number difference states.  At resonance, the Josephson pendell\"osung oscillations between the states $n=0$ and $n=-1$ occur with a period
\begin{equation}
\tau_{\mbox{pendell{\"o}sung}}= 2 \pi/ \Lambda \ .
\end{equation}
This has a different dependence upon $\Lambda$ than the Josephson oscillation period (as given by Equation 
\eqref{eq:plasmonperiod}) and, combined with the resonance behaviour, would provide an experimental signature of the Bragg analogue. Although we have only plotted the $\vert n \vert =1$ and $\vert n \vert =2$ Bragg resonances, corresponding, respectively, to the tunnelling oscillation of one and two atoms between the wells, it is possible to isolate resonances involving the precise transfer of larger numbers of atoms. 

\section{Band Structure, Bloch oscillations and Schr\"{o}dinger cats in the Josephson Junction}
\label{sec:band}

\begin{figure}[t]
\centering
\subfigure[]{\includegraphics[width=0.45\columnwidth]{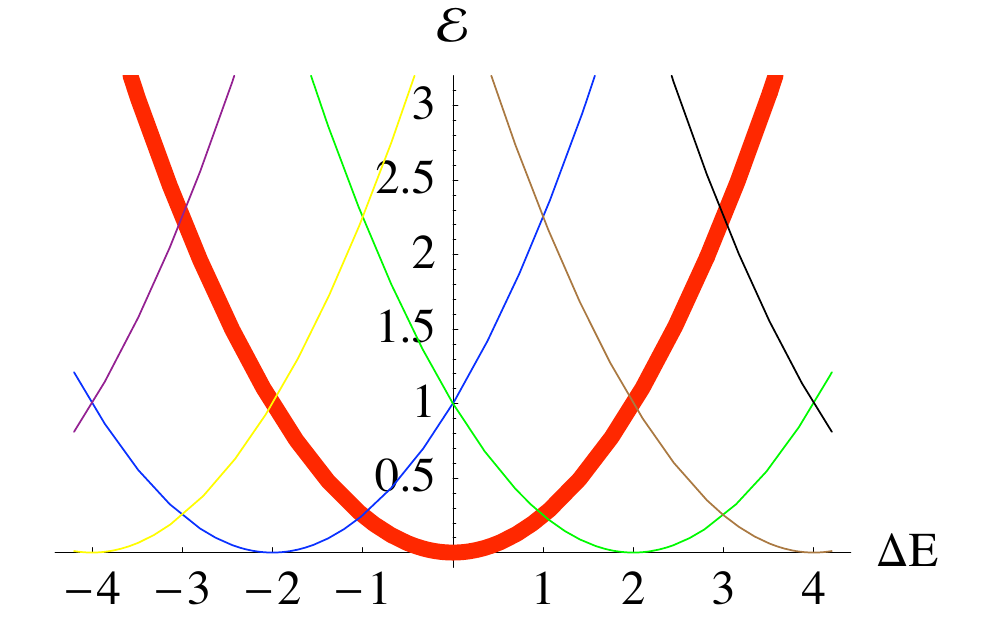}}
\subfigure[]{\includegraphics[width=0.45\columnwidth]{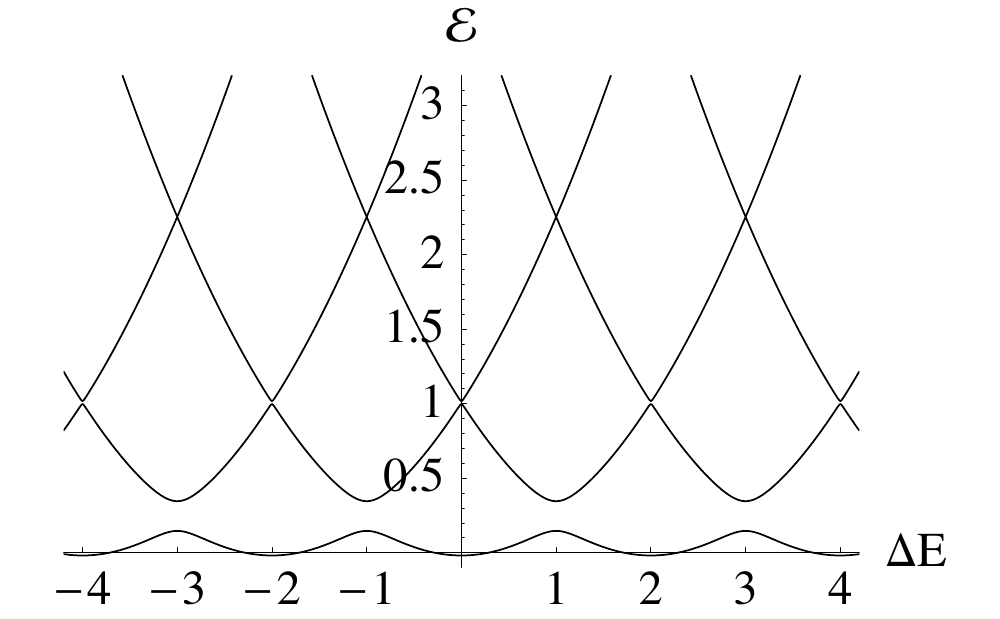}}
\caption{\scriptsize{Energy bands of a quantum particle in a sinusoidal potential as a function of the dimensionless quasimomentum $\Delta E=2 q/K$. (a) Shows the energies of the relevant momentum basis states in the absence of the sinusoidal potential and (b) shows the actual energy bands in the presence of the sinusoidal potential with $\Lambda=0.2$.
In particular, the thick red parabola in (a) is the dispersion relation of a free particle. This free particle corresponds to the incident wave in the diffraction problem discussed in Section \ref{sec:bragg}. The other parabolas also represent free particles but their momenta have been shifted by an integer number of units of $\hbar K$.  Together these states form the natural basis states inside the sinusoidal potential (see text).  
The energy level structure shown in (b) is calculated using  the Raman-Nath equations (\ref{eq:RNbandstructure}).  We see that the parabolic basis states are coupled such that their intersections become avoided crossings.  Bragg resonances lie at the places where the thick red parabola crosses the other parabolas. Bloch oscillations correspond to an adiabatic evolution along a single band as $\Delta E$ is varied. 
 }}
\label{fig:bandstructure}
\end{figure}

It is instructive to re-cast our discussion of diffraction in a lattice in terms of band theory \cite{ashcroft+mermin}. We shall see that this gives a deeper understanding of the Bragg resonance phenomena in a double-well and also leads naturally to analogies to a class of adiabatic phenomena related to Bloch oscillations.
According to the Bloch theorem, the stationary eigenfunctions for a quantum particle in the lattice potential $V(x)=-V_{0} \cos K x$ can be factorized 
\begin{equation}
\psi^{q,j}(x)=\exp (\mathrm{i}q x) U^{q,j}(x)
\end{equation} 
in terms of a plane wave part $\exp (\mathrm{i}q x)$ and a spatially dependent amplitude part  $U^{q,j}(x)$ which is periodic with the same period as the lattice: $U^{q,j}(x+2 \pi/K)=U^{q,j}(x)$. The wave function $\psi^{q,j}(x)$ depends on two quantum numbers: $q$
which gives the quasimomentum $\hbar q$, and the band index $j$.
In the diffraction problem presented above the quasimomentum  is nothing but the $x$-component of the momentum of the incident wave. The band index $j$ is also already familiar to us: we use the same symbol we used to label the eigenenergies and eigenvectors of the Josephson problem because it has an identical meaning here. However, because in the Josephson problem the wave function must be $2 \pi$-periodic, there we had to set $q=0$, thus seemingly loosing the full richness of the general problem of a quantum particle in a periodic potential.
Substituting  $\psi^{q,j}(x)$ into the time-independent Schr\"{o}dinger equation yields
\begin{equation}
\frac{\hbar^{2}}{2m}\left(\mathrm{i} \frac{\mathrm{d}}{\mathrm{d} x}+q \right)^{2}U^{q,j}(x)
-V_{0} \cos Kx \ U^{q,j}(x) = \epsilon^{q,j} U^{q,j}(x). \label{eq:schrodingerband}
\end{equation}
The eigenenergy $\epsilon^{q,j}$, which is a function of both the band index $j$ and the quasimomentum $q$, has the well known band structure shown in Figure \ref{fig:bandstructure}.
The periodicity of $U^{q,j}(x)$ means that we can expand it as a Fourier series
\begin{equation}
U^{q,j}(x)= \sqrt{\frac{K}{2 \pi}} \sum_{n=-\infty}^{\infty} B_{n}^{q,j}\exp(\mathrm{i}nKx).
\label{eq:Uexpansion}
\end{equation}
Substituting this into (\ref{eq:schrodingerband}) yields 
\begin{equation}
\left(n^2+ 2 \frac{q}{K} \, n \right) B_{n}^{q,j}- \frac{\Lambda}{2} (B_{n+1}^{q,j}+B_{n-1}^{q,j})=(\mathcal{E}^{q,j}-\frac{q^{2}}{K^{2}})B_{n}^{q,j}. \label{eq:RNbandstructure}
\end{equation}
In this formula $\Lambda$ is the parameter defined in Equation (\ref{eq:lambadiffraction}), and the dimensionless energy $\mathcal{E}^{q,j}$ is given by $\mathcal{E}^{q,j}=2 m \epsilon^{q,j} / \hbar^{2} K^2$. Comparing (\ref{eq:RNbandstructure}) with the RN Equations (\ref{rec_rel2}) for the asymmetric double-well, we see that the two are identical if: (a) we assign $2 q/K=\Delta E$, just as we already did in  (\ref{eq:deltaEdiffraction}), i.e. we recognize that the tilt $\Delta E$ is really equivalent to a quasimomentum, and (b) we put
\begin{equation}
\mathcal{E}^{q,j}-\frac{q^{2}}{K^{2}}=\mathcal{E}^{q,j}-\frac{\Delta E^2}{4}=E^{j}_{a} \ .
\label{eq:redefinition}
\end{equation}
Recall that $E^{j}_{a}$ is the eigenvalue corresponding to the $j$th eigenvector of the asymmetric double-well and so is a function of $\Delta E$.  

The fact 
that the tilt asymmetry $\Delta E$ between the wells allows us to introduce a term that plays the role of a quasimomentum into the Josephson problem even though the wave function is strictly $2 \pi$-periodic is of considerable significance; it is crucial for many of the phenomena discussed so far and also for Bragg resonances and Bloch oscillations (see below). In the Josephson problem the quasimomentum is introduced directly into the hamiltonian (see Equation \eqref{jHa}) rather than via the wave function: it is a parameter rather than a dynamical variable.

An example of the band structure of the eigenenergies $\mathcal{E}^{q,j}$ obtained by solving the RN equations (\ref{eq:RNbandstructure}) is shown in Figure \ref{fig:bandstructure}. In the absence of the periodic potential ($\Lambda=0$) the eigenstates are $\exp(\mathrm{i}qx)$ corresponding to a free particle. Their dispersion relation $\hbar^{2} q^{2}/2m$ is plotted as the thick red parabola in Figure \ref{fig:bandstructure}(a).  Also shown in Figure \ref{fig:bandstructure}(a) are the displaced parabolas $\hbar^{2} (q + nK)^{2}/2m$ centered at the positions $q= -nK$ which are the dispersion relations of the momentum eigenstates $\exp[\mathrm{i}(q + n K)x]$, where $n=0, \pm1, \pm 2, \pm 3 \ldots$. These are the basis states used  in the expansion (\ref{eq:Uexpansion}) of the wave function inside in the periodic potential. Inside the periodic potential the momentum basis states  are coupled, which is the physical content of the the RN equation (\ref{eq:RNbandstructure}). For shallow lattices ($\Lambda<1$) this coupling is weak and is only significant close to the points where the free parabolas
would otherwise have crossed and caused a degeneracy. The effect of the coupling is to turn the degeneracies into
avoided crossings, thereby forming continuous energy bands as a function of $q$, separated by gaps of forbidden energies, as can be seen in 
 Figure  \ref{fig:bandstructure}(b). Technically, each band is a superposition of an infinite number of basis states although when the potential is weak  only a small number make a significant contribution at any particular value of $q$. However, the particular combination required depends on $q$. Take, for example, the first Brillouin zone (BZ) defined as the region $-K/2 \le q < K/2$. When $\Lambda$ is small the first band in the first BZ can be adequately described  using only the $n=0$ state $ \exp[\mathrm{i}qx]$ and the two $\vert n \vert=1$ states  $\exp[\mathrm{i}(q-K)x]$ and $\exp[\mathrm{i}(q+K)x]$, which are needed to take into account the coupling between the free particle states at the edges of the first BZ. For instance, at the right hand edge of the first BZ, where $q=K/2$, there is a strong coupling between between the $n=0$ and $n=-1$ free particle states $ \exp[\mathrm{i}qx]$ and $\exp[\mathrm{i}(q-K)x]$, respectively,  whose energies would otherwise be degenerate. Using degenerate perturbation theory one finds that the band splitting at the avoided crossing between the first and second bands is given by 
\begin{equation}
E^{j=2}_{a}-E^{j=1}_{a}=\Lambda. 
\label{eq:bandgap}
\end{equation}
On the left hand side of the first Brillouin zone at $q=-K/2$ it is the $n=0$ and $n=1$ free particle states $ \exp[\mathrm{i}qx]$ and $\exp[\mathrm{i}(q+K)x]$, respectively, that are strongly coupled. 

Bragg resonances have a simple interpretation in this energy band picture: they correspond to conservation of energy and momentum. Referring to  Figure \ref{fig:bandstructure}(a), Bragg scattering takes place where the dispersion relation of the incident wave (thick red parabola) crosses those of the other basis states. We see that 
Bragg resonances correspond to having an incident quasimomentum given by $q_{\mathrm{B}}=(n/2) K$ or, equivalently, $\Delta E_{\mathrm{B}}=n$ where $n$ is a postive or negative integer. Bragg scattering therefore takes place at the edges of the Brillouin zones. Consider a wave $ \exp[\mathrm{i}qx]$ (ignoring the trivial dependence on $z$) incident somewhere in the first BZ. If it enters with a value of $q$ not close to $q=\pm K/2$  then, according to the above discussion, when $\Lambda$ is small this wave coincides with the eigenfunction describing the first band and it propagates through the lattice unmodified and hence undeflected. The weak potential is not capable of coupling the wave to higher bands unless the wave enters at a Bragg angle. In the case that the wave is incident close to either of the first Bragg angles $\Delta E = \pm 1$ the wave function $ \exp[\mathrm{i}qx]$ of the incident wave is seen to be a superposition of the first and second band eigenfunctions. The beating between the two bands leads to the  pendell\"osung. 
Similarly, at $\Delta E=\pm 2$ there are Bragg resonances between the incident wave and the second and third bands and so on for larger angles of incidence. The magnitude of the band gap at the avoided crossings gives an indication of the strength of the corresponding Bragg resonance. Generalizing (\ref{eq:bandgap}) we find that the band gap between the $j+1$ and the $j$th bands scales as $\Lambda^{j}$ and so the higher resonances become weaker when $\Lambda <1$. In the case of atomic diffraction this can be understood physically by noting that $2 j$ photons must be exchanged between the atoms and the laser beams at the $j$th Bragg resonance.

\begin{figure}[t]
\centering
\subfigure[]{\includegraphics[width=0.45\columnwidth]{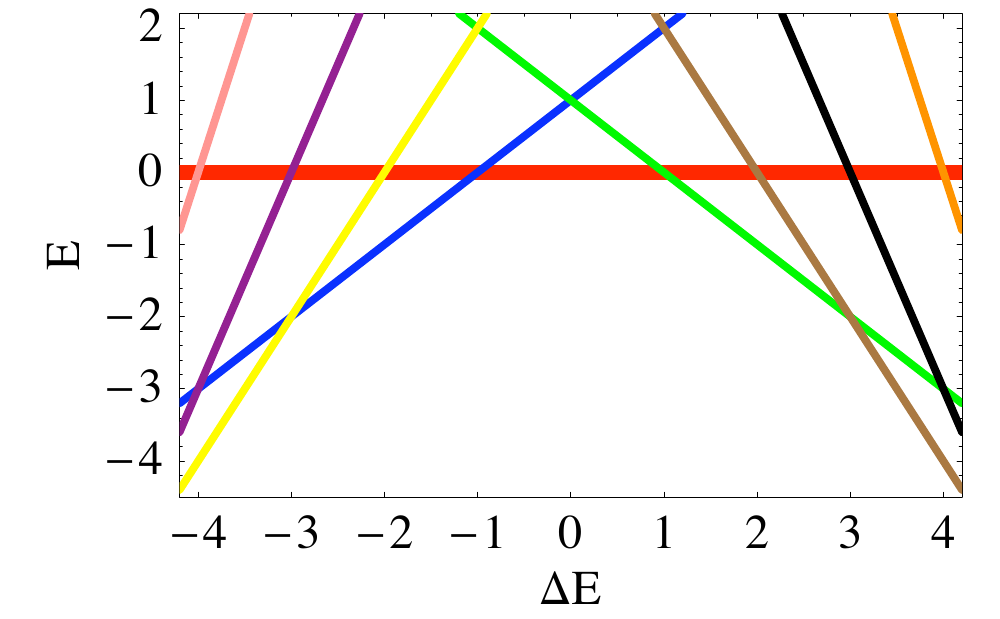}}
\subfigure[]{\includegraphics[width=0.45\columnwidth]{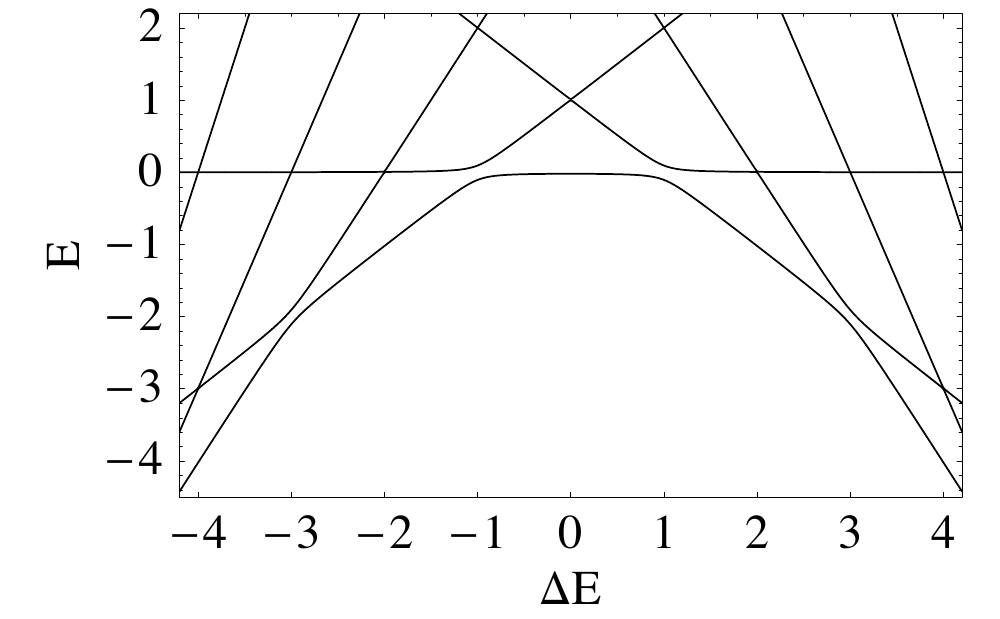}}
\caption{\scriptsize{Energy band structure of the Josephson double-well system.
In the Josephson double-well problem the role of the quasimomentum is played by the tilt asymmetry $\Delta E$ between the double-wells.
Each line in (a) corresponds to the energy of a particular number difference eigenstate as given by Equation (\ref{eq:numberstateenergies}). These states are the eigensolutions of the double-well problem in the absence of tunnelling ($\Lambda=0$). (b) shows the actual energy level structure in the presence of tunnelling ($\Lambda=0.2$). Note that every apparent level crossing in (b) is actually an avoided crossing.  The bottom curve is the ground band $E^{1}_{a}$, the next up is for $E^{2}_{a}$, the next is for $E^{3}_{a}$ etc. The eigenenergies $E^{j}_{a}$ of the asymmetric double-well for a particular value of $\Delta E$ correspond to vertical slices through this band structure.
The thick red horizontal line in (a) is the energy of the $n=0$ number difference eigenstate and is equivalent to the thick red parabola shown in Figure \ref{fig:bandstructure}(a). In fact, using the transformation (\ref{eq:redefinition}) between $\mathcal{E}^{q,j}$ and $E^{j}_{a}$ we find that the entire band structure shown here is equivalent to that shown in Figure \ref{fig:bandstructure}. }}
\label{fig:bandstructureRN}
\end{figure}

The foregoing analysis suggests a band structure interpretation of tunnelling resonances in the Josephson problem. The energy eigenvalues $E^{j}_{a}$ of the RN equations (\ref{rec_rel2}) as a function of the tilt asymmetry $\Delta E$ between the wells  have the structure shown in Figure \ref{fig:bandstructureRN}.  In the quantum regime $\Lambda$ is small meaning that the interaction energy $E_{C}$ dominates the tunnelling energy $E_{J}$. In this limit it makes sense to choose the basis states to be the number difference eigenstates 
$\exp[\mathrm{i}n \phi ]$, which are analogous to the momentum eigenstates we used in the diffraction problem, and indeed the RN equations are written in this basis. However, unlike in the diffraction problem, the quasimomentum phase factors $\exp[\mathrm{i} q x ]=\exp[\mathrm{i} \Delta E/2 \, \phi]$ are not included in the basis states because they occur in the hamiltonian not the wave function, which must be $2\pi$-periodic as noted above. Furthermore, the energies associated with the number difference  eigenstates are not parabolic but are linear in $\Delta E$ 
\begin{equation}
E(n, \Delta E)=  \  n^2+ n \, \Delta E \, . 
\label{eq:numberstateenergies}
\end{equation}
These energies are shown as the straight lines in Figure \ref{fig:bandstructureRN}(a).
As can be seen in Figure \ref{fig:bandstructureRN}(b), there is a band structure  associated with the Josephson double-well problem, but it is distorted by gravity. When we plot the bands in terms of their ``true'' energies $E^{j}_{a}$ (rather than $\mathcal{E}^{q,j}$) the forbidden energy gaps in the spectrum disappear except
 for one, just below $E=0$, although this itself becomes extremely narrow for large $\Delta E$, and in any case our hamiltonian eventually breaks down at large tilts where the condition $\Delta E \ll N$ is violated. However, with the trivial redefinition (\ref{eq:redefinition}) of the energy from $E^{j}_{a}$ to $\mathcal{E}^{q,j}$ the standard band structure shown in Figure \ref{fig:bandstructure} is recovered.

Let us examine in more detail the band structure of the Josephson double-well problem as illustrated in Figure \ref{fig:bandstructureRN} for small $\Lambda$. For small tilts ($\Delta E \ll 1$) the first band is adequately described by 
the number difference eigenfunction $\exp[\mathrm{i} 0 \phi ]$ corresponding to zero atom number difference between the wells and represented by the thick red horizontal line in Figure \ref{fig:bandstructureRN}(a).  Similarly, the second band in this range consists of a superposition of the states $\exp[\mathrm{i} \phi ]$ and $\exp[-\mathrm{i} \phi ]$, which describe cases where one particle has been transferred to the left side and one has been transferred to the right side, respectively.
Near the Bragg resonance at the right hand edge of the first BZ where $\Delta E =1$, the states  $\exp[\mathrm{i} 0 \phi ]$ and $\exp[-\mathrm{i} \phi]$ are strongly coupled leading to the avoided crossing shown in Figure \ref{fig:bandstructureRN}(b). By analogy with the diffraction case, the first Josephson Bragg resonance occurs when the double-well system is \emph{suddenly} tilted from $\Delta E=0$ to $\Delta E=1$. This projects the initial $n=0$ state over the superpositions of the $n=0$ and $n=1$ states that make up the eigenfunctions giving the first two bands at $\Delta E=1$.  A single atom will then oscillate back and forth between the two wells. 

The successively higher Bragg resonances, i.e.\ the places where the thick red line in Figure \ref{fig:bandstructureRN}(a) crosses the lines located successively further away from the origin, correspond to places where the $n=0$ balanced state couples strongly to states with successively higher values of $n$. 
When the high $n$ state is macroscopically large, $n=P$, say, where $P \gg 1$, the system oscillates between having a population difference of precisely zero and precisely P atoms. Denoting by $\vert n \rangle$ the number difference kets,  at a Bragg resonance the double-well is in a superposition of the symmetric and antisymmetric Schr\"{o}dinger cat states $\Psi_{\pm}^{\mathrm{SC}}=(\vert 0 \rangle \pm  \vert P \rangle)/\sqrt{2}$, which are themselves superpositions of macroscopically distinguishable states. The lower of the two bands making up the avoided crossing corresponds to the symmetric state and the higher to the antisymmetric state. In order to realize a single Schr\"odinger cat state, as opposed to a superposition, i.e.\ either $\Psi_{+}^{\mathrm{SC}}$ or $\Psi_{-}^{\mathrm{SC}}$, it is necessary to occupy just one band. One way to achieve this is via Bloch oscillations rather than Bragg resonances.

Bloch oscillations are in a sense the conjugate phenomenon to Bragg scattering because they rely on adiabatic evolution rather than sudden projection.
Conventionally, Bloch oscillations occur when a constant external force $F$ is applied to a quantum particle in a periodic potential, e.g.\ an electron in a crystal lattice subject to a constant electric field.  It transpires that in a periodic potential it is the quasimomentum which obeys Newton's second law under the action of the force
\begin{equation}
q(t)=q_{0}-F t/ \hbar
\end{equation}
a result known as Bloch's acceleration theorem \cite{bloch28,zener34,holthaus96}. In this formula $q_{0}$ is the value of $q$ at $t=0$.
Thus, under the influence of the external force the quasimomentum travels along the band at a linear rate, and if this motion is adiabatic the system remains in the same band. The  periodicity of the band structure means that rather than uniformly accelerating the particle undergoes an oscillatory motion.
Non-adiabatic corrections to this evolution involve Landau-Zener transitions to other bands at the avoided crossings. By contrast, Bragg scattering is inherently non-adiabatic and involves suddenly exciting a superposition of two bands.

In the case of the double-well Josephson problem an analogy to Bloch oscillations occurs when the tilt asymmetry is slowly changed. Imagine starting at $\Delta E=0$ in the first band in Figure \ref{fig:bandstructureRN} and slowly increasing $\Delta E$, thereby sweeping the system along the first band. The initial state therefore has the same number of atoms in each well. If the sweep proceeds adiabatically the system remains in the first band. At the first avoided crossing it smoothly evolves into the new ground state so that it is now in a state where exactly one particle has been transferred into the lower well. If the adiabatic sweep continues we move successively through states with more and more atoms in the lower well. This can continue until all the particles are in the lower well. 
In Figure \ref{fig:bandstructureRN}(b) it appears that the avoided crossings between the first and second band become smaller and smaller further from the origin, and so the sweep must become slower and slower to remain adiabatic, but this is an illusion. From the transformation (\ref{eq:redefinition}) we see that the bands gaps are identical to those in Figure \ref{fig:bandstructure}(b) where one can see that they are independent of $\Delta E$.
Note, however, that our approximate hamiltonian (\ref{jHa}) is only valid as long as $\Delta E \ll N$. 
When this condition is violated the band gaps will become a function of $\Delta E$ and so the adiabaticity condition will in general depend on $\Delta E$ for large asymmetries.   

If the avoided crossing traversals are not entirely adiabatic we mix in some Bragg scattering-like character into the evolution. Consider a non-adiabatic traversal of the avoided crossing between the first BZ and second BZ: the system is then put into a superposition of the first two bands (i.e.\  a superposition of the balanced state and the state with one particle transferred) and the system will be set into oscillation. Non-adiabatic traversals imply that the final state when $\Delta E$ becomes large will not be the ground state with all the atoms in one well but rather an excited state with atoms oscillating between the wells.

If one wishes to excite a single Schr\"{o}dinger cat state $\Psi^{\mathrm{SC}}$ in the double-well it is necessary to use the time oscillating asymmetry method of exciting a single eigenstate as described in Section \ref{sec:timemodulated}. Consider starting, as before, in the precisely balanced state $\vert 0 \rangle$ at $\Delta E=0$. If one were to simply increase $\Delta E$ adiabatically then at the first avoided crossing one generates the state $(\vert 0 \rangle + \vert -1 \rangle)/\sqrt{2}$. Remaining in this first band only generates superpositions between the successive neighbouring parabolas shown in Figure \ref{fig:bandstructure}(a) which only differ by a particle number difference of a single particle. To generate a superposition of two states differing by a macroscopically large particle number one should use the time oscillating asymmetry to excite a higher band. In fact, remaining on the $\Delta E =0$ axis generates superpositions of the form  
 $(\vert n \rangle + \vert -n \rangle)/\sqrt{2}$. However, if one tries to directly excite to an avoided crossing where two bands nearly touch there is a considerable risk of exciting both bands. Therefore, a superior method would be to first adiabatically sweep the tilt asymmetry whilst still in the first band to the point $\Delta E = 0.5$ where the two higher lying bands are maximally separated, apply the time oscillating tilt to excite just one of them, then adiabatically sweep $\Delta E$ back to zero to reach the high lying avoided crossing.

Finally, we briefly discuss some points concerning the realization of Josephson Bragg scattering and adiabatic phenomena such as Bloch oscillations. Bragg scattering requires that the initial state is a single population difference state. One way this can be achieved is if the double well begins in its ground state in the quantum limit where $\Lambda \ll 1$ (low tunnelling rate regime). Then, if the tilt is suddenly changed to one of the Bragg angles $\Delta E_{\mathrm{B}}=n$ the initial state is projected over the two bands with which it is resonant (see Figure \ref{fig:bandstructure}). Because the Bragg resonances have a finite width there is some tolerance to errors in $\Delta E$. Higher Bragg resonances involving the tunnelling of a large but precise number of atoms are harder to achieve because the higher resonances are weaker when $\Lambda < 1$ as mentioned above.  
Going to larger $\Lambda$ has the effect of mixing in a larger number of population difference basis states into each band. This means that one can no longer claim, for example, that the ground band at $\Delta E=0$ solely consists of the $n=0$ state. 
 When it comes to Bloch oscillations driven by a sweep in $\Delta E$, adiabaticity is most likely to be maintained in the semiclassical (meanfield) limit of large $\Lambda$ where the tunnelling ensures a large splitting between states at the avoided crossings. However, if one is interested in achieving a precise number difference of atoms between the two wells using this adiabatic method then, for the reasons already mentioned, it pays to have a small value of $\Lambda$. All of these considerations need to be set in the context of the very significant experimental challenge of cooling the double well system to its ground state \cite{gati2}. This is the desirable initial state for demonstrating Bragg scattering, Bloch oscillations and also Schr\"{o}dinger cat states. The robustness of these phenomena to finite temperature effects will be the subject of future work.

\section{Conclusions}

We have analysed the atomic Josephson junction from the point of view that the parameter $1/\sqrt{\Lambda}=\sqrt{E_{c}/(2 E_{J})}$ is proportional to Planck's constant. This provides a simple way to predict when the dynamics obeys the classical (mean-field) Gross-Pitaevskii theory and when it must be quantised. If the system is set into motion by suddenly removing an asymmetry between the wells both low energy Josephson plasmons and high energy rotor  excitations are accessible, depending on the magnitude of the asymmetry. For small $\Lambda$ the system is very quantum and both excitations rapidly deviate from the classical mean-field predictions, the former by a change in frequency and the latter by undergoing collapses and revivals. We give an expression for the period of the revivals, see Equation \eqref{eq:revivaltime} and the surrounding discussion. 

Between the low energy and high energy regimes lies a classical separatrix. We find that the quantum and classical evolution always quickly diverge from each other for motion close to the separatrix. This is true even when in other respects the system is expected to behave classically, i.e.\ $\Lambda$ is large. We interpret this divergence as being due to quantum tunnelling and show that quantum and classical must deviate after one quarter of the classical period. 

A second method for exciting the system is to have a tilt asymmetry which is periodically modulated in time. By tuning the modulation frequency this method allows the excitation of a single eigenstate which is of course a non-classical state. In the quantum limit $\Lambda \rightarrow 0$ this method provides a  way of generating states of the form $(\vert n \rangle \pm \vert -n \rangle)/\sqrt{2}$ where $n$ is the difference in the number of particles between the two wells. When $n$ becomes large these are Schr\"{o}dinger cat states involving macroscopically distinguishable superpositions.

Finally, we have discussed at length the analogy between the asymmetric (tilted) double-well problem and the diffraction of waves by a periodic lattice, including an analysis in terms of band structure familiar from solid state physics. Bragg scattering in the diffraction problem corresponds to tunnelling resonances in the Josephson problem where a precise number of atoms oscillate between the wells in close analogy with the pendell\"osung oscillations. These resonances are not present in the classical Gross-Pitaevskii theory. 
The role of the quasimomentum in the band structure analysis is played by the energy difference between the two wells. Bragg resonances correspond to a sudden tilt of the double wells to a precise final value (the ``Bragg angle'') whereas a slow steady increase of the tilt asymmetry is analogous to Bloch oscillations.

\section{Acknowledgements}

We thank M.J. Farrar, B. Prasanna Venkatesh, S. Giovanazzi, and M.K. Oberthaler for enlightening discussions, as well as two anonymous referees for useful suggestions.
This research was funded by the Natural Sciences and Engineering Research Council of Canada (NSERC). G.K. thanks NSERC for an Undergraduate Student Research Award.

\section*{References}

\begin{thebibliography}{10}

\bibitem{b_josephson}
{Josephson B D 1962 \emph{Phys. Lett.} \textbf {1} 251}

\bibitem{josephson_eff}
{Barone A and Paterno G 1982 \emph {Physics and Applications of the Josephson Effect} (Wiley, New York)}

\bibitem{pereverzev}
{Pereverzev S, Loshak A, Backhaus S, Davis J C and Packard R E 1997 \emph{Nature} \textbf{388} 449}

\bibitem{sukhatme}
{Sukhatme K, Mukharsky Y, Chui T and Pearson D 2001 \emph{Nature} \textbf{411} 280}


\bibitem{cataliotti01}
{Cataliotti F S, Burger S, Fort C, Maddaloni P, Minardi F, Trombettoni A, Smerzi A and Inguscio M 2001\emph{Science} \textbf{293} 843}

\bibitem{shin04}
{Shin Y, Saba M, Pasquini T A, Ketterle W, Pritchard D E and Leanhardt A E 2004 \emph{Phys. Rev. Lett.} \textbf{92} 050405}

\bibitem{albiez05}
{Albiez M, Gati R Foelling J, Hunsmann S, Cristiani M and Oberthaler M K 2005 \emph{Phys. Rev. Lett.} \textbf{95} 010402}

\bibitem{levy07}
{Levy S, Lahoud E, Shomroni I and Steinhauer J 2007 \emph{Nature} \textbf{449} 579}


\bibitem{bose_ein}
Pitaevskii L P and Stringari S 2003
\newblock  \textit{Bose-Einstein Condensation}
\newblock (Clarendon Press, Oxford)


\bibitem{javanainen86}
{Javanainen J 1986 \emph{Phys. Rev. Lett.} \textbf{57} 3164}



\bibitem{dalfovo96}
{Dalfovo F, Pitaevskii L P, Stringari S 1996 \emph{Phys. Rev. A} \textbf{54} 4213}

\bibitem{jack96}
{Jack M W, Collet M J, Walls D F 1996 \emph{Phys. Rev. A}  \textbf{54} R4625 }

\bibitem{smerzi97}
{Smerzi A, Fantoni S, Giovanazzi S and Shenoy R S 1997 \emph{Phys. Rev. Lett.} \textbf{79} 4950}

\bibitem{zapata98}
{Zapata I, Sols F and Leggett A J 1998 \emph{Phys. Rev. A} \textbf{57} R28}

\bibitem{ruostekoski98}
{Ruostekoski J and Walls D 1998 \emph{Phys. Rev. A} \textbf{58} R50}

\bibitem{giovanazzi00}
{Giovanazzi S, Smerzi A and Fantoni S 2000 \emph{Phys. Rev. Lett.} \textbf{84} 4521}

\bibitem{leggettreview01}
{Leggett A J 2001 \emph{Rev. Mod. Phys.} \textbf{73} 307}




\bibitem{giovanazzi08}
{Giovanazzi S, Esteve J and Oberthaler M K 2008 \textit{N. J. Phys} \textbf{10} 045009}


\bibitem{mathfunctions}
{Abramowitz M and Stegun I 1964 \textit{Handbook of Mathematical Functions} 
(National Bureau of Standards, Washington)}

\bibitem{leggettleshouches}
{Leggett A J 1987 in \textit{Chance and Matter}, edited by J. Souletie et al., Les Houches 1986, Session XLVI (North-Holland, Amsterdam)}


\bibitem{gati2}
{Gati R and Oberthaler M K 2007 \textit{J. Phys. B} \textbf{40} R61}

\bibitem{RN35}
{Raman C V and Nagendra Nath N S 1935 \textit{Proc. Ind. Acad. Sci. A} \textbf{2} 406}

\bibitem{berry66}
{Berry M V 1966 \textit{The Diffraction of Light by Ultrasound} (New York, Academic)}

\bibitem{martin87}
{Martin P J, Gould P L, Oldaker B G , Miklich A H, and Pritchard D E 1987 \textit{Phys. Rev. A} \textbf{36} 2495}

\bibitem{odell01}
O'Dell D H J 2001 \textit{J. Phys. A} \textbf{34}, 3897

\bibitem{farrar}
{Farrar M J 2007 \textit{Private communication}}

\bibitem{odell}
{O'Dell D H J 1998 \textit{PhD thesis} unpublished}

\bibitem{hooley}
{Hooley C and Quintanilla J 2004 \textit{Phys. Rev. Lett.} \textbf{93} 080404}

\bibitem{smerzi00}
{Smerzi A and Raghavan S 2000 \textit{Phys. Rev. A} \textbf{61} 063601}

\bibitem{doncheski03}
{Doncheski M A and Robinett R W 2003 \textit{Ann. Phys.} \textbf{308} 578}

\bibitem{styer01}
{Styer D F 2001 \textit{Am. J. Phys.} \textbf{69} 56}







\bibitem{quantum_mech}
{Bransden B H and Joachain C J 1994 \textit{Quantum Mechanics} (John Wiley and Sons, New York)}

\bibitem{nienhuis03}
{Haroutyunyan H L and Nienhuis G 2003  \textit{Phys. Rev. A}  \textbf{67} 053611}

\bibitem{odell98}
{Berry M V and O'Dell D H J 1998 \textit{J. Phys. A} \textbf{31}, 2093}

\bibitem{martin88}
{Martin P J, Oldaker B G, Miklich A H, Pritchard D E 1988 \textit{Phys. Rev. Lett.} \textbf{60} 515}

\bibitem{oberthaler96}
{Oberthaler M K, Abfalterer R, Bernet S, Schmiedmayer J, and Zeilinger A 1996 \textit{ Phys. Rev. Lett.} \textbf{77}, 4980}

\bibitem{shorebook}
{Shore B W 1990 \textit{The Theory of Coherent Atomic Excitation} (Wiley, New York), p1005 }

\bibitem{ashcroft+mermin}
{Ashcroft N W and Mermin N D 1976 \textit{Solid State Physics} (Thomson Learning, New York)}


\bibitem{bloch28}
{Bloch F 1928 \textit{Z. Phys.} \textbf{52}, 555}

\bibitem{zener34}
{Zener C 1934 \textit{Proc. R. Soc. A} \textbf{145}, 523}


\bibitem{holthaus96}
{Holthaus M and Hone D W 1996 \textit{Phil. Mag. B} \textbf{74}, 105 }

\end{thebibliography}

\end{document}